\documentclass[floatfix,reprint,nofootinbib,amsmath,amssymb,aps,
]{revtex4-1}

\usepackage{aas_macros}
\usepackage{graphicx}
\usepackage{dcolumn}
\usepackage{bm}
\usepackage{mathtools}
\usepackage[normalem]{ulem}
\usepackage{xcolor}
\definecolor{colorLink}{rgb}{0.9,0,0} 
\definecolor{colorCite}{rgb}{0,0.7,0} 
\definecolor{colorURL} {rgb}{0,0,0.8} 
\usepackage[colorlinks=true,linktocpage=true,linkcolor=colorLink,citecolor=colorCite,urlcolor=colorURL]{hyperref}

\newcommand{\beq}{\begin{equation}}
\newcommand{\eeq}{\end{equation}}
\newcommand{\beqa}{\begin{eqnarray}}
\newcommand{\eeqa}{\end{eqnarray}}
\def\dng{\delta_{\textsc{n}_\textup{g}}}
\def\ng{\textup{N}_\textup{g}}
\def\nr{\textup{N}_\textup{r}}

\def\d12{\delta_{\textup{I}_{12}}}
\def\C{\textbf{C}}
\def\L{\mathcal{L}}
\def\k{{\bf k}}
\def\p{{\bf p}}
\def\q{{\bf q}}
\def\x{{\bf x}}
\def\e{{\bm \epsilon}}
\def\I{\textup{I}}

\def\ga{\mathrel{\mathpalette\fun >}}
\def\fun#1#2{\lower3.6pt\vbox{\baselineskip0pt\lineskip.9pt
        \ialign{$\mathsurround=0pt#1\hfill##\hfil$\crcr#2\crcr\sim\crcr}}}

\def\kMpc{\, h \, {\rm Mpc}^{-1}}
\newcommand{\Ddel}{\delta_{\rm D}}
\newcommand{\dt}{\boldsymbol{\cdot}}

\begin{document}

\title{Galaxy Power Spectrum Multipoles Covariance in Perturbation Theory}

\author{Digvijay Wadekar}
\email{jay.wadekar@nyu.edu}
\author{Rom\'{a}n Scoccimarro}
\affiliation{Center for Cosmology and Particle Physics, Department of Physics, New York University, New York, NY 10003, USA}

\date{October 7, 2019}

\begin{abstract}

We compute the covariance of the galaxy power spectrum multipoles in perturbation theory, including the effects of nonlinear evolution, nonlinear and nonlocal bias, radial redshift-space distortions, arbitrary survey window and shot noise. We rewrite the power spectrum FKP estimator in terms of the usual windowed galaxy fluctuations and the  fluctuations in the number of galaxies inside the survey volume. We show that this leads to a stronger super-sample covariance than assumed in the literature and causes a substantial leakage of Gaussian information. We decompose the covariance matrix into several contributions that provide an insight into its behavior for different biased tracers. We show that for realistic surveys, the covariance of power spectrum multipoles is already dominated by shot noise and super survey mode coupling  in the weakly non-linear regime. Both these effects can be accurately modeled analytically, making a perturbative treatment of the covariance very compelling. Our method allows for the covariance to be varied as a function of cosmology and bias parameters very efficiently, with survey geometry entering as fixed kernels that can be computed separately using fast fourier transforms (FFTs). We find excellent agreement between our analytic covariance and that estimated from BOSS DR12 Patchy mock catalogs in the whole range we tested, up to $k=0.6$ h/Mpc. This bodes well for application to future surveys such as DESI and Euclid. The \textsc{CovaPT} code that accompanies this paper is available at \href{https://github.com/JayWadekar/CovaPT}{\textcolor{blue}{https://github.com/JayWadekar/CovaPT}}.







\end{abstract}

\maketitle
\section{Introduction}
Large-scale structure surveys provide precise constraints on parameters of cosmological models. In order to extract cosmological constraints from clustering measurements, one needs to have an estimate of their covariance properties, as gravitational clustering, bias, redshift-space distortions, survey geometry, and shot noise lead to non-trivial covariances. 

A common way to estimate the covariance matrix of a quantity is to measure its covariance over an ensemble of numerical  simulations~\cite{ScoZalHui9912,MeiWhi99,HamRimSco0609,TakYosTak0907,LiHuTak1404,BloCorAli1501,VilHahMas1909}. However, a reliable estimate of covariance can be computationally expensive, requiring many thousands of realizations~\cite{TayJoaKit1306,BloCorAli1501,VilHahMas1909}.  Simulating thousands of survey volume sized $\mathcal{O} (\textup{Gpc}^3)$ $N$-body simulations is computationally prohibitive and therefore several faster  algorithms have been developed for making mock galaxy  catalogs~\cite{BonMye9603,ScoShe02,ManScoPer1301,TasZalEis1306,WhiTinMcB1401,KitRodChu1603,IzaCroFos1607,LipSanCol1810,BloCroSef1905,ColSefMon1811}, that typically make approximations at small nonlinear scales.   Furthermore, a numerical approach is prone to sampling noise and this can cause numerical instabilities when inverting covariance matrices to perform likelihood analyses,  as a result often the uncertainties in the final constraints need to be artificially inflated~\cite{HarSimSch0703,DodSch1309,PerRosSan1404,SelHea1602}. As volume of upcoming surveys such as LSST~\cite{LSST_Aba1211}, DESI~\cite{DESI_Agh1611}, Euclid~\cite{Euclid_Ame1206}, WFIRST~\cite{WFIRST_Dor1804} increases, we need simulations with even larger volumes. Hence it is very desirable to find alternative ways of computing covariance matrix elements.  
 
There are some approaches in the literature that attempt to bypass the need to simulate survey volume sized mock catalogs. One type of methods try to obtain the covariance matrix directly from survey data. Such methods either involve splitting the data into smaller samples or use techniques like bootstrap or jackknife directly on the data. However, sub-sampling the data does not properly capture the influence of super-survey modes on the data and the sub-samples are not entirely independent which leads to biases in the covariance~\cite{NorBauGaz0906,FriSeiEif1603,LacKun1708}.  In addition, some semi-analytic procedures try to estimate the covariance matrix from fewer mocks or data~\cite{PopSza0809,HowPer1712,PeaSam1603,OcoEisVar1611,Joa1703,HalTay1902,OcoEis18,PhiEis19}, but such methods do not clearly take into account  all the physical effects that affect galaxy clustering. 

On the other hand, the physical ingredients that generate covariance are by now theoretically well understood: a dominant Gaussian contribution at large scales, survey geometry characterized by its  window function, shot noise,  non-Gaussian clustering contributions that couple the scales of interest among themselves~\cite{ScoZalHui9912,BerSchSol1606,BerSchSol1606b,MohSelVla1704,TarNisJeo20,HarPen12,HarPen13} and non-Gaussian clustering contributions that couple the scales of interest to Fourier modes of wavelengths longer than the survey size~\cite{HamRimSco0609,SefCroPue0607,PutWagMen1204,TakHu1306}.

An analytic approach to the covariance matrix does not have all the aforementioned drawbacks but it involves two major challenges. The first is to treat the highly non-trivial 3D survey window of a realistic redshift-space galaxy survey. The calculations involving the survey window are further complicated by the fact that the line-of-sight (LOS) cannot be assumed to be fixed for a wide angle survey. The second challenge involves properly describing non-Gaussianity because the covariance of the power spectrum (a two point function) involves a four point function (the trispectrum)  and its calculation  involves modeling non-linear gravitational evolution, galaxy bias, redshift-space distortions and shot noise. In addition, these two challenges are coupled to each other, resulting in high-dimensionality in the integrals one needs to compute. 

However there is a regime in which these challenges are less daunting than it seems at first sight: when galaxies are not too far from the Poisson dominated limit and result from  the evolution of Gaussian primordial fluctuations. In this case, which fortunately is a very good approximation in practice, most of the signal that is extracted from galaxy surveys originates at  scales smaller than the survey size where clustering is stronger (unlike fluctuations that evolve from local primordial non-Gaussianity~\cite{DalDorHut0806}),  all the way down to scales that become  dominated by shot noise. The crucial quantity is therefore the density of the sample (characterized by the number density $\bar{n}$) in units of the clustering strength (characterized by the power spectrum $P$), i.e. the dimensionless $\bar{n}P$ combination. 

Current and upcoming redshift surveys are typically designed such that $ \bar{n}P \sim$~few at the BAO scale (see e.g. Fig.~2 of \cite{FonMcDMos1405}). As a result of this, as scales enter in the nonlinear regime they also become shot noise dominated, $ \bar{n}P<1$. This in turn greatly simplifies an analytic approach since complicated physics like nonlinear evolution and velocity dispersion  become subdominant compared to shot noise. On the other hand, shot noise contributions are straightforward to describe  in the Poisson approximation from the knowledge of the radial selection function of the survey. 

The dominance of shot noise makes the calculation of the covariance matrix using Perturbation Theory (PT) very compelling, since tree-level predictions may suffice if loop corrections, when important, are overwhelmed by shot noise. In addition, the dimensionality of the integrals in the PT calculation can be reduced drastically at scales smaller than the survey size, where the cosmology dependence can be decoupled from the survey geometry. This makes the use of an analytic covariance extremely efficient for likelihood analyses. Also, having an analytic covariance in equal footing with the mean of the observable being  predicted allows for a self-consistent likelihood analysis, where the covariance can be changed as wished in terms of cosmological and bias parameters (something that is computationally prohibitive using the traditional technique of mock catalogs). Finally, the analytic approach provides a physical insight into the relative contribution of various effects.
 
We will focus on the covariance of multipoles of the galaxy power spectrum in this work, providing a similar treatment of the bispectrum covariance in a forthcoming paper. We present a first attempt to calculate the full (diagonal and non-diagonal) covariance matrix of power spectrum multipoles including nonlinear evolution, galaxy bias, and radial redshift distortions up to cubic order in (tree-level) PT. A significant amount of work has already been done along these lines in the literature, most notably on the effect of  coupling between long-wavelength survey-size modes and the short modes of interest which can dominate the extra diagonal elements of the covariance. This was first pointed out in~\cite{HamRimSco0609} (who called the effect, \emph{beat-coupling}, hereafter BC) and ~\cite{SefCroPue0607} showed its importance in cosmological parameter estimation from joint analyses of the power spectrum and bispectrum. It was pointed out in~\cite{PutWagMen1204} that the contribution of  beat modes to the covariance is suppressed if density fluctuations are normalized by the mean overdensity in the survey region, and called this suppression the \emph{local-average effect}. 

Both effects were later reformulated by~\cite{TakHu1306} in terms of the response of the power spectrum to the long modes and this approach was coined as the \emph{super-sample covariance} (SSC). The effect of the beat modes was quantified in~\cite{LiHuTak1404,LiHuTak1411,BalSelSen1609,BarNelPil1904}   using separate universe simulations, while~\cite{AkiTakLi1704,AkiTak1803,LiSchSel1802} extended the SSC approach to include effects of large-scale tidal fields. Although in this paper we limit ourselves to modeling the covariance for spectroscopic surveys, it is worth mentioning some of the recent work on building analytic models for photometric surveys for the covariance of various projected observables like the galaxy angular power spectrum, weak lensing and cluster counts \cite{TakSpe1407, SchTakSpe1412, KraEif1709, HikOguHam1903, BarKraSch1806, BarSch1706, BarSch1709, LacRos1608,LacLimAgu18, SinMan17, Lac1806,LacGra19,Lac20,LeeDvo2001}. Because photometric surveys do not suffer from redshift-space distortions but are sensitive to smaller scales, the challenges are different from galaxy redshift surveys. But interestingly the response function approach can be extended to the fully non-linear regime by inferring the responses from measurements in separate universe simulations, see e.g.~\cite{BarNelPil1904}. This can potentially be used to model the covariance of cross observables in spectroscopic and photometric surveys.

The outline of the paper is as follows. We start with the FKP estimator and define the notations used in Sec. \ref{sec:notation}. For simplicity, we begin our discussion of the power spectrum covariance in real space in Sec.~\ref{sec:RealSpaceFormalism}, where we first look at the effects of the survey window. After that we present the calculation of the trispectrum in Sec.~\ref{sec:Trispectrum} and introduce the local average effect in Sec.~\ref{sec:LA}. We then move to redshift-space in Sec.~\ref{sec:RedshiftSpaceFormalism}, where we first model the Gaussian covariance in Sec.~\ref{sec:GaussCova} and then compute the non-Gaussian contributions by generalizing the beat-coupling and the local-average effects in sections~\ref{sec:RSD_LA} and~\ref{sec:RSD_Trispectrum} respectively. We then introduce discreteness and recalculate our redshift-space results in Sec.~\ref{sec:ShotNoise}, highlighting the choice of the true shot noise estimator as opposed to that in FKP in Sec.~\ref{sec:TrueSN}. We compare our approach to others in the literature in Sec.~\ref{sec:Literature_comparison}. Finally we compare our results to the MultiDark-Patchy mock catalog measurements in Sec.~\ref{sec:Patchy_compare}. We conclude in Sec.~\ref{sec:Conclusions}. 

A number of appendices present supplementary and in some cases derive the more technical results presented in the main text: in Appendix~\ref{apx:RSDkernels}, we discuss the PT redshift-space kernels used in our tree-level calculation;  in Appendix \ref{apx:SN_exact_gaussian}, we derive the exact equations for the shot noise contribution to the Gaussian covariance; in Appendix~\ref{apx:LOS_BeatTerms}, we derive the effect of  beat modes on the covariance using radial redshift-space distortions; in Appendix~\ref{apx:FKP_SN_analytic} we estimate the extra covariance of the FKP shot noise estimator; in Appendix~\ref{apx:CovaError}, we estimate the error on the covariance elements analytically; in Appendix~\ref{apx:LowZ_comparison}, we extend the comparison of our predictions to the MultiDark-Patchy mocks to the low-$z$ bin as opposed to the high-$z$ bin presented in the main text.

\section{Basics}
\subsection{Notation}
\label{sec:notation}
To simplify the equations in this paper, we use the following shorthand notation for configuration space (using variable \x) and Fourier-space (using any variable other than \x) integrals
\begin{equation}
\int_{\x} \rightarrow \int d^3 \x\, , \ \ \ \ \ \ \ \ \int_{\k} \rightarrow \int \frac{d^3 \k}{(2\pi)^3}\, .
\end{equation}
When we bin over the $i^\textup{th}$ spherical \k-shell with width $\Delta k$ and bin center at $k_i$ and  volume $V_{k_i}=\frac{4\pi}{3}[(k_i+\frac{\Delta k}{2})^3-(k_i-\frac{\Delta k}{2})^3]$. The average over the k-shell is denoted by
\begin{equation}
\int_{\hat{\k}_i} \rightarrow \int \frac{d^3\k}{V_{k_i}}\, .
\label{eq:ShellAvg}\end{equation}
Unless otherwise noted, we do not use the thin-shell approximation, in which the average over the shell is taken to be equal to the solid-angle average. 
For definiteness, we use, as an example of a realistic redshift survey, the high-$z$ bin ($0.5\leq z < 0.75$) of the SDSS-BOSS DR12 NGC survey window for all the results in this paper, and also show results for the low-$z$ bin ($0.25\leq z < 0.5$) in Appendix~\ref{apx:LowZ_comparison}. The survey window has no symmetry at all, but roughly speaking has the shape of a  cone truncated due to the radial selection function of a given redshift bin and an azimuthal dependence based on the survey mask \cite{ReiHoPad16}.
We now introduce the following notation 
\begin{equation}
W_{ij}(\x) \equiv \bar{n}^i(\x) w^j(\x),\ \ \  \ \ \ \I_{ij} \equiv \int _{\x} \bar{n}^i(\x)w^j(\x)\, ,
\label{WandI}
\end{equation}
where $\I_{ij}$ can be interpreted as the window $W_{ij}$ normalization factors. The window functions in Fourier space $W_{ij}(\k) \equiv \int_{\x} W_{ij}(\x) e^{-i \k\cdot \x}$
obey the following identities which will be used extensively in this paper
\begin{subequations}
\begin{equation}
\int_{\textbf{q}'}  W_{11}(\textbf{q}') W_{11}(\textbf{q}-\textbf{q}') = \int_{\x} W_{22}(\x) e^{-i \textbf{q}\cdot\x} = W_{22}(\textbf{q}),
\end{equation}
\begin{equation}\begin{split}
&\int_{\textbf{q}'} \textbf{q}' W_{11}(\textbf{q}') W_{11}(\textbf{q}-\textbf{q}') = \frac{1}{i}\int_{\x} (\nabla W_{11}(\x)) W_{11}(\x) e^{-i \textbf{q}\cdot\x}\\
&=\frac{1}{i} \int_{\x}\frac{1}{2} (\nabla W^2_{11}(\x)) e^{-i \textbf{q}\cdot\x}=\frac{1}{2}\textbf{q} W_{22}(\textbf{q}).
\end{split}\end{equation}
\label{eq:WindowIdentities}
\end{subequations}
These identities generalize the results of \cite{TakHu1306} to go beyond a simple top-hat window function. 

\subsection{FKP estimator}
Because of selection effects, redshift surveys do not have a spatially-independent galaxy mean density. One must therefore carefully define local density fluctuations whose power spectrum one is interested in. For this purpose, we use throughout the FKP estimator~\cite{FelKaiPea9405} corresponding to a galaxy survey where a particular galaxy catalog has $\ng$ galaxies and the radial and angular selection functions are characterized by a random catalog with $\nr$ objects (typically $\nr/\ng \gg 1$ to reduce shot noise in the random catalog).
For simplicity, let us start with the case of continuous fields and we will introduce discreteness later in Sec.~\ref{sec:ShotNoise}. 
Using the  survey selection function $\bar{n}(\x)$ and the FKP  weight  $w(\x)$, 
we can write a continuous form of the standard FKP estimator for the galaxy overdensity as
\begin{equation}
\hat{\delta}^\textup{FKP}(\x) \equiv \frac{w(\x)[n_g(\x)-\alpha n_r(\x)]}{\big[\alpha \int_{\x'}\, w^2(\x') \bar{n}(\x')n_r(\x')\big]^{1/2}}\, ,
\label{FKP1}
\end{equation}
where $n_g$ and $n_r$ are the number densities of the particular galaxy catalog and its random catalog respectively, $\alpha\equiv \ng/\nr$ and
\beq
w(\x)=\frac{1}{1+\bar{n}(\x)P_0}
\label{eq:weight_FKP}\eeq
is the FKP weight function obtained by minimizing the power spectrum variance in the limit of Gaussian fluctuations at scales which are smaller than the size of the survey~\cite{FelKaiPea9405}.

At this stage it is worth mentioning a couple of points about Eq.~(\ref{FKP1}). First, the normalization given by its denominator is chosen to convert the power spectrum of the numerator into an unbiased estimate of the power spectrum of density fluctuations (with some caveats that we shall discuss below). This denominator is in practice evaluated by summing over the discrete objects in the random catalog, converting $\int d^3x\, n_r(\x) \to \sum_r$. Second, it is important to note that since $\alpha\equiv \ng/\nr$, $\alpha$ itself is a random variable (containing modes longer than the survey size responsible for $\ng$ being different in equivalent survey volumes): in the numerator $\alpha$ guarantees that the estimator has zero average\footnote{More precisely zero average is obtained when $\alpha$ is defined as $\alpha = \int_{\x} w\, n_g \ / \int_{\x} w\, n_r$, which in turn changes $\dng$ in Eq.~(\ref{deltaNg}) by $W_{10}\to W_{11}$ and $I_{10}\to  \I_{11}$. In practice this redefinition of $\alpha$ is very close to $\ng/\nr$, which we adopt in this paper.} (as it must be since the galaxy mean density is estimated from the data itself), while in the denominator it scales the sum over the 
\clearpage
 random catalog to the actual density observed\footnote{The typical procedure of estimating the power spectrum covariance is to use thousands of simulated galaxy catalogs but only a single random catalog and rescale the random catalogue by the factor $\alpha\equiv \ng/\nr$ to match the number density of the corresponding galaxy catalogues. A more realistic procedure would instead be to simulate a separate random catalogue for each of the simulated galaxy catalogues, in which case no such rescaling by $\alpha$ is necessary (see \cite{deMRuh1908}). This is however never done in practice for calculating covariance but it can change the contribution of super-survey modes and will be discussed in a future paper.}. The latter point has important consequences in the covariance matrix, as it renormalizes the effect of super-survey modes on the power spectrum covariance.

To make the connection of $\alpha$ to the super-survey modes explicit, we use that $\alpha\, n_r(\x)$ differs from the selection function $\bar{n}(\x)$ precisely by such modes described by the density perturbation (weighted by the selection function) at the survey scale $\dng$ 

\begin{equation}
\alpha\, n_r(\x) = \ \bar{n}(\x) \frac{\int_{\x'} \bar{n}(\x')(1+\delta(\x'))}{\int_{\x'} \bar{n}(\x')} \equiv \ \bar{n}(\x) \left(1+\dng \right).
\end{equation}
Using Eq.~(\ref{WandI}), the density perturbation at the survey scale then reads
\beq
\dng \equiv \frac{\int_{\x} \delta(\x) W_{10}(\x)}{\I_{10}}
\label{deltaNg}
\eeq

Note that in a redshift survey $\dng$ corresponds to the {\em galaxy density perturbation in redshift space}, thus $\dng$ includes nonlinear evolution, bias, shot noise and anisotropies due to redshift-space distortions. Therefore the FKP estimator can be written in a compact form,
\begin{equation}
\begin{split}
&\hat{\delta}^\textup{FKP}(\x)= 
\frac{1}{\sqrt{\I_{22}}}\frac{W_{11}(\x) \delta(\x)-W_{11}(\x)\dng}{(1+\dng)^{1/2}}.
\end{split}\label{eq:1.1}\end{equation}
The term $W_{11}(\x)\dng$ has only Fourier content at very low-$k$ (comparable to the survey size) and as such we will neglect it in what follows. This term gives rise to the well-known integral constraint~\cite{PeaNic9111,TegHamStr98,BeuSeoSai1704} and it is negligible at the scales where we are interested in computing an accurate covariance matrix  ($k \ga 0.01 \kMpc$ for the BOSS DR12 sample we consider here), although these contributions are fairly easy but cumbersome to add to our expressions derived below.

 As we encounter the product $ W_{11}(\mathbf{x})\delta(\mathbf{x})$ frequently in our expressions, we use the shorthand, 
 \beq
 \delta_W(\mathbf{x}) \equiv W_{11}(\mathbf{x})\delta(\mathbf{x})
 \label{deltaW}
 \eeq
 for the windowed galaxy fluctuations. Using all the defined notation, the FKP estimator finally reads in our approximation to Eq.~(\ref{eq:1.1}),
\begin{equation}
\hat{\delta}^\textup{FKP}(\x) \simeq \frac{1}{\sqrt{\I_{22}}} \frac{\delta_W (\x)}{(1+\dng)^{1/2}}\, ,
\label{eq:delta_FKP}
\end{equation}
Our strategy to compute the power spectrum covariance matrix now becomes fairly straightforward. We just calculate the covariance of the power spectrum estimator corresponding to Eq.~(\ref{eq:delta_FKP}). If we were to ignore the effect of fluctuations in $\ng$ ($\dng=0$), it simply corresponds to computing the windowed trispectrum on top of the standard Gaussian covariance.

\section{Covariance in Real Space}
\label{sec:RealSpaceFormalism}
We first discuss the calculation of the power spectrum covariance in real space to simplify the understanding of the various effects contributing to it. We then include redshift-space distortions of all these effects in the next Sec.~\ref{sec:RedshiftSpaceFormalism}, before we add shot noise in Sec.~\ref{sec:ShotNoise}.
\subsection{Window convolved Power and Covariance}
In the absence of survey windows, the density field is statistically homogeneous and the Fourier space two-point function becomes

\begin{equation}
\langle \delta(\mathbf{k}_1)\, \delta(\mathbf{k}_2)\rangle = (2\pi)^3 \delta_D(\mathbf{k}_1+\mathbf{k}_2) P(\mathbf{k}_1),
\end{equation}
where $P(\k_1)$ is the true power spectrum. This motivates the estimator of the power spectrum monopole $\widehat{P}(k)$ to be written as a shell average in the form, 
\begin{equation}
\widehat{P}(k)\propto \int_{\hat{\k}} \delta (\k) \delta (-\k)\ .
\end{equation}
Since we work in real space in this section, we only consider the power spectrum monopole here and discuss higher-order multipoles in redshift space in the next section. In the presence of a survey window, the continuous form of the FKP monopole power spectrum estimator is
\begin{equation}
\widehat{P}^\textup{FKP} (k) \equiv\int_{\hat{\k}} |\delta^\textup{FKP}(\k)|^2 = \frac{1}{\I_{22}}\int_{\hat{\k}} \frac{|\delta_W(\k)|^2 }{(1+\dng)}\ .
\label{eq:trash_P_FKP}
\end{equation}

Here we momentarily ignore $\ng$ fluctuations, setting $\dng=0$ in Eq.~(\ref{eq:trash_P_FKP}), postponing the more general case including the effects of $\dng$ to Sec. \ref{sec:LA}. In this simpler case the power spectrum estimator is just
\beq
\widehat{P}_W(k) = \frac{1}{\I_{22}}\int_{\hat{\k}} |\delta_W(\k)|^2,
\label{PWnoNg}
\eeq
and using Eq.~(\ref{deltaW}) in Fourier space the expectation value of the estimator is
\begin{equation}
\begin{split}
\langle \widehat{P}_W(k)\rangle =  &\frac{1}{\I_{22}}\int_{\hat{\k}}\int_{\k'}\, \left | W_{11}(\k') \right |^2 P(\k-\k')\\
\cong &\frac{1}{\I_{22}}\int_{\hat{\k}}\, P(k)\ \int_{\k'}\, \left | W_{11}(\k') \right |^2\\
= &P(k)\, \frac{1}{\I_{22}} \int_{\k'} \left | W_{11}(\k') \right |^2 = P(k)\ ,
\end{split}
\label{eqn:convolved_power}
\end{equation}
where we assumed that the power spectrum $P(\k)$ varies slowly inside the bin $k$ and we used $\I_{22}  \equiv \int_{\x} W_{11}^2(\x)=\int_{\k} |W_{11}(\k)|^2$. We also assumed that the radius of the bin is larger than the Fourier space window size ($|\mathbf{k}'|\ll |\k|$). In other words, since we are interested in $k$ that corresponds to scales much smaller than the survey size, we ignore convolution of the power spectrum with the survey window throughout this paper (although this convolution effect is important to accurately model the power spectrum at low-$k$ \cite{BeuSeoSai1704}, ignoring this effect is a good approximation to model the covariance and greatly simplifies our expressions).

Therefore the covariance from the simplified power spectrum estimator in Eq.~(\ref{PWnoNg}) is simply
\begin{equation}
\textbf{C}(k_1,k_2)\ \equiv \ \langle \widehat{P}_W(k_1)\widehat{P}_W(k_2) \rangle - \langle \widehat{P}_W(k_1) \rangle\langle\widehat{P}_W(k_2) \rangle\ ,
\end{equation}
where
\begin{equation}
\begin{split}
\langle& \widehat{P}_W(k_1)\widehat{P}_W(k_2) \rangle\\
&= \frac{1}{\I_{22}^2} \int_{\hat{\k}_1,\hat{\k}_2} \langle \delta_W(\mathbf{k}_1) \delta_W(-\mathbf{k}_1) \delta_W(\mathbf{k}_2) \delta_W(-\mathbf{k}_2)\rangle \\
&= \frac{1}{\I_{22}^2} \int_{\hat{\k}_1,\hat{\k}_2,\p_1,\p'_1,\p_2,\p'_2} W_{11}(\mathbf{k}_1-\p_1) W_{11}(-\mathbf{k}_1-\p'_1)\\ 
& \times W_{11}(\mathbf{k}_2-\p_2) W_{11}(-\mathbf{k}_2-\p'_2) \langle \delta(\p_1) \delta(\p'_1) \delta(\p_2) \delta(\p'_2)\rangle.
\end{split}
\label{eqn:cova_simplest}
\end{equation}
The covariance can now be decomposed into its Gaussian (denoted by G) and non-Gaussian (due to the trispectrum, denoted by T)  parts as
\begin{equation}
\textbf{C}(k_1,k_2)\ = \textbf{C}^\textup{G}(k_1,k_2)+\textbf{C}^\textup{T}(k_1,k_2)\, ,
\end{equation}
where
\beqa
\textbf{C}^\textup{G}(k_1,k_2) = \frac{2}{\I_{22}^2}&\ & \int_{\hat{\k}_1,\hat{\k}_2} \langle \delta_W(\mathbf{k}_1) \delta_W(\mathbf{k}_2)\rangle \nonumber \\
& &\ \ \ \ \ \  \times \langle \delta_W(-\mathbf{k}_1) \delta_W(-\mathbf{k}_2)\rangle,
\label{CGrealspace}
\eeqa
and
\begin{equation}
\textbf{C}^\textup{T}(k_1,k_2) = \frac{1}{\I_{22}^2} \int_{\hat{\k}_1,\hat{\k}_2} \langle \delta_W(\mathbf{k}_1) \delta_W(-\mathbf{k}_1) \delta_W(\mathbf{k}_2) \delta_W(-\mathbf{k}_2)\rangle_c.
\label{eq:C_trisp_real}
\end{equation} 
The Gaussian part primarily contributes to the diagonal elements of the covariance matrix and the trispectrum part primarily to the non-diagonal elements.
A detailed calculation for the Gaussian covariance will be discussed in redshift space where it is less trivial (see~Sec.~\ref{sec:GauCova}), we first discuss the trispectrum part in what follows.
\subsection{Non-Gaussian Covariance: Trispectrum Contribution}
\label{sec:Trispectrum}
We write the connected 4-point term in Eq.~(\ref{eq:C_trisp_real}) in terms of the trispectrum as
\vspace{-0.1cm} 
\begin{widetext}
\onecolumngrid
\begin{equation}
\begin{split}
\textbf{C}^\textup{T}(k_1,k_2) = &\frac{1}{\I_{22}^2}\int_{\hat{\k}_1,\hat{\k}_2,\p_1,\p'_1,\p_2,\p'_2} W_{11}(\p_1) W_{11}(\p'_1)  W_{11}(\p_2)  W_{11}(\p'_2)  \\
&\ \ \ \ \ \ \ \ \ \ \ \ \ \ \ \ \ \ \ \ \ \ \ \ \ \times (2\pi)^3 \delta_D^3(\p_1+\p'_1+\p_2+\p'_2)\ T(\k_1 -\p_1,-\k_1 -\p'_1,\k_2 -\p_2,-\k_2 -\p'_2)\ .
\end{split}
\end{equation}
\begin{figure}
    \centering
        \includegraphics[scale=0.43,keepaspectratio=true]{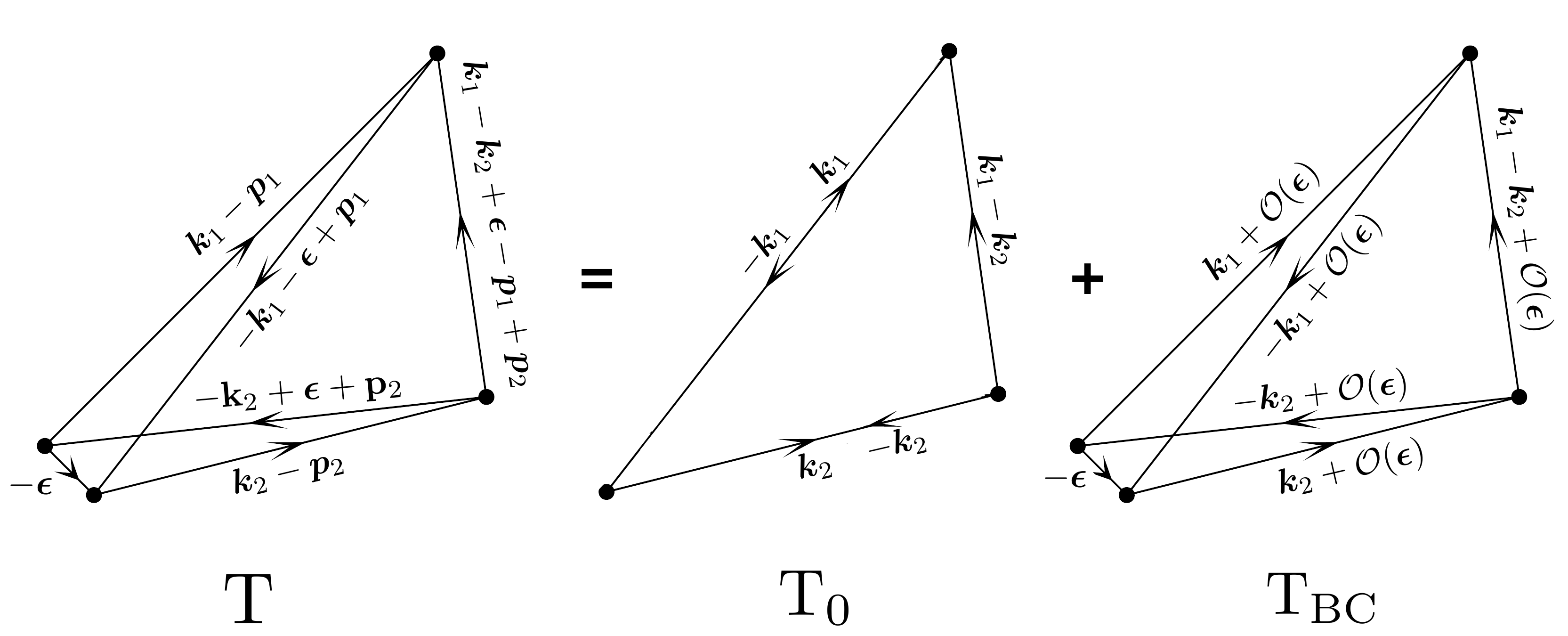}
    \caption{The trispectrum contribution to the covariance can be split into a regular part T$_0$ (only contribution in an infinite universe or for periodic boundary conditions, Eq.~\ref{T0def}) and a beat-coupling (BC) part T$_\textup{BC}$ (Eq.~\ref{TBCdef}) that describes the coupling of small-scale modes $\k_i$ to the large-scale mode $\e$.}
    \label{fig:BC}
\end{figure}
The fourier modes that are arguments of window functions are restricted to survey scales. It is therefore convenient to 
introduce a wave-vector that characterizes such modes, known as {\em the beat mode}~\cite{HamRimSco0609}, $\e \equiv (\p_1+\p'_1) = -(\p_2+\p'_2)$ and write


\begin{equation}
\textbf{C}^\textup{T}(k_1,k_2) = 
\frac{1}{\I_{22}^2}\int_{\hat{\k}_1,\hat{\k}_2,\e,\p_1,\p_2} W_{11}(\p_1)W_{11}(\p_2)W_{11}(\e-\p_1)W_{11}(-\e-\p_2)\ 
T(\k_1 -\p_1,-\k_1 -\e+\p_1,\k_2 -\p_2,-\k_2 +\e+\p_2).
\label{eq:trisp_real}
\end{equation}

\end{widetext}
  \twocolumngrid
Fig.~\ref{fig:BC} illustrates the importance of the beat mode $\e$, which characterizes the non-zero volume of the tetrahedron describing the six magnitudes on which the trispectrum depends on~\cite{HamRimSco0609}. 


 Setting such beat modes to zero collapses the tetrahedron
  to two coplanar triangles in which only
  small-scale modes play a role: this is the `regular' trispectrum contribution $T_0$ which is the only contribution that survives in the infinite volume limit or in the case of periodic boundary conditions. Expanding in the beat modes generates the coupling to large-scales ({\em beat coupling}, denoted by $T_{\rm BC}$), which describes that our estimator of the power spectrum in Eq.~(\ref{PWnoNg}) does know about modes much longer than $k$ due to convolution with the survey window. Such trispectrum contributions can actually dominate the non-Gaussian covariance.

Another way of thinking about $T_{\rm BC}$~\cite{SefCroPue0607} is that in a finite volume of size $\epsilon^{-1}$ it is impossible (due to the uncertainty principle) to determine $k$ to better than $\epsilon$, thus our $\widehat{P}_W$ estimator at $k$ has true modes in the range $k\pm \epsilon$ and thus its covariance will necessarily have contributions through second-order in PT coming from $P(\epsilon)$. The terms in the trispectrum proportional to $P(\epsilon)$ define  $T_{\rm BC}$ and through Eq.~(\ref{eq:trisp_real}) the beat-coupling covariance $\textbf{C}^\textup{BC}$.
 
In this paper, we will only consider tree-level PT contributions to the trispectrum, neglecting loop corrections which become important at small scales where shot noise is dominant for galaxy redshift survey applications, as we shall see below. In tree-level PT the trispectrum reads~\cite{Fry84}

\begin{widetext}
\begin{equation}
\begin{split}
 T(\k_1 ,&\k_2,\k_3,\k_4)\\ =&\ 4\, Z_1(\k_1)P_{\rm L}(\k_1)Z_1(\k_2)P_{\rm L}(k_2) P_{\rm L}(\k_{13})Z_2 (\k_1,-\k_{13})Z_2 (\k_2,\k_{13})+ \textup{cyclic (12 snake terms)}\\
 &+ Z_1(\k_1)P_{\rm L}(\k_1)Z_1(\k_2) P_{\rm L}(\k_2)Z_1(\k_3) P_{\rm L}(\k_3)[Z_3(\k_1,\k_{2},\k_3)+\textup{perm. (6 terms)}]+ \textup{cyclic (4 star terms)}\, ,
\end{split}
\label{trispPT}
\end{equation}
where $P_{\rm L}$ is the linear {\em matter} power spectrum and the kernels $Z_n$~\cite{ScoCouFri9906} are presented in Appendix~\ref{apx:RSDkernels} in the most general redshift-space version (however, setting $f=0$ yields the real-space kernels, in which case e.g. $Z_1=b_1$, with $b_1$ the linear bias parameter). The regular trispectrum is given by
\beqa
T_0 \equiv T(\k_1 ,-\k_1,\k_2,-\k_2) &=& \bigg[8P_{\rm L}^2(\k_1)Z_1^2(\k_1)P_{\rm L}(\k_1+\k_2) Z_2^2(-\k_1,\k_1+\k_2)+ (\k_1\leftrightarrow \k_2)\bigg] \nonumber \\
&& +16P_{\rm L}(\k_1)Z_1(\k_1)P_{\rm L}(\k_2)Z_1(\k_2)P_{\rm L}(\k_1+\k_2)Z_2(-\k_1,\k_1+\k_2) Z_2(-\k_2,\k_1+\k_2) \nonumber \\
&& + \bigg[12\, Z_1^2(\k_1) P_{\rm L}^2(\k_1) Z_1(\k_2) P_{\rm L}(\k_2)Z_3(\k_1,-\k_1,\k_2) + (\k_1\leftrightarrow \k_2)\bigg]
\label{T0def}
\eeqa
and its corresponding covariance $\textbf{C}^{\textup{T}_0}$ follows by just plugging  this result into Eq.~(\ref{eq:trisp_real}), giving the usual regular tree-level non-Gaussian covariance~\cite{ScoZalHui9912}. On the other hand, the beat-coupling trispectrum is (with $P(\epsilon)=b_1^2 P_{\rm L}(\epsilon)$ at tree-level)
\beqa
T_{\rm BC} &\simeq& 4 P(\epsilon)\, \bigg[P_{\rm L}(\k_1-\p_1) Z_2(\k_1-\p_1,\e)+ P_{\rm L}(-\k_1-\e+\p_1) Z_2(-\k_1-\e+\p_1,\e)\bigg] 
\nonumber \\ & & \ \ \ \times \, 
\bigg[P_{\rm L}(\k_2-\p_2) Z_2(\k_2-\p_2,-\e)+ P_{\rm L}(-\k_2-\e+\p_2) Z_2(-\k_2+\e+\p_2,-\e)\bigg]\ \ \ \ \ 
\label{TBCdef}
\eeqa
Upon substitution of Eq.~(\ref{TBCdef}) into Eq.~(\ref{eq:trisp_real}), it follows that the second term in the first square bracket of Eq.~(\ref{TBCdef}) is the same as the first term in the first square bracket upon the change of variables $\k_1\rightarrow-\k_1$ \& $\p_1 \rightarrow \e-\p_1$. Following the same procedure for the second square bracket of Eq.~(\ref{TBCdef}) we obtain 
\beqa
\textbf{C}^\textup{BC}(k_1,k_2)&= &\frac{16}{\I_{22}^2}\int_{\e} P(\epsilon)
\bigg \{\int_{\hat{\k}_1,\p_1}W_{11}(\p_1)W_{11}(\e-\p_1) P_{\rm L}(\k_1-\p_1) Z_2(\k_1-\p_1,\e) \bigg \} \nonumber \\
&&\times \ \bigg \{ \int_{\hat{\k}_2,\p_2} W_{11}(\p_2) W_{11}(-\e-\p_2)  P_{\rm L}(\k_2-\p_2) Z_2(\k_2-\p_2,-\e) \bigg \}\ .
\label{eq:1}
\eeqa
Since the structure within curly brackets appears frequently, let us work it out in some detail. We cannot naively ignore the wave-vectors $\e, \p_i$ because of the poles in the kernel $F_2(\k_i-\p_i, \e)$ inside $Z_2$~\cite{LiHuTak1404}; let us explicitly show the solution of one of the terms to illustrate this point
\begin{equation}
\begin{split}
&\bigg \{\int_{\hat{\k}_1,\p_1}W_{11}(\p_1)W_{11}(\e-\p_1) P_{\rm L}(\k_1-\p_1) F_2(\k_1-\p_1,\e)\bigg \}\\
&\simeq \int_{\hat{\k}_1,\p_1} W_{11}(\p_1) W_{11}(\e-\p_1) P_{\rm L}(k_1)\left(1-\frac{\partial\ln P_{\rm L}}{\partial\ln k_1}\frac{\k_1\cdot \p_1}{k_1^2}\right) \left[\frac{5}{7}+\frac{1}{2}(\k_1\cdot\e-\p_1\cdot\e)\left(\frac{1}{\epsilon^2}+\frac{1}{k_1^2}\right)+\frac{2}{7}\frac{(\k_1\cdot\e-\p_1\cdot\e)^2}{\epsilon^2k_1^2} \right]\\
& =P_{\rm L}(k_1)\int_{\p_1} W_{11}(\p_1) W_{11}(\e-\p_1) \left[\frac{17}{21}-\frac{1}{2}\frac{\p_1\cdot\e}{\epsilon^2}-\int_{\hat{\k}_1} \frac{1}{2}\ \frac{\partial\ln P_{\rm L}}{\partial\ln k_1}\ \frac{(\k_1\cdot\p_1)(\k_1\cdot\e)}{k_1^2\epsilon^2}\right]\ ,
\end{split}
\label{eq:BC_noRSD1}
\end{equation}
where we used the standard expression for the  $F_2$ kernel (see Eq.~\ref{eq:ModeCouplingKernels}) and neglected higher order terms such as $\mathcal{O} (p_i/k_i)$ and $\mathcal{O} (\epsilon/k_i)$. This stems from our assumption that the window size is much larger than the size of the modes being measured ($\sim 1/k_i$) and therefore $|\p_i|\ll k_i$ and $|\e|\ll k_i$ as $\p_i$ \& $\e$ appear as arguments of window functions. This means that for biased tracers in real space we have (after doing the  solid-angle integration over $\hat{\k}_1$)
\begin{equation}
\begin{split}&
\bigg \{\int_{\hat{\k}_1,\p_1}W_{11}(\p_1)W_{11}(\e-\p_1) P_{\rm L}(\k_1-\p_1) Z_2(\k_1-\p_1,\e)\bigg \}
\\&
\simeq P_{\rm L}(k_1)\int_{\p_1} W_{11}(\p_1) W_{11}(\e-\p_1) \left[\frac{17}{21}b_1+\frac{b_2}{2}-\frac{2}{3}\gamma_2-\frac{b_1}{6}\frac{\p_1\cdot\e}{\epsilon^2} \ \gamma(k_1)\right]\ ,
\end{split}
\label{eq:BC_noRSD1b}
\end{equation}
\end{widetext}
where $b_2$ and $\gamma_2$ describe quadratic bias and $\gamma(k)$ characterizes the spectral index $n_{\rm eff}$
\beq
\gamma(k)\equiv \frac{d \ln k^3P_{\rm L}}{d \ln k}(k) \equiv n_{\rm eff}(k) +3 
\label{gammaEff} 
\eeq
We now use the window identities in Eq.~(\ref{eq:WindowIdentities}) to get 
\begin{equation}
\begin{split}
\bigg \{ \int_{\hat{\k}_1,\p_1}&W_{11}(\p_1)W_{11}(\e-\p_1) P_{\rm L}(\k_1-\p_1) Z_2(\k_1-\p_1,\e)\bigg \}\\
=&\frac{1}{4} P(k_1)  W_{22}(\e)\left[\frac{C_{21}(k_1)}{b_1}+ \frac{2\,b_2^{\rm sph}}{b_1^2} \right]\ .
\end{split}\label{eq:BC_noRSD2}
\end{equation}
where $b_2^{\rm sph}\equiv b_2 -4\gamma_2/3$ is the local quadratic bias parameter in the spherical approximation~\cite{EggScoSmi1906} and 
\beq
C_{21}(k) \equiv \frac{68}{21}- \frac{\gamma(k)}{3}
\label{C21def}
\eeq
is the Fourier space analog of the two-point hierarchical skewness coefficient at tree-level in PT~\cite{Ber9608}. Using Eq.~(\ref{eq:BC_noRSD2})  in Eq.~(\ref{eq:1}) and defining 
\beq
\sigma^2_{22} \equiv {1\over \I_{22}^2}  \int_{\e} P(\epsilon) |W_{22}(\e)|^2
\label{sigma22}
\eeq
we have the final result for the beat-coupling covariance in real space,
\beqa
\textbf{C}^\textup{BC}(k_1,k_2) &=& \left[ {C_{21}(k_1)\over b_1}+ {2\,b_2^{\rm sph}\over b_1^2} \right] \left[ {C_{21}(k_2)\over b_1}+ {2\,b_2^{\rm sph}\over b_1^2}\right] \nonumber \\ & 
&\times \, \sigma^2_{22}\  P(k_1) P(k_2),
\label{eqn:trispec}
\eeqa

For unbiased tracers ($b_1=1$, $b_2^{\rm sph}=0$) this result reduces to that in~\cite{HamRimSco0609}, which neglected $\gamma$ in Eq.~(\ref{C21def}) since they worked in the nonlinear regime where $\gamma(k)\approx 0$. The importance of this term was stressed by~\cite{LiHuTak1404}, who call it a `dilation effect'. The calculation above shows that this arises in a similar way to spectral index corrections to the skewness~\cite{Ber94,Ber9608,BerColGaz0209} and is part of the beat-coupling trispectrum. In fact, each square bracket in Eq.~(\ref{eqn:trispec}) corresponds to the {\em galaxy} $C_{21}$ coefficient defined with respect to of galaxy fluctuations. 

\vspace{-0.2cm}
\subsection{Local average due to $\textup{N}_\textup{g}$ fluctuations ($\delta_{\textup{N}_\textup{g}} \neq 0$)}
\label{sec:LA}
In a  galaxy survey, the galaxy number density averaged over the survey volume will generally differ from the true average galaxy number density in the universe, which causes the total number of galaxies inside the survey volume ($\ng$) to be a random variable whose fluctuations $\dng$ must be taken into account. The effect of using the local average of $\ng$ determined from the survey itself (rather than the unknown global average) on the power spectrum covariance matrix was first pointed out  by~\cite{PutWagMen1204}. In this section, we generalize their results to include the effect of a realistic survey window for the FKP estimator in Eq.~(\ref{eq:trash_P_FKP}).

Before we delve into the covariance itself, let us start by evaluating the expectation value of the FKP estimator in the presence of $\ng$ fluctuations, from Eq.~(\ref{eq:trash_P_FKP}) we have
\begin{equation}
\begin{split}
&\langle \widehat{P}^\textup{FKP} (k) \rangle= \frac{1}{\I_{22}}\int_{\hat{\k}} \bigg \langle\frac{|\delta_W(\k)|^2 }{(1+\dng)}\bigg \rangle\\
&\simeq P(k) - \frac{1}{\I_{22}} \int_{\hat{\k}} \langle |\delta_W(\k)|^2 \dng\rangle + \frac{1}{\I_{22}} \int_{\hat{\k}}  \langle |\delta_W(\k)|^2 \dng^2\rangle \\
&=P(k) - \frac{1}{\I_{22}} \int_{\hat{\k}} \langle |\delta_W(\k)|^2 \dng\rangle +P(k)\ \langle \dng^2\rangle+ \ldots\, ,
\end{split}
\label{meanPdNg}
\end{equation}
where we have used $\langle |\delta_W(\k)|^2\rangle = P(k)$ as derived in Eq.~(\ref{eqn:convolved_power}) from the narrow window approximation. The bispectrum term becomes, using Eq.~(\ref{deltaNg}) for $\dng$ and tree-level PT for the bispectrum
\begin{equation}
\begin{split}
&\int_{\hat{\k}} \langle|\delta_W(\k)|^2 \dng\rangle\\
=&\frac{1}{\I_{10}} \int_{\hat{\k},\e,\p} W_{10}(-\e) W_{11}(\p) W_{11}(\e-\p) B(\e, \k-\p)\\
\simeq&\frac{1}{\I_{10}}\int_{\hat{\k},\e,\p} W_{10}(-\e) W_{11}(\p) W_{11}(\e-\p)\\
\times &\Big[ 2\, Z_2(\e,\k-\p)\ P(\epsilon)P_{\rm L}(\k-\p)\\
&+ 2\, Z_2(\e,-\k+\p-\e)\ P(\epsilon)P_{\rm L}(\k-\p+\e)\\
&+ b_2b_1^2\, P_{\rm L}(\k-\p)P_{\rm L}(\k-\p+\e)\Big]
\label{3ptFirst}
\end{split}
\end{equation}
where in the last term of the bispectrum we have used that $Z_2(\k-\p,-\k+\p-\e) \simeq {b_2/2} + {\cal O}(\epsilon^2/k^2)$. 
Upon changing variable $\p\rightarrow 2\k-\p+\e$ in the second $Z_2$ term, it becomes the same as the first $Z_2$ term. Further, the term proportional to $b_2$ is suppressed as compared to the others since typically $P(\e)\gg P(\k)$ (we will keep the $b_2$ term for completeness when adding redshift-space distortions in Sec.~\ref{sec:RSD_LA}, although it is still subdominant.). Therefore we can write
\begin{equation}\begin{split}
&\int_{\hat{\k}} \langle|\delta_W(\k)|^2 \dng\rangle
\simeq \frac{4}{\I_{10}}\int_{\e} W_{10}(-\e) P(\epsilon)\\
&\times \bigg \{ \int_{\hat{\k},\p} W_{11}(\p) W_{11}(\e-\p) Z_2(\e,\k-\p)P_{\rm L}(|\k-\p|) \bigg \}\, ,
\label{3ptSecond}
\end{split}\end{equation}
where the expression in the curly bracket has been worked out earlier in Eq.~(\ref{eq:BC_noRSD2}), leading to 
\begin{equation}\begin{split}
{1\over \I_{22}}\int_{\hat{\k}} \langle|\delta_W(\k)|^2 \dng\rangle
\simeq&\, \sigma_{10 \times 22}^2\, \left[ {C_{21}(k)\over b_1}+ {2\,b_2^{\rm sph}\over b_1^2}\right]P(k)  \, ,
\label{eq:2}\end{split}\end{equation}
where 
\beq
\sigma_{22 \times 10}^2 \equiv \frac{1}{\I_{10}\I_{22}}\int_{\e} P(\epsilon) W_{22}(\e) W_{10}(-\e)\, .
\label{sigma10x22}
\eeq
Defining in a similar way, 
\begin{equation}
\langle \dng^2 \rangle \equiv \sigma_{10}^2 = \frac{1}{\I_{10}^2}\int_{\e} P(\epsilon) |W_{10}(\e)|^2,
\label{sigma10}
\end{equation}
we obtain for the power spectrum
\beq
\langle\widehat{P}^\textup{FKP} (k) \rangle=P(k) \left[1-\sigma_{22 \times 10}^2 \left( {C_{21}(k)\over b_1}+ {2\,b_2^{\rm sph}\over b_1^2} \right) +\sigma_{10}^2 \right]
\label{meanPKFPreal}
\eeq
This signals that the FKP estimator is not unbiased in the presence of $\ng$ fluctuations, but in practice this bias is negligible (and in fact much smaller than the effects of window convolution that we already neglected in Eq.~\ref{eqn:convolved_power}). Indeed, for a realistic survey window such as BOSS DR12, $\sigma^2_{22} \sim \sigma^2_{10} \sim \sigma_{22 \times 10}^2  \sim 10^{-5} \, \sigma_8^2$. Although the contribution of such rms terms to the power spectrum is negligible, these effects give the leading order contribution to the covariance as we  shall see later. We now focus on the effect of $\dng$ on the covariance matrix,  working to leading order in PT
\vspace{0.2cm}
\begin{widetext}
\begin{equation}
\begin{split}
\textbf{C}(k_1,k_2) \equiv& \langle \widehat{P}^\textup{FKP}(k_1)\widehat{P}^\textup{FKP}(k_2) \rangle - \langle \widehat{P}^\textup{FKP}(k_1) \rangle\langle\widehat{P}^\textup{FKP}(k_2) \rangle\\
=& \frac{1}{\I_{22}^2}\int_{\hat{\k}_1,\hat{\k}_2} \left[ \left\langle \frac{|\delta_W(\k_1)|^2|\delta_W(\k_2)|^2)}{(1+\dng)^2}\right\rangle- \left\langle\frac{|\delta_W(\k_1)|^2}{(1+\dng)}\right\rangle \left\langle\frac{|\delta_W(\k_2)|^2}{(1+\dng)}\right\rangle  \right]\\
\simeq& \frac{1}{\I_{22}^2}\int_{\hat{\k}_1,\hat{\k}_2} \bigg[\langle|\delta_W(\k_1)|^2 |\delta_W(\k_2)|^2\rangle - \langle|\delta_W(\k_1)|^2\rangle \langle|\delta_W(\k_2)|^2\rangle\bigg] \\ 
&+\frac{1}{\I_{22}^2}\int_{\hat{\k}_1,\hat{\k}_2} \bigg[(-2) \langle|\delta_W(\k_1)|^2 |\delta_W(\k_2)|^2 \dng\rangle
+ 3 \langle|\delta_W(\k_1)|^2 |\delta_W(\k_2)|^2 \dng^2\rangle  \\
 & + \langle|\delta_W(\k_1)|^2 \dng\rangle \langle|\delta_W(\k_2)|^2 \rangle   + \langle|\delta_W(\k_1)|^2 \rangle \langle|\delta_W(\k_2)|^2 \dng\rangle  -2 \langle|\delta_W(\k_1)|^2\rangle \langle|\delta_W(\k_2)|^2\rangle \langle\dng^2\rangle\bigg]\\
=&  \textbf{C}^\textup{G}(k_1,k_2)+\textbf{C}^\textup{T}(k_1,k_2)\\ 
&-\frac{1}{\I_{22}^2}\int_{\hat{\k}_1,\hat{\k}_2} \bigg[\langle|\delta_W(\k_1)|^2 \dng\rangle \langle|\delta_W(\k_2)|^2 \rangle + \langle|\delta_W(\k_1)|^2 \rangle \langle|\delta_W(\k_2)|^2 \dng\rangle-\langle|\delta_W(\k_1)|^2\rangle \langle|\delta_W(\k_2)|^2\rangle \langle\dng^2\rangle \bigg]\\
=& \textbf{C}^\textup{G}(k_1,k_2)+\textbf{C}^\textup{T}(k_1,k_2)+\textbf{C}^\textup{LA}(k_1,k_2)
\end{split}
\label{eq:cov_point_terms}
\end{equation}
where we denoted the contribution of the local average terms due to $\dng$ to the covariance as LA. We also simplified the 5-point and 6-point terms to include only the leading order contributions as per the following equations,

\begin{equation}
\begin{split}
 \langle|\delta_W(\k_1)|^2 |\delta_W(\k_2)|^2 \dng\rangle \simeq&  \langle|\delta_W(\k_1)|^2 \dng\rangle \langle |\delta_W(\k_2)|^2  \rangle + \langle|\delta_W(\k_1)|^2 \rangle \langle |\delta_W(\k_2)|^2  \dng\rangle\, ,\\
 \langle|\delta_W(\k_1)|^2 |\delta_W(\k_2)|^2 \dng^2\rangle 
 \simeq& \langle|\delta_W(\k_1)|^2 \rangle \langle |\delta_W(\k_2)|^2 \rangle \langle \dng^2 \rangle\, .
 \end{split}
\end{equation}
We can use the earlier result in Eq.~(\ref{eq:2}) for computing the bispectrum terms and split the trispectrum contribution into regular and beat coupling contributions, taking advantage that the latter (Eq.~\ref{eqn:trispec}) has similar structure
 to the LA contribution, so that the full real-space covariance becomes

\begin{equation}
\begin{split}
\textbf{C}(k_1,k_2) =& \textbf{C}^\textup{G}(k_1,k_2)+ \textbf{C}^{\textup{T}_0}(k_1,k_2)+P(k_1)P(k_2)\left[\sigma_{22}^2 \left( {C_{21}(k_1)\over b_1}+ {2\,b_2^{\rm sph}\over b_1^2} \right) \left( {C_{21}(k_2)\over b_1}+ {2\,b_2^{\rm sph}\over b_1^2}\right)\right.\\
&\left. -\sigma_{22 \times 10}^2 \left( {C_{21}(k_1)\over b_1}+ {2\,b_2^{\rm sph}\over b_1^2} \right)  -\sigma_{22 \times 10}^2\left( {C_{21}(k_2)\over b_1}+ {2\,b_2^{\rm sph}\over b_1^2}\right)+ \sigma^2_{10} \right]
\end{split} \label{eq:CovaNOrsdSimplified}
\end{equation}
For a top-hat window $\sigma_{22}^2=\sigma_{22 \times 10}^2=\sigma^2_{10}= \sigma_{\rm TH}^2$, this reduces to a compact expression
\beq
\textbf{C}(k_1,k_2) = \textbf{C}^\textup{G}(k_1,k_2)+ \textbf{C}^{\textup{T}_0}(k_1,k_2)+P(k_1)P(k_2)\, \sigma_{\rm TH}^2 
\left( {C_{21}(k_1)\over b_1}+ {2\,b_2^{\rm sph}\over b_1^2} -1 \right) \left( {C_{21}(k_2)\over b_1}+ {2\,b_2^{\rm sph}\over b_1^2} -1\right).
\label{eq:CovaLArealspace}\eeq
\end{widetext}
This result {\em does not} agree with that in~\cite{PutWagMen1204,TakHu1306} who only considered a top-hat survey window, since their result has a $-2$ instead of a $-1$ subtraction. The reason for this is that the FKP estimator is different from the power spectrum estimator considered by~\cite{PutWagMen1204,TakHu1306}. As a result, their results overestimate the suppression of the beat-coupling covariance due to $\ng$ fluctuations when applied to realistic surveys. We discuss this in more detail in Sec.~\ref{sec:Literature_comparison}.

\section{Covariance in Redshift Space}
\label{sec:RedshiftSpaceFormalism}

Let us now extend the above results to redshift space.
We proceed in similar fashion to the real space case: we begin the discussion with the Gaussian covariance for the power spectrum multipoles in Sec.~\ref{sec:GaussCova}, then include the trispectrum contributions in Sec.~\ref{sec:RSD_Trispectrum}, and finally include the effects of $\ng$ fluctuations in Sec.~\ref{sec:RSD_LA}.

\subsection{Gaussian Covariance}
\label{sec:GaussCova}

Clustering in redshift-space depends on the velocity along the line-of-sight (LOS) direction, which changes with the location of galaxies in the sky. In what follows we do not assume that the LOS is a fixed global vector (i.e. the plane-parallel approximation), although we will keep terms that are leading order in $(kd)^{-1}$, where $d$ is the distance from the observer to the galaxies ($kd \gg 1$ in typical surveys).

\subsubsection{Power spectrum multipoles}

Let us consider the window convolved power spectrum multipoles. It was shown in~\citep{YamNakKam0602,SamBraPer1510,CasWhi1709}  that a very good approximation for the changing LOS direction corresponding to a galaxy pair is to set the LOS to be  along either one of the galaxies, i.e. $\mathcal{L}_\ell (\hat{\k}\, \cdot\,  \textup{LOS}) \rightarrow \mathcal{L}_\ell (\hat{\k}\cdot\hat{\x}_i)$ where $i=1,2$ labels the two galaxies in a pair. This leads to computationally more efficient power spectrum multipoles estimators that are factorizable~\cite{BiaGilRug1505,Sco1510}. 

We follow the notation in~\cite{Sco1510}, the  estimator for power spectrum multipoles is 
\begin{equation}
\widehat{P}_\ell (\k)\equiv\frac{(2\ell+1)}{\I_{22}}\int_{\hat{\k}} F_\ell (\k) F_0 (-\k)\, ,
\label{PellEstim}
\end{equation}
where 
\begin{equation}
F_\ell (\k)\equiv\int_{\x}W_{11}(\x)e^{-i\k\cdot\x} \L_\ell(\hat{\k}\cdot\hat{\x})\delta(\x)\, .
\label{eq:fl_estimator}\end{equation}
The expectation value of the power spectrum is then 
\begin{equation}\begin{split}
\langle \widehat{P}_\ell (k) \rangle = \frac{(2\ell+1)}{\I_{22}}& \int_{\hat{\k}} \int_{\x,\x'} e^{-i \k \cdot (\x-\x')} \langle \delta(\x)\delta(\x') \rangle\\
&\times W_{11}(\x)W_{11}(\x') \mathcal{L}_\ell(\hat{\x} \cdot \hat{\k})
\end{split}\end{equation}
Now, $\langle \delta(\x)\delta(\x') \rangle = \xi(\textbf{s};\x_+)$, where $\xi$ is  the redshift-space two-point function, 
 $\textbf{s}=\x'- \x$ is the relative coordinate, and the bisector $\x_+\equiv (\x+\x')/2$ characterizes the LOS, which as pointed out above can be taken to be instead either of the galaxies, e.g. $\x$. We can therefore write
\beqa
\langle \widehat{P}_\ell (k) \rangle = \frac{(2\ell+1)}{\I_{22}} \int_{\hat{\k}} \int_{\x,\textbf{s}} &e^{-i \k \cdot \textbf{s}}&\, \xi(\textbf{s};\x) W_{11}(\x)
W_{11}(\x-\textbf{s}) \nonumber \\ & &\times \mathcal{L}_\ell(\hat{\x} \cdot \hat{\k})\, .
\label{eq:3.1}
\eeqa
Now introducing the redshift-space local power spectrum~\cite{Sco1510}
\beq
P_{\rm local}(\k;\x) \equiv \int_{\textbf{s}} \xi(\textbf{s};\x) e^{-i \k \cdot \textbf{s}}
\label{Plocal}
\eeq
we can rewrite the following integral from Eq.~(\ref{eq:3.1})
\begin{equation}
\begin{split}
\int_{\textbf{s}}& e^{-i \k \cdot \textbf{s}} \xi(\textbf{s};\x) W_{11}(\x-\textbf{s})\\
=& \int_{\textbf{s},\k',\textbf{q}} e^{-i \k \cdot \textbf{s}}  P_{\rm local}(\k';\x) W_{11}(\textbf{q}) e^{i[\textbf{s}\cdot\k'+(\x-\textbf{s})\cdot \textbf{q}]}\\ 
=& \int_{\textbf{q}}  P_{\rm local}(\k+\textbf{q};\x) W_{11}(\textbf{q}) e^{i\x\cdot \textbf{q}} \simeq P_{\rm local}(\k;\x) W_{11} (\x)\, ,
\end{split} \label{eq:xi(s)_simplify}
\end{equation}
where we have again used $\k \gg \textbf{q}$ since $\textbf{q}$ is constrained to be small by the window function. Substituting Eq.~(\ref{eq:xi(s)_simplify}) into Eq.~(\ref{eq:3.1}) we get  (recall $W_{22}(\x)=W_{11}^2(\x)$ from Eq.~\ref{WandI})
\beqa
\langle \widehat{P}_\ell (k) \rangle
&= &\frac{(2\ell+1)}{\I_{22}} \int_{\hat{\k},\x} \mathcal{L}_\ell(\hat{\x}\cdot \hat{\k}) P_{\rm local}(\k;\x) W_{22}(\x) \nonumber \\
&\simeq& \frac{(2\ell+1)}{\I_{22}} \sum_{\ell'} P_{\ell'} (k) \int_{\hat{\k},\x} \mathcal{L}_{\ell'}(\hat{\x}\cdot \hat{\k}) \mathcal{L}_\ell(\hat{\x}\cdot \hat{\k}) \nonumber \\ &&
\ \ \ \ \ \ \  \ \ \ \ \ \ \ \ \ \ \ \ \ \ \ \ \ \ \ \ \ \ \ \times W_{22}(\x) \nonumber \\
&=& \frac{P_{\ell} (k)}{\I_{22}} \int_{\x} W_{22}(\x)= P_{\ell} (k)\, .
\label{meanPell}
\eeqa
where in the second line we have used the expansion of the local power spectrum in multipoles,
\beqa
P_{\rm local}(\k;{\bf d}) &=& \sum_{\ell'}  \mathcal{L}_{\ell'}(\hat{{\bf d}}\cdot \hat{\k}) \, P_{\ell'} (k; kd) \nonumber \\
& =& \sum_{\ell'}  \mathcal{L}_{\ell'}(\hat{{\bf d}}\cdot \hat{\k}) \, P_{\ell'} (k) + {\mathcal O}(kd)^{-2}
 \label{PlocalLeg}
\eeqa
 and used that to leading order in $(kd)^{-1}$ ($d$ being the distance to galaxies), the multipoles $P_\ell(k)$ can be replaced by their $kd \to \infty$ limit, which reduce to those in the plane-parallel approximation~\cite{Sco1510,ReiBerPit1601,CasWhi1709}.  The approximation in Eq.~(\ref{PlocalLeg}) corresponds to approximating the PT kernels in redshift space to be those in the plane-parallel approximation but with distortions acting along a line of sight direction that is radial, see Appendix~\ref{apx:LOS_BeatTerms} for details. We will use this approximation throughout this paper. 
%

\subsubsection{Covariance}
\label{sec:GauCova}
We follow similar steps for the covariance; using the estimator  in Eq.~(\ref{PellEstim}), the continuous (i.e. neglecting discreteness) Gaussian covariance reads
\begin{widetext}
\begin{equation}
\begin{split}
\textbf{C}^\textup{G}_{\ell_1\ell_2} (k_1,k_2)=&\frac{(2\ell_1+1)(2\ell_2+1)}{\I_{22}^2}\bigg[\int_{\hat{\k}_1,\hat{\k}_2} \langle F_{\ell_1}(\k_1) F_{0}(-\k_1) F_{\ell_2}(\k_2) F_{0}(-\k_2) \rangle \bigg] - \langle \widehat{P}_{\ell_1} (k_1) \rangle\ \langle \widehat{P}_{\ell_2} (k_2) \rangle \\
=&\frac{(2\ell_1+1)(2\ell_2+1)}{\I_{22}^2}\int_{\hat{\k}_1,\hat{\k}_2} \bigg[ \langle F_{\ell_1}(\k_1) F_{0}(-\k_2)\rangle \langle F_{\ell_2}(\k_2) F_{0}(-\k_1) \rangle+ \langle F_{\ell_1}(\k_1) F_{\ell_2}(-\k_2)\rangle \langle F_{0}(\k_2) F_{0}(-\k_1) \rangle\bigg]
\end{split}\label{eq:3.2}
\end{equation}
where we have split the 4-point function using Gaussianity. Expanding using Eq.~(\ref{eq:fl_estimator}),
\begin{equation}
\begin{split}
&\textbf{C}^\textup{G}_{\ell_1\ell_2} (k_1,k_2)=\frac{(2\ell_1+1)(2\ell_2+1)}{\I_{22}^2} \int_{\hat{\k}_1,\hat{\k}_2} \int_{\x_1,\x'_1,\x_2,\x'_2} e^{-i\k_1 \cdot (\x_1-\x'_1)- i\k_2 \cdot (\x_2-\x'_2)}\\
&\times \langle \delta(\x_1)\delta(\x'_2) \rangle \langle \delta(\x'_1)\delta(\x_2) \rangle W_{11}(\x_1)W_{11}(\x'_1)W_{11}(\x_2)W_{11}(\x'_2) \mathcal{L}_{\ell_1}(\hat{\x}_1 \cdot \hat{\k}_1)\Big[ \mathcal{L}_{\ell_2}(\hat{\x}_2 \cdot \hat{\k}_2) + \mathcal{L}_{\ell_2}(-\hat{\x}'_2 \cdot \hat{\k}_2) \Big ]
\end{split}
\label{stillSYM}
\end{equation}
Introducing again relative coordinates $\textbf{s}_1\equiv  \x_1-\x'_2$  and $\textbf{s}_2\equiv  \x_2- \x'_1$, and approximating the LOS as before (using in addition that $\ell_2$ is even) we have
\begin{equation}
\begin{split}
\textbf{C}^\textup{G}_{\ell_1\ell_2}& (k_1,k_2)=\frac{(2\ell_1+1)(2\ell_2+1)}{\I_{22}^2} \int_{\hat{\k}_1,\hat{\k}_2} \int_{\x_1,\x_2,\textbf{s}_1,\textbf{s}_2} e^{-i(\k_1-\k_2)\cdot(\x_1-\x_2)} e^{-i\k_1 \cdot \textbf{s}_2 -i \k_2 \cdot \textbf{s}_1}\ \xi(\textbf{s}_1;\x_1)\, \xi(\textbf{s}_2;\x_2)\\
&\times W_{11}(\x_1)W_{11}(\x_1-\textbf{s}_1) W_{11}(\x_2)W_{11}(\x_2-\textbf{s}_2) \mathcal{L}_{\ell_1}(\hat{\x}_1 \cdot \hat{\k}_1)\Big[ \mathcal{L}_{\ell_2}(\hat{\x}_2 \cdot \hat{\k}_2) + \mathcal{L}_{\ell_2}(\hat{\x}_1 \cdot \hat{\k}_2) \Big]
\end{split}\label{nolongerSYM}
\end{equation}
Using Eq.~(\ref{eq:xi(s)_simplify}) to simplify the integrals over  $\textbf{s}_1$ and $\textbf{s}_2$, and then using the multipole expansion to leading order  in  $(kd)^{-1}$ (as discussed following Eq.~(\ref{meanPell})), we finally obtain
\begin{equation}
\begin{split}
 \textbf{C}^\textup{G}_{\ell_1\ell_2} (k_1,k_2)=&\frac{(2\ell_1+1)(2\ell_2+1)}{\I_{22}^2} \int_{\hat{\k}_1,\hat{\k}_2,\x_1,\x_2} P_{\rm local}(\k_2;\x_1) P_{\rm local}(\k_1;\x_2)\\
 &\ \ \ \ \ \ \ \ \ \ \ \ \  \ \ \ \ \ \ \ \ \ \times W_{22}(\x_1) W_{22}(\x_2)\, e^{-i(\x_1-\x_2)\cdot(\k_1-\k_2)}\mathcal{L}_{\ell_1}(\hat{\x}_1 \cdot \hat{\k}_1)\Big[ \mathcal{L}_{\ell_2}(\hat{\x}_2 \cdot \hat{\k}_2) + \mathcal{L}_{\ell_2}(\hat{\x}_1 \cdot \hat{\k}_2) \Big]\\
\simeq& \sum_{\ell'_1,\ell'_2}P_{\ell'_1}(k_1)P_{\ell'_2}(k_2) \bigg\{ \frac{(2\ell_1+1)(2\ell_2+1)}{\I_{22}^2}\int_{\hat{\k}_1,\hat{\k}_2,\x_1,\x_2} W_{22}(\x_1) W_{22}(\x_2)\, e^{-i(\x_1-\x_2)\cdot(\k_1-\k_2)}\\
&\ \ \ \ \ \ \ \ \ \ \ \ \  \ \ \ \ \ \ \ \ \ \ \ \ \ \times \mathcal{L}_{\ell_1}(\hat{\x}_1 \cdot \hat{\k}_1) \mathcal{L}_{\ell'_1}(\hat{\x}_2 \cdot \hat{\k}_1) \mathcal{L}_{\ell'_2}(\hat{\x}_1 \cdot \hat{\k}_2) \Big[ \mathcal{L}_{\ell_2}(\hat{\x}_2 \cdot \hat{\k}_2) + \mathcal{L}_{\ell_2}(\hat{\x}_1 \cdot \hat{\k}_2) \Big] \bigg\}\, ,\\
\equiv& \sum_{\ell'_1,\ell'_2} P_{\ell'_1}(k_1)\, P_{\ell'_2}(k_2) \ \mathcal{W}^{(1)}_{\ell_1,\ell_2,\ell'_1,\ell'_2} (k_1,k_2) 
\end{split}\label{eq:C_GaussFinal}\end{equation}
\end{widetext}
where the power spectrum multipoles $P_\ell(k)$ in Eq.~(\ref{eq:C_GaussFinal}) correspond to those in the plane-parallel approximation and we have encoded the integral over the survey volume and the volume of $k$-bins as a kernel $\mathcal{W}^{(1)}$.
Let us now outline the procedure we use to compute $\mathcal{W}^{(1)}$ using the survey random catalog. We first decompose the Legendre polynomials as a contraction of tensors on $\hat{\x}_i$ with $\hat{\k_i}$. We then calculate relevant FFTs of the random catalog corresponding to the configuration space integrals (see Eqs.~(17-20) of \cite{Sco1510} for a similar procedure used on overdensity instead of the window terms). We then sum over the modes of the FFTs in the respective $k$-bins to calculate the $\hat{\k}_i$ integrals.

It is important to note that the cosmology and galaxy bias dependence of the Gaussian covariance in Eq.~(\ref{eq:C_GaussFinal}) is factorized. Therefore to recalculate the covariance for a different set of cosmology or bias parameters is computationally trivial as it amounts to simply computing the power spectrum multipoles for that set of parameters (the kernel $\mathcal{W}^{(1)}$ which depends only on the survey geometry need not be recalculated). This factorization between geometry and cosmology has been noted in the literature before, in real space in~\cite{HamRimSco0609,PutWagMen1204,BarKraSch1806} and in redshift space in~\cite{GriSanSal1604,LiSinYu1811} (we compare in more detail our results with previous literature in Sec.~\ref{sec:Literature_comparison}). This factorization allows us to include  velocity dispersion effects and loop corrections in the Gaussian covariance quite easily through the theoretical model for $P_\ell(k)$. Therefore our model for the Gaussian covariance will be accurate up to the scales where we can model the power spectrum theoretically. Note however that for our non-Gaussian covariance calculation in Sec. \ref{sec:RSD_Trispectrum}, we keep the calculation at tree-order, which is formally of the same order as one-loop in the Gaussian covariance.

Another important point to note is that the Gaussian covariance is not generally diagonal but leaks to a few $k$-bins on the either side of the diagonal. This happens because the survey window has a finite width in $k$-space and the width is not negligible as compared to the width of the $k$-bins typically used in an analysis. 
For covariances typically used in fisher forecasts, one encounters the approximation that the Gaussian covariance is diagonal (i.e $\mathcal{W}_{\ell_1,\ell_2,\ell'_1,\ell'_2} (k_1,k_2)=0$ when $k_1 \neq k_2$). This approximation is applicable only when the width of the window in $k$-space is much narrower than the $k$-bin width and we can substitute $|\k_1-\k_2|\rightarrow 0$ in Eq.~(\ref{eq:C_GaussFinal}) (see \cite{LiSinYu1811} for a detailed analysis of this approximation). However, such a diagonal approximation is not true in general; we show later in Fig.~\ref{fig:SN_contrib} that the Gaussian covariance indeed contributes to nearly a couple matrix elements on the each side of the diagonal.
Another approximation used in covariances for Fisher forecasts is assuming that the LOS is fixed to a particular direction $\hat{\textbf{n}}$ (which would amount to $\hat{\k}\cdot \hat{\x}_i \rightarrow \hat{\k}\cdot \hat{\textbf{n}}$ in terms in Eq.(\ref{eq:C_GaussFinal})).
 We however {\em do not} assume this global plane parallel approximation while calculating $\mathcal{W}^{(1)}$. In fact doing so will result in significant errors beyond the monopole autocovariance (e.g. about 40\% in the quadrupole autocovariance) for the same reason as the errors in a global plane parallel approximation analysis to measure the power spectrum multipoles $\ell \geq 2$ themselves.

\begin{figure}
    \includegraphics[scale=0.7,keepaspectratio=true]{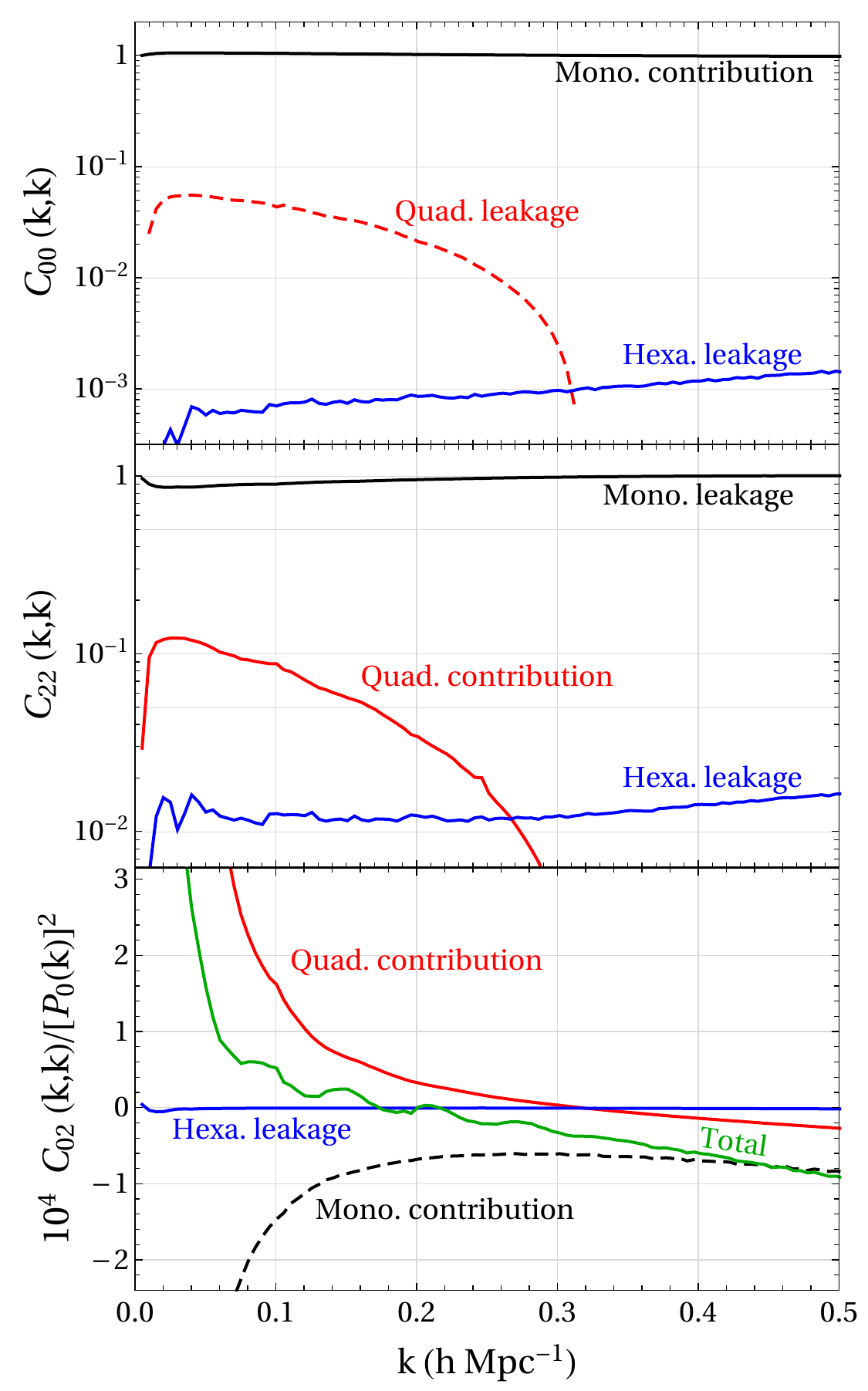}
    \caption{Fractional (total) contribution of power spectrum multipoles to the diagonal elements of the Gaussian part of the auto (cross) covariance matrix. We used the BOSS DR12 window for $0.5<z<0.75$ to calculate these contributions. Dashed lines represent negative contributions. Notice that the quadrupole auto-covariance is almost entirely made up by leakage of monopole power at high-$k$.}
    \label{fig:LeakageWindow}
\end{figure}

 \begin{figure}
\includegraphics[scale=0.8,keepaspectratio=true]{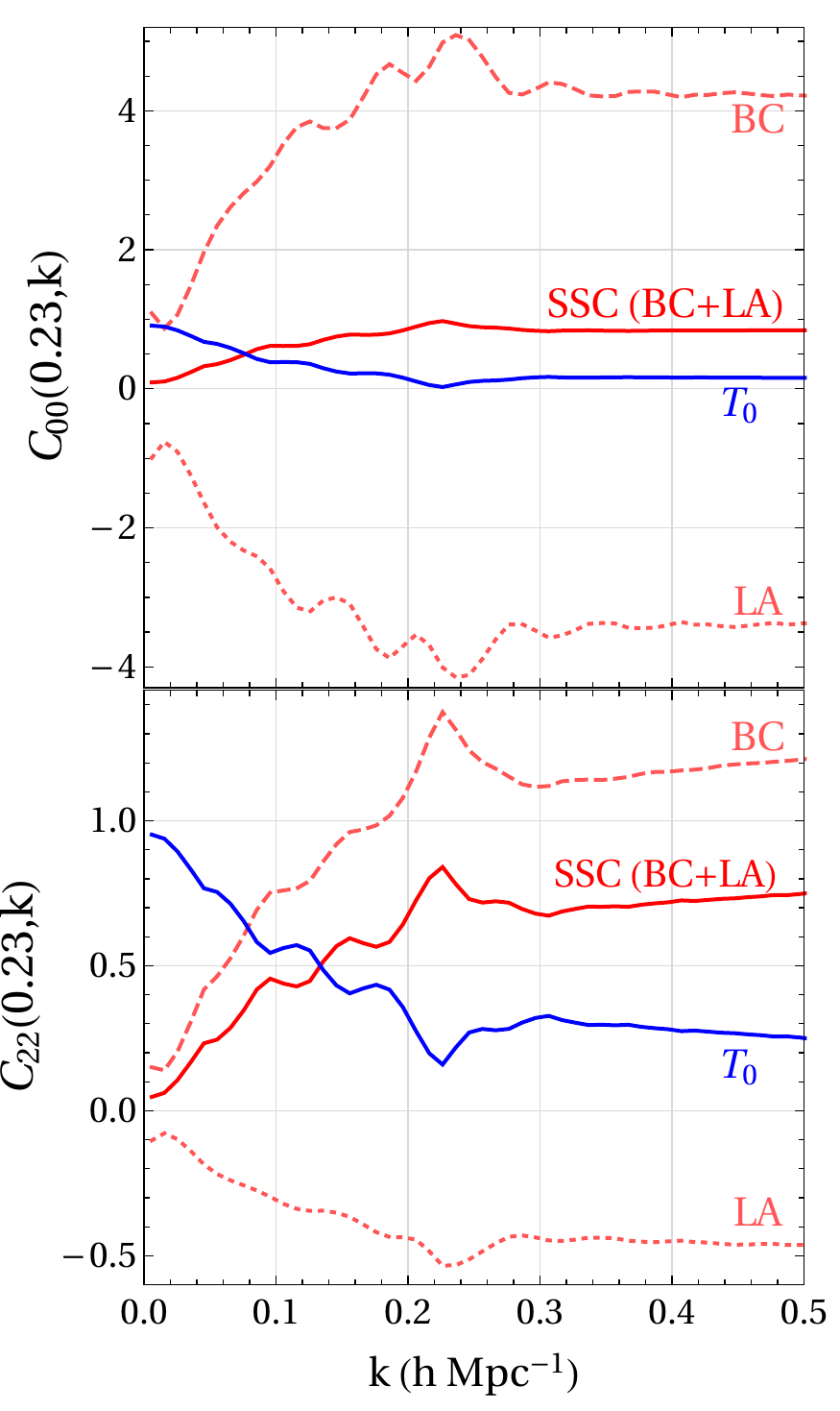}
\caption{ Fractional contribution of the ordinary trispectrum (T$_0$) and the total super-survey mode contribution (BC+LA) to the non-Gaussian part (BC+LA+T$_0$) of the autocovariance matrix of the monopole (top) and quadrupole (bottom). The contribution from the super-survey modes (referred to as super sample covariance (SSC) in the literature) can be broken down into two parts: the beat-coupling terms in the trispectrum (BC) and the local average (LA) effect which is caused due to fluctuations in the total number of galaxies $\ng$ inside the survey volume. Because the super-survey modes (which are in the linear regime) dominate over T$_0$ at high-$k$, PT can be used to model the non-Gaussian covariance fairly accurately.}
\label{fig:ED_components}
\end{figure}

Fig.~\ref{fig:LeakageWindow} shows the fractional contributions to Eq.~(\ref{eq:C_GaussFinal}) for the monopole autocovariance  (top panel), quadrupole autocovariance  (middle panel) and monopole-quadrupole cross-covariance (bottom panel)  along the diagonal as a function of $k$. 
We see that in all cases, the auto-covariances are dominated by the monopole term in Eq.~(\ref{eq:C_GaussFinal}), that is the $\ell_1'=\ell_2'=0$ term (black lines in Fig.~\ref{fig:LeakageWindow}). The quadrupole terms (where the {\em largest} of $\ell_1'$ and $\ell_2'$ is 2) are always subdominant for autocovariances and hexadecapole terms (where the {\em largest} of $\ell_1'$ and $\ell_2'$ is 4) even more so. The same holds for the hexadecapole autocovariance (not shown). We label off diagonal contributions $\ell'_i \neq \ell_j$ in Eq.~(\ref{eq:C_GaussFinal}) as `leakage' in Fig.~\ref{fig:LeakageWindow}. It is worth mentioning that there are some important contributions which are missed if one neglects the window shape and uses the diagonal approximation mentioned earlier, for. e.g., the $\C_{02}$ case in Fig.~\ref{fig:LeakageWindow} has a monopole contribution only because of the anisotropy of the survey window.

The main conclusion from these results on the autocovariances is that monopole contributions always dominate the contribution to the Gaussian covariance, which is a welcome result since the monopole power is the multipole least affected by velocity dispersion and dominated by shot noise at high-$k$ and therefore the simplest multipole to predict. This makes  predictions quite reliable even at scales beyond the validity of PT, as we shall see in Sec.~\ref{sec:Patchy_compare}. The work~\cite{BeuSeoSai1704} already noted that the quadrupole error is dominated by the monopole power based on the results of~\cite{TarNisSai1009,YooSel1502} for a finite box (rather than a nontrivial survey window).  

We also see from Fig.~\ref{fig:LeakageWindow} that the cross-covariance $C_{02}$ monopole and quadrupole terms both matter and they partially cancel each other, leading to a changing sign in the net result. A similar situation holds for $C_{24}$  while for $C_{04}$ there is a much smaller cancellation and the net result is mostly  monopole dominated and positive definite.

The formula for the Gaussian covariance in real space can be obtained trivially from Eq.~(\ref{eq:C_GaussFinal}) by replacing the monopole power spectrum by the linear power and setting all higher order multipoles to zero, i.e. $P_\ell(k)\rightarrow P_{\rm L}(k)\, \delta^K_{\ell,0}$. In order to test the accuracy of Eq.~(\ref{eq:C_GaussFinal}) in this limit for a realistic window function, we generated $4 \times 10^6$  mocks using continuous Gaussian random fields and imposed the BOSS DR12 survey window. We found Eq.~(\ref{eq:C_GaussFinal}) to agree with the measurements within $2\%$ independently of $k$. 


\subsection{Non-Gaussian Covariance: Trispectrum Contribution}
\label{sec:RSD_Trispectrum}

Let us now focus on the non-Gaussian contribution to covariance, which is the dominant contribution towards the off-diagonal covariance matrix elements~\cite{ScoZalHui9912}, at least far away from the diagonal where the Gaussian contribution to the covariance is negligible. 

For the non-gaussian covariance, we will describe the redshift space distortions only to leading order in PT (see Appendix~\ref{apx:RSDkernels}) and we will mostly ignore velocity dispersion effects (i.e. so-called ``fingers of god").
Proceeding along the same lines as for the redshift-space Gaussian covariance, our aim is to obtain the non-Gaussian covariance in the limit where $kd\gg 1$, where $d$ is the distance to the galaxies. In this limit, however, the redshift-space kernels can be taken to leading order in $(kd)^{-1}$ to be those in the plane-parallel approximation but with the local line of sight (LOS) pointing in the radial direction. However, since working out the windowed trispectrum within the plane-parallel approximation is much simpler, we start in this section by assuming a fixed LOS $\hat{\textbf{n}}$ across the survey region. We quote how the results change at the end in the realistic case of radial distortions to leading order in $(kd)^{-1}$ (presented in more detail in Appendix \ref{apx:LOS_BeatTerms}).

From this section onwards we change our notation in Eq.~(\ref{eq:ShellAvg}) for the shell averaging integral to have the following form for simplifying the equations under the plane parallel approximation 
\begin{equation}
\int_{\hat{\k}_{\ell_1}} \rightarrow \int \frac{d^3\k}{V_{k_i}} (2\ell_1+1) \mathcal{L}_{\ell_1}(\hat{\k}_1\cdot\hat{\textbf{n}})\ ,
\end{equation}
where we made our notation compact because the shell $k_1$ is always associated with the Legendre polynomial of degree $\ell_1$. For the non-Gaussian contribution to the covariance the thin shell approximation $\int d^3\k/V_{k} \rightarrow \int d\Omega_\k/(4\pi)$ works fairly well for the bin size we have chosen. A perceptive reader would notice that we had argued after Eq.~(\ref{eq:C_GaussFinal}) that the thin-shell approximation is inappropriate for calculation of the Gaussian covariance; This is because the Gaussian covariance integral in Eq.~(\ref{eq:C_GaussFinal}) very strongly depends on the difference $|\k_1-\k_2|$. So we needed to take into account the exact location of wave-vectors $\k_1$ and $\k_2$ inside their respective shells to accurately calculate $|\k_1-\k_2|$. On the other hand, as we will see later in this section, the non-Gaussian covariance depends on $P_{\rm L}(|\k_1|)$, $P_{\rm L}(|\k_2|)$ and their derivatives with respect to the wave-vectors. Such terms show a negligible change over the bin-width that we use ($k_b\sim 0.005$ h/Mpc). Therefore the thin-shell approximation is valid for calculating the non-Gaussian covariance and will be assumed in what follows.

We first write expressions involving only the beat-coupling trispectrum (see Eq.~\ref{TBCdef}) 
\beq
T_{\textup{BC}}\equiv T(\k_1 -\p_1,-\k_1 -\e+\p_1,\k_2 -\p_2,-\k_2 +\e+\p_2),
\eeq
and calculate its contribution to the covariance matrix following similar substitutions as in Sec. (\ref{sec:Trispectrum}) to simplify the covariance integral as
\begin{widetext}
\begin{equation}
\begin{split}
\textbf{C}^{\textup{BC}}_{\ell_1\ell_2}(k_1,k_2) =& \frac{1}{\I_{22}^2}\int_{\hat{\k}_{\ell_1},\hat{\k}_{\ell_2}, \e,\p_1,\p_2}  W_{11}(\p_1)W_{11}(\p_2)W_{11}(\e-\p_1)W_{11}(-\e-\p_2)\\
& \times 16 P_{\rm L}(\epsilon)\, [P_{\rm L}(|\k_1-\p_1|) Z_1(\k_1-\p_1) Z_2(\k_1-\p_1,\e)]\,
[P_{\rm L}(|\k_2-\p_2|)Z_1(\k_2-\p_2) Z_2(\k_2-\p_2,-\e)]\\
=&\frac{1}{\I_{22}^2}\int_{\e} P_{\rm L}(\epsilon) \bigg \{ 4\int_{\hat{\k}_{\ell_1},\p_1}W_{11}(\p_1) W_{11}(\e-\p_1) P_{\rm L}(|\k_1-\p_1|) Z_1(\k_1-\p_1) Z_2(\k_1-\p_1,\e)\bigg \}\\ & \times \bigg \{4\int_{\hat{\k}_{\ell_2},\p_2}W_{11}(\p_2)W_{11}(-\e-\p_2)P_{\rm L}(|\k_2-\p_2|)Z_1(\k_2-\p_2) Z_2(\k_2-\p_2,-\e)\bigg \}\, .
\end{split}
\end{equation}
The terms in curly brackets are encountered frequently in our expressions so we define a new kernel $\mathcal{Z}_{21}$ as (generalizing Eq.~(\ref{eq:BC_noRSD2}) to the redshift-space case\footnote{Note that unlike Eq.~(\ref{eq:BC_noRSD2}), we factorize here the linear matter spectrum rather than the galaxy power spectrum. Bias and redshift distortions are included in $\mathcal{Z}_{21}$.})
\begin{equation}
\bigg \{4\int_{\hat{\k}_{\ell_1},\p_1} W_{11}(\p_1) W_{11}(\e-\p_1) P_{\rm L}(|\k_1-\p_1|) Z_1(\k_1-\p_1) Z_2(\k_1-\p_1,\e) \bigg \} \equiv P_{\rm L}(k_1) W_{22}(\e)\mathcal{Z}_{21} (k_1,\ell_1,\hat{\e}\cdot\hat{\textbf{n}})\, ,
\label{eq:Z12_definition}
\end{equation}
\end{widetext}
where $\mathcal{Z}_{21}(k_1,\ell_1,\hat{\e}\cdot\hat{\textbf{n}})$ can be calculated using the window identities in Eq.~(\ref{eq:WindowIdentities}) and using similar steps as in Sec. \ref{sec:Trispectrum} (see Appendix \ref{apx:RSDkernels} for a detailed derivation). $\mathcal{Z}_{21}$ depends only on $\hat{\e}$ since we neglect terms ${\cal O}(\e^2)$. 
The covariance due to the beat mode is then 
\begin{equation}
\begin{split}
\textbf{C}&^\textup{BC}_{\ell_1\ell_2} (k_1,k_2)= \frac{1}{\I_{22}^2}P_{\rm L}(k_1)P_{\rm L}(k_2)\\
&\times \int_{\e} P_{\rm L}(\epsilon) |W_{22}(\e)|^2  \mathcal{Z}_{21}(k_1,\ell_1,\hat{\e}\cdot\hat{\textbf{n}})\mathcal{Z}_{21}(k_2,\ell_2,-\hat{\e}\cdot\hat{\textbf{n}})\, .
\end{split}
\label{CBCl1l2}
\end{equation}

Let us briefly show that the expression in Eq.~(\ref{CBCl1l2}) is equivalent to the super sample covariance (SSC) approach presented in \cite{LiSchSel1802} for unbiased tracers.
Expanding the kernel $\mathcal{Z}_{21}$ in multipoles
\begin{equation}
\mathcal{Z}_{21}(k_1,\ell_1,\hat{\e}\cdot\hat{\textbf{n}})=\sum_{L_1} \mathcal{Z}'_{21}(k_1,\ell_1,L_1)\ \L_{L_1}(\hat{\e}\cdot\hat{\textbf{n}})
\label{Z21multipole}
\end{equation}
Eq.~(\ref{CBCl1l2}) becomes
\begin{equation}
\begin{split}
\textbf{C}&^{\textup{BC}}_{\ell_1\ell_2} (k_1,k_2)\\
=& \sum_{L_1L_2}
\left[\frac{1}{\I_{22}^2}\int_{\e} P_{\rm L}(\epsilon) |W_{22}(\e)|^2 \L_{L_1}(\hat{\e}\cdot\hat{\textbf{n}}) \L_{L_2}(-\hat{\e}\cdot\hat{\textbf{n}})  \right]\\
&\times \Big\{P_{\rm L}(k_1) \mathcal{Z}'_{21}(k_1,\ell_1,L_1)\Big\} \Big\{P_{\rm L}(k_2)\mathcal{Z}'_{21}(k_2,\ell_2,L_2)\Big\}. 
\end{split}
\label{CBCl1l2k1k2}
\end{equation}
Therefore we can rewrite the beat-coupling covariance as a product of the responses of power spectra (we have checked that the expressions in the curly brackets agree with the power spectrum responses presented in \cite{LiSchSel1802} when we ignore galaxy bias.)
In the sum over $L_1,L_2$ in Eq.~(\ref{CBCl1l2k1k2}) we include terms corresponding to the monopole, quadrupole and hexadecapole in this paper. The choice is based on the fact that our answers converge by inclusion of multipoles until the hexadecapole. 

Eq.~(\ref{CBCl1l2k1k2}) is our final result for the beat coupling covariance of power spectrum multipoles in the plane-parallel approximation. However, as it stands, this is not yet applicable to a realistic redshift survey, as it does not make sense to calculate multipoles of the window $|W_{22}|^2$ at the beat mode $\e$ with respect to a fixed line-of-sight $\hat{\textbf{n}}$, since by definition such modes subtend a large angle on the sky for wide-angle surveys (see Fig.~\ref{fig:VaryingLOS} below). 

Appendix \ref{apx:BC_radialRSD} discusses in detail how this result must be modified to account for the LOS variation effect. While the derivation is a little involved, the final result is not difficult to understand. Indeed it corresponds to changing the above plane-parallel expression in the following way
\vspace{0.05cm}
\begin{widetext}
\beq\begin{split}
 &\int_{\e} P_{\rm L}(\epsilon) |W_{22}(\e)|^2 \L_{L_1}(\hat{\e}\cdot\hat{\textbf{n}}) \L_{L_2}(-\hat{\e}\cdot\hat{\textbf{n}})\longrightarrow \int \frac{4\pi \epsilon^2 d\epsilon}{(2\pi)^3} P_{\rm L}(\epsilon)  \bigg \{\int_{\hat{\e},\x,\x'} W_{22}(\x)W_{22}(\x') e^{-i \e\cdot(\x-\x')} \L_{L_1}(\hat{\e}\cdot\hat{\x}) \L_{L_2}(\hat{\e}\cdot\hat{\x}')\bigg \},
\end{split}\label{eq:LOS_BC}\eeq
where the term in the curly brackets can be easily calculated by FFT techniques from the random catalog of the survey. 
We now turn to the contribution of the regular trispectrum $T_0$ to the redshift-space covariance. We start from Eq.~(\ref{eq:trisp_real}) and neglect $\e, \p_1$ \& $\p_2$ as compared to $\k_1$ \& $\k_2$, this gives

 \begin{equation}
\begin{split}
\textbf{C}^{\textup{T}_0}_{\ell_1\ell_2}(k_1,k_2) =&\frac{1}{\I_{22}^2} \int_{\hat{\k}_{\ell_1},\hat{\k}_{\ell_2}, \e}  |W_{22}(\e)|^2 \bigg\{\bigg[8P_{\rm L}^2(\k_1)Z_1^2(\k_1)P_{\rm L}(\k_1+\k_2) Z_2^2(-\k_1,\k_1+\k_2)+ (\k_1\leftrightarrow \k_2)\bigg]\\
&\ +16P_{\rm L}(\k_1)Z_1(\k_1)P_{\rm L}(\k_2)Z_1(\k_2)P_{\rm L}(\k_1+\k_2)Z_2(-\k_1,\k_1+\k_2) Z_2(-\k_2,\k_1+\k_2)\\
&\ + \bigg[12\, Z_1^2(\k_1) P_{\rm L}^2(\k_1) Z_1(\k_2) P_{\rm L}(\k_2)Z_3(\k_1,-\k_1,\k_2) + (\k_1\leftrightarrow \k_2)\bigg]\bigg\} \\
=&\frac{\I_{44}}{\I_{22}^2} \int_{\hat{\k}_{\ell_1},\hat{\k}_{\ell_2}}  P_{\rm L}(\k_1+\k_2)\bigg[\bigg(8P_{\rm L}^2(\k_1)Z_1^2(\k_1) Z_2^2(-\k_1,\k_1+\k_2)+ (\k_1\leftrightarrow \k_2)\bigg)\\
&\ +16P_{\rm L}(\k_1)Z_1(\k_1)P_{\rm L}(\k_2)Z_1(\k_2)Z_2(-\k_1,\k_1+\k_2) Z_2(-\k_2,\k_1+\k_2)\bigg]\\
&\ + \frac{\I_{44}}{\I_{22}^2} \int_{\hat{\k}_{\ell_1},\hat{\k}_{\ell_2}}
\bigg[12\, Z_1^2(\k_1) P_{\rm L}^2(\k_1) Z_1(\k_2) P_{\rm L}(\k_2)Z_3(\k_1,-\k_1,\k_2) + (\k_1\leftrightarrow \k_2)\bigg]
\end{split}\label{eq:CovaT0}
\end{equation}
\end{widetext}
Fig.~\ref{fig:ED_components} shows a comparison between the contribution due to the beat-coupling trispectrum (BC) and due to the regular trispectrum (T$_0$) to the monopole (top) and quadrupole (bottom) autocovariance. Note that the vertical scale is in terms of {\em relative} contributions, with the sum adding to unity. We will discuss this figure further in the next section, where we discuss the local average (LA) effect due the $\ng$ fluctuations and see that there are large cancellations to the the beat coupling contribution. As a result, the regular trispectrum contribution dominates over that of the beat/super-survey modes at low-$k$ and therefore cannot be naively neglected for a redshift survey (unlike the case for a weak lensing survey where the LA effect is absent and the $T_0$ term is subdominant \cite{BarSch1706,BarSch1709}).

\subsection{Non-Gaussian Covariance: $\ng$ fluctuations}
\label{sec:RSD_LA}

Similar to what we discussed already in the absence of redshift distortions in Sec. \ref{sec:LA}, taking into account $\ng$ fluctuations through its local average in the observed volume leads to a reduction in the beat coupling effect from large-scale modes. This effect follows from the dependence of the redshift-space FKP estimator in Eq.~(\ref{eq:delta_FKP}) on $\ng$ fluctuations $\dng$. We will now consider the effect of $\dng$ fluctuations in redshift space including nonlinear bias and evolution. 

As presented in the previous section \ref{sec:RSD_Trispectrum}, we first discuss the effect of redshift distortions first in the plane-parallel approximation where the derivation is easier, then quote the more general results with the radial distortions which are necessary to properly treat the beat modes. The variance of $\ng$ fluctuations in redshift space in the plane-parallel approximation is,
\begin{equation}
\begin{split}
\langle \dng^2 \rangle &= \sigma_{10}^2 = \frac{1}{\I_{10}^2}\int_{\e} P_{\rm L}(\epsilon) |W_{10}(\e)|^2 Z_1^2(\e)\, ,
\end{split}
\end{equation}
which for a varying LOS generalizes to (see Appendix~\ref{apx:LA_radialRSD} for a derivation)
\begin{equation}
\begin{split}
\langle \dng^2 \rangle =\frac{1}{\I_{10}^2} \int_{\e} P_{\rm L}(\epsilon) \bigg \{ &\int_{\x_1,\x_2} W_{10}(\x_1)\, W_{10}(\x_2) e^{-i\e\cdot(\x_1-\x_2)}\\
&\times Z_1^{\hat{\x}_1}(\e) \, Z_1^{\hat{\x}_2}(\e)  \bigg \}
\end{split}\label{eq:dngRMS}
\end{equation}
where 
\beq
Z_1^{\hat{\x}}(\k) \equiv b_1+f\, (\hat{\k}\cdot\hat{\x})^2
\label{Z1approx}
\eeq
is the linear PT redshift-space kernel, and in the leading approximation corresponds to the plane-parallel kernel but with distortions acting along the radial LOS. The quantity in curly brackets can be calculated efficiently by FFTs from the random catalog of the survey.
The expectation value of the FKP power spectrum multipoles follows from the estimator in Eq.~(\ref{eq:delta_FKP}), expanded up to quadratic order in $\dng$, and after using Eq.~(\ref{meanPell}) is given by
\begin{equation}
\begin{split}
\langle \widehat{P}^\textup{FKP}_{\ell} (k) \rangle = &P_\ell(k) - \frac{1}{\I_{22}} \int_{\hat{\k}_\ell} \langle |\delta_W(\k)|^2 \dng\rangle + P_\ell(k)\, \sigma^2_{10}\, .
\end{split}
\end{equation}
Similar to Eq.~(\ref{meanPdNg}), we need to evaluate the  3-point term, but now with redshift distortions. After using the tree-level PT bispectrum we get

\begin{widetext}
\begin{equation}\begin{split}
\frac{1}{\I_{22}} & \int_{\hat{\k}_{\ell_1}} \langle|\delta_W(\k_1)|^2 \dng\rangle =\frac{1}{\I_{22} \I_{10}} \int_{\hat{\k}_{\ell_1}, \e,\p} W_{10}(-\e) W_{11}(\p) W_{11}(\e-\p) B(\e, \k_1-\p)\\
=&\frac{2}{\I_{22} \I_{10}} \int_{\hat{\k}_{\ell_1}, \e,\p} W_{10}(-\e) W_{11}(\p) W_{11}(\e-\p)\Big[ Z_2(\e,\k_1-\p)Z_1(\e)Z_1(\k_1-\p)P_{\rm L}(\epsilon)P_{\rm L}(|\k_1-\p|)\\
&+ Z_2(\e,-\k_1+\p-\e)Z_1(\e)Z_1(\k_1-\p+\e)P_{\rm L}(\epsilon)P_{\rm L}(|\k_1-\p+\e|)\\
&+ Z_2(\k_1-\p,-\k_1+\p-\e)Z_1(\k_1-\p)Z_1(\k_1-\p+\e)P_{\rm L}(q)P_{\rm L}(|\k_1-\p+\e|)\Big]\, ,
\label{3ptTermRSD}
\end{split}\end{equation}
%
Using similar substitutions as before in going from Eq.~(\ref{3ptFirst}) to~(\ref{3ptSecond}) we have
\begin{equation}
\begin{split}
&\frac{1}{\I_{22}} \int_{\hat{\k}_{\ell_1}} \langle|\delta_W(\k_1)|^2 \dng\rangle = \frac{2}{\I_{22} \I_{10}} \int_{\hat{\k}_{\ell_1}, \e,\p} W_{10}(-\e) W_{11}(\p) W_{11}(\e-\p)\\
&\times \Big[2Z_2(\e,\k_1-\p)Z_1(\e)Z_1(\k_1-\p)P_{\rm L}(\epsilon)P_{\rm L}(|\k_1-\p|)\\
&\ \ \ + Z_2(\k_1-\p,-\k_1+\p-\e)Z_1(\k_1-\p)Z_1(\k_1-\p+\e)P_{\rm L}(|\k_1-\p|)P_{\rm L}(|\k_1-\p+\e|)\Big]
\end{split}
\label{eqn:rsd_bisp}
\end{equation}
Again, since $|\p| \simeq |\e| \ll |\k_1|$, the  second term in square brackets can be approximated to lowest order as $Z_2(\k_1-\p,-\k_1+\p-\e)= b_2/2 +{\cal O}(\epsilon^2/k_1^2)$ and we get
\begin{equation}
\frac{b_2}{\I_{22} \I_{10}} \int_{\hat{\k}_{\ell_1}, \e,\p} W_{10}(-\e) W_{11}(\p) W_{11}(\e-\p) Z^2_1(\k_1)P_{\rm L}^2(k_1) = b_2\left[ \int_{\hat{\k}_{\ell_1}}Z_1^2(\k_1)\right] P_{\rm L}^2(k_1) \frac{\I_{32}}{\I_{22}\I_{10}}\, ,
\end{equation}
whereas the first term in square brackets of Eq.~(\ref{eqn:rsd_bisp}) can be written after using the kernel $\mathcal{Z}_{21}$  defined in Eq.~(\ref{eq:Z12_definition}) as
\begin{equation}
\begin{split}
&\frac{1}{\I_{22}\I_{10}} \int_{\e} W_{10}(-\e)P_{\rm L}(\epsilon)Z_1(\e)\bigg\{4 \int_{\hat{\k}_{\ell_1},\p_1} W_{11}(\p_1) W_{11}(\e-\p_1)Z_2(\e,\k_1-\p_1) Z_1(\k_1-\p_1)P_{\rm L}(|\k_1-\p_1|)\bigg\}\\
&=\frac{P_{\rm L}(k_1)}{\I_{10}\I_{22}} \int_{\e} W_{10}(-\e)P_{\rm L}(\epsilon)Z_1(\e) W_{22}(\e)\mathcal{Z}_{21}(k_1,\ell_1,\hat{\e}\cdot\hat{\textbf{n}})\ .
\end{split}\label{eq:3-pointRSD}\end{equation}
Putting these two results together we have then, 
\begin{equation}
\frac{1}{\I_{22}} \int_{\hat{\k}_{\ell_1}} \langle|\delta_W(\k_1)|^2 \dng\rangle = \frac{P_{\rm L}(k_1)}{\I_{10}\I_{22}} \int_{\e}W_{10}(-\e)W_{22}(\e)\left(P_{\rm L}(\epsilon)Z_1(\e) \mathcal{Z}_{21}(k_1,\ell_1,\hat{\e}\cdot\hat{\textbf{n}})+P_{\rm L}(k_1)b_2\left[\int_{\hat{\k}_{\ell_1}}Z_1^2(\k_1)\right]\right)\ ,
\label{eq:3-pointSolnRSD}
\end{equation}
%
and the expectation value of the FKP power spectrum becomes
\begin{equation}\begin{split}
\langle \widehat{P}^\textup{FKP}_{\ell} (k) \rangle = &P_\ell(k)\bigg[1 - \frac{1}{\I_{10}\I_{22}} \frac{\int_{\e}W_{10}(-\e)W_{22}(\e)P_{\rm L}(\epsilon)Z_1(\e) \mathcal{Z}_{21}(k_1,\ell_1,\hat{\e}\cdot\hat{\textbf{n}})}{\int_{\hat{\k}_{\ell_1}}Z_1^2(\k_1)}-\frac{\I_{32}}{\I_{10}\I_{22}}b_2P_{\rm L}(k) + \sigma^2_{10}\bigg]\, .
\label{meanPKFPredshift}
\end{split}\end{equation}
This generalizes Eq.~(\ref{meanPKFPreal}) to redshift space, including now also the subleading term proportional to $b_2$. As it was the case before for real space, the relative bias of the FKP estimator is $\mathcal{O}(10^{-5}$) and remains negligible in redshift space. 
Having done the power spectrum, the expressions for the contribution of $\ng$ fluctuations to the covariance are similar (see discussion in Sec.~\ref{sec:LA}) and using the split in 
Eq.~(\ref{eq:cov_point_terms}), we get 
\begin{equation}
\begin{split}
\textbf{C}^\textup{LA}_{\ell_1\ell_2}(k_1,k_2) = P_{\ell_1}(k_1)P_{\ell_2}(k_2)\sigma_{10}^2&-\frac{P_{\rm L}(k_1) P_{\ell_2}(k_2)}{\I_{10}\I_{22}} \int_{\e}W_{10}(-\e)W_{22}(\e)[P_{\rm L}(\epsilon)Z_1(\e) \mathcal{Z}_{21}(k_1,\ell_1,\hat{\e}\cdot\hat{\textbf{n}})+P_{\ell_1}(k_1)b_2]\\
&-\frac{P_{\rm L}(k_2) P_{\ell_1}(k_1)}{\I_{10}\I_{22}} \int_{\e}W_{10}(-\e)W_{22}(\e)[P_{\rm L}(\epsilon)Z_1(\e) \mathcal{Z}_{21}(k_2,\ell_2,\hat{\e}\cdot\hat{\textbf{n}})+P_{\ell_2}(k_2)b_2]
\label{CLAredshift}
\end{split}
\end{equation}
which generalizes the local average contribution in Eq.~(\ref{eq:CovaNOrsdSimplified}) to redshift-space.

As discussed earlier, this result, as it stands, is not applicable to a wide redshift survey, which requires us to take into account the changing LOS across the survey volume for beat modes, see Appendix~\ref{apx:LA_radialRSD}. This corresponds to replacing in Eqs.~(\ref{meanPKFPredshift}) and~(\ref{CLAredshift})
\begin{equation}
\begin{split}
&W_{10}(-\e)W_{22}(\e)Z_1(\e) \mathcal{Z}_{21}(k_1,\ell_1,\hat{\e}\cdot\hat{\textbf{n}}) \longrightarrow 
 \sum_{L_1} \mathcal{Z}'_{21}(k_1,\ell_1,L_1) \bigg \{\int_{\x,\x'} W_{22}(\x)W_{10}(\x') e^{-i \e\cdot(\x-\x')} \L_{L_1}(\hat{\e}\cdot\hat{\x})\ Z_1^{\hat{\x}'}(\e) \bigg \}
\end{split}\label{eq:LOS_BC2}
\end{equation}
where we have used the multipole expansion in Eq.~(\ref{Z21multipole}). Again, the expression in  curly brackets can be calculated by FFTs from the random catalog specifying the survey geometry.
 \end{widetext}
Fig.~\ref{fig:ED_components} shows the result of this calculation for the monopole (top) and quadrupole (bottom) autocovariances. As was the case in real space (see Eq.~\ref{eq:CovaNOrsdSimplified}),
 the LA contribution partially cancels the beat coupling contribution, resulting in a residual due to super-survey modes (BC+LA).
 Note that the residual is dominant over the regular trispectrum contribution ($T_0$) at high-$k$.
  This is a welcome result for a perturbative treatment of the off-diagonal covariance because the effect of the super-survey modes is straightforward to model in PT while the loop corrections which dominate in the trispectrum at high-$k$  are relatively more difficult to model.



\section{Shot noise}
\label{sec:ShotNoise}

\subsection{Covariance Contributions}
\label{sec:SNcova}

Real surveys are made up of discrete objects (galaxies) rather than continuous fields, this gives rise to a significant contribution to the covariance in the form of shot noise (SN). We now show the effect of introducing discreteness into all the terms that were derived in the previous sections.
Note that we have only included shot noise terms in the Poissonian approximation in this work. There is however a constant offset from Poisson shot noise in the galaxy power spectrum, which can be interpreted as being due to the halo exclusion effect \cite{SheLem99,SmiScoShe07,BalSelSmi13} (at high-$k$, there is also an additional scale-dependent effect of  $\mathcal{O}(k^2)$ \cite{DesJeoSch18,BalSelSmi13}). We will later show in Sec.~\ref{sec:Patchy_compare} that we have a very good agreement with the SDSS Patchy mocks by only including the Poissonian SN contribution. This hints at the fact that the non-Poissonian effects may not be well-modeled in the approximate techniques used to make mocks. The non-Poissonian shot noise terms can be included in the analytic covariance calculation if desired.

In this section, we label the discrete terms by the index `$d$' (e.g. $\delta^d$, $P^d$) whereas for quantities in the continuous limit we use e.g. $\delta$ and~$P$ in order to avoid confusion. We also label the objects in the galaxy catalog
 by $g$ and the objects in the random catalog by $r$.
In the discrete case, we must subtract from the FKP power spectrum estimator (written as a discrete form of Eq.~(\ref{eq:trash_P_FKP})), the SN coming from self-pairs, leading to
\begin{widetext}
\begin{equation}
\begin{split}
\widehat{P}^\textup{FKP}(\k) \equiv& \frac{1}{\alpha\, \I^r_{22}} \Big[ \bigg( \sum_{i}^{\ng} -\alpha \sum_{i}^{\nr} \bigg)\bigg( \sum_{j}^{\ng}-\alpha \sum_{j}^{\nr} \bigg) w_i w_{j}e^{-i \k\cdot(\x_i-\x_{j})}- \bigg( \sum_{j}^{\ng} +\alpha^2 \sum_{j}^{\nr} \bigg) w^2_j \Big]\\
 =&\frac{1}{\I_{22}(1+\dng^d)} \bigg( \sum_{i}^{\ng} -\alpha \sum_{i}^{\nr} \bigg)\bigg( \sum_{j(\neq i)}^{\ng} -\alpha \sum_{j(\neq i)}^{\nr} \bigg) w_i w_{j}e^{-i \k\cdot(\x_i-\x_{j})} \equiv \frac{\widehat{P}^d(\k)}{\I_{22}(1+\dng^d)}\, ,
\end{split}\label{eq:4.1}\end{equation}
\end{widetext}
where $\bar{\alpha}\equiv\langle \ng \rangle /\nr$ (and is typically $\ll 1$) and we defined $\widehat{P}^d (\k)$ in the last equality. The $\ng$ fluctuations in the discrete case are given by
\beq
\dng^d \equiv {1\over \I_{10}}\bigg(\sum_i^{\ng}-\bar{\alpha} \sum_{i}^{\nr}\bigg)\ 1
\label{dngD}
\eeq
Note that the SN subtraction in Eq.~(\ref{eq:4.1}) is different from that in the FKP paper~\cite{FelKaiPea9405} where the {\em expected} rather than the \emph{true} SN is subtracted. This subtle difference has important consequences for the covariance, see Sec.~\ref{sec:TrueSN} below for a detailed discussion. A simple diagrammatic way to express the discrete estimator in Eq.~(\ref{eq:4.1}) is
\beq
 \widehat{P}^\textup{FKP} \equiv (\, \dt \quad \dt\ + \ \underbracket{\dt\quad \dt} \,)\ - \ \underbracket{\dt\quad \dt}\ = \ (\,\dt \quad \dt\,)
\eeq
where the dots denote the positions of the two galaxies and the underbrackets or overbrackets denote the cases when the two or more objects are the same (i.e. self-pairs with $i=j$ in Eq.~\ref{eq:4.1}). The true shot noise term therefore cancels out the the contribution of such a cases.
The expectation value of the FKP estimator, expanding in $\dng^d\ll 1$ to leading order,
\begin{widetext}
\begin{equation}
\int_{\hat{\k}_{\ell_1}} \langle \widehat{P}^\textup{FKP}(\k) \rangle  \simeq  \frac{1}{\I_{22}} \int_{\hat{\k}_{\ell_1}} \bigg[ \langle \widehat{P}^d(\k) \rangle - \langle\widehat{P}^d(\k) \,\dng^d\rangle + \langle\widehat{P}^d(\k) \rangle \langle (\dng^d)^2 \rangle \bigg]\, .
\label{PellMeanSN}
\end{equation}

The variance in the total number of galaxies can now be explicitly written as 
\begin{equation}
\begin{split}
\langle (\dng^d)^2 \rangle =& \frac{1}{\I^2_{10}} \bigg\langle \bigg(\sum_{j}^{\ng}-\bar{\alpha} \sum_{j}^{\nr} \bigg) \bigg( \sum_{i}^{\ng} -\bar{\alpha} \sum_{i}^{\nr} \bigg)\bigg\rangle=\frac{1}{\I^2_{10}} \bigg\langle \bigg(\sum_{j}^{\ng}+\bar{\alpha}^2 \sum_{j}^{\nr} \bigg)\bigg\rangle+\frac{1}{\I^2_{10}} \bigg\langle \bigg(\sum_{j}^{\ng}-\bar{\alpha} \sum_{j}^{\nr} \bigg) \bigg( \sum_{i(\neq j)}^{\ng} -\bar{\alpha} \sum_{i(\neq j)}^{\nr} \bigg)\bigg\rangle\\
=&\frac{1+\bar{\alpha}}{\I_{10}}+\frac{1}{\I^2_{10}}\int_{\x,\x'} W_{10}(\x) W_{10}(\x') \langle \delta(\x) \delta(\x')  \rangle 
\\
=&\frac{1+\bar{\alpha}}{\I_{10}}+\frac{1}{\I^2_{10}} \int \frac{4\pi \epsilon^2  d\epsilon}{(2\pi)^3} P_{\rm L}(\epsilon) 
\int_{\hat{\e},\x,\x'} W_{10}(\x)W_{10}(\x') e^{-i \e\cdot (\x-\x')} \ Z_1^{\hat{\x}}(\e)\, Z_1^{\hat{\x}'}(\e)\, .
\end{split}\label{eq:4.2}
\end{equation}
The final ingredient needed for evaluating Eq.~(\ref{PellMeanSN}) is the 3-point term,
\begin{equation}
\begin{split} \langle \widehat{P}^d(\k_1)\, \dng^d\rangle=& \bigg\langle \frac{1}{\I_{10}} \bigg(\sum_{j}^{\ng}-\bar{\alpha} \sum_{j}^{\nr} \bigg) \bigg( \sum_{i}^{\ng} -\alpha \sum_{i}^{\nr} \bigg)\bigg( \sum_{i'(\neq i)}^{\ng} -\alpha \sum_{i'(\neq i)}^{\nr} \bigg) w_i w_{i'} e^{-i \k_1\cdot (\x_i-\x_{i'})} \bigg\rangle\, .
\end{split}
\end{equation}
There are two SN terms in the above equation corresponding to $j=i$ and $j=i'$ which can be diagrammatically represented as\\ $\dt\ \times (\, \dt \quad \dt\, ) = \dt\quad \dt \quad \dt\ + \ \underbracket{\dt\quad \dt} \quad \dt\ + \  \underbracket{\dt\quad \dt \quad \dt}\ $  and can be explicitly calculated as 
\begin{equation}
\begin{split} \langle \widehat{P}^d(\k_1)\, \dng^d\rangle \simeq& \frac{2}{\I_{10}} \bigg\langle \bigg( \sum_{i}^{\ng} +\bar{\alpha}^2 \sum_{i}^{\nr} \bigg)\bigg( \sum_{i'(\neq i)}^{\ng} -\bar{\alpha} \sum_{i'(\neq i)}^{\nr} \bigg) w_i w_{i'} e^{-i \k_1\cdot (\x_i-\x_{i'})} \bigg\rangle\\
&+\frac{1}{\I_{10}} \bigg\langle \bigg(\sum_{j}^{\ng}-\bar{\alpha} \sum_{j}^{\nr} \bigg) \bigg( \sum_{i(\neq j)}^{\ng} -\bar{\alpha} \sum_{i(\neq j)}^{\nr} \bigg)\bigg( \sum_{i'(\neq j)(\neq i)}^{\ng} -\bar{\alpha} \sum_{i'(\neq j)(\neq i)}^{\nr} \bigg) w_i w_{i'} e^{-i \k_1\cdot (\x_i-\x_{i'})} \bigg\rangle\\
=& \frac{2}{\I^2_{10}}\int_{\x,\x'} W_{11}(\x) W_{11}(\x') \langle(1+\bar{\alpha} + \delta(\x)) \delta(\x') \rangle e^{-i \k_1\cdot(\x-\x')}\\
&+\frac{1}{\I^2_{10}}\int_{\x_1,\x_2,\x_3} W_{10}(\x_1) W_{11}(\x_2) W_{11}(\x_3)\langle \delta(\x_1) \delta(\x_2) \delta(\x_3) \rangle e^{-i \k_1\cdot(\x_2-\x_3)}\\
=& \frac{2}{\I_{10}}\langle |\delta_W(\k_1)|^2 \rangle + \langle |\delta_W(\k_1)|^2\dng\rangle = 2\frac{\I_{22}}{\I_{10}}Z_1^2(\k_1) P_{\rm L}(k_1)+ \langle |\delta_W(\k_1)|^2\dng\rangle\, ,
\end{split}\label{eq:SNresponse}
\end{equation}
where the last term corresponding to the squeezed bispectrum in the continuous case has already been calculated in Eq.~(\ref{eq:3-pointSolnRSD}). Overall, for the case of discrete tracers, the expectation value of the FKP estimator is given by subtracting the term $P_\ell(k)[(1-\bar{\alpha})/\I_{10}]$ from the the expectation value in the continuous case in Eq.~(\ref{meanPKFPredshift}). As $\I_{10}\equiv\ng$ is very large in our case, the expectation value remains unbiased.

Let us now write down the covariance, again expanding in $\dng^d\ll 1$ to leading order,
\begin{equation}
\begin{split}
\textbf{C}&_{\ell_1 \ell_2}(k_1,k_2)= \frac{1}{\I_{22}^2}\int_{\hat{\k}_{\ell_1},\hat{\k}_{\ell_2}} \left[ \left\langle \frac{\widehat{P}^d(\k_1)\widehat{P}^d(\k_2))}{(1+\dng^d)^2}\right\rangle- \left\langle\frac{\widehat{P}^d(\k_1)}{(1+\dng^d)}\right\rangle \left\langle\frac{\widehat{P}^d(\k_2)}{(1+\dng^d)}\right\rangle  \right]\\
 &= \textbf{C}_{\ell_1,\ell_2}^\textup{G}(k_1,k_2) -\frac{1}{\I_{22}^2}\int_{\hat{\k}_{\ell_1},\hat{\k}_{\ell_2}} \bigg[\langle\widehat{P}^d(\k_1) \,\dng^d\rangle \langle\widehat{P}^d(\k_2) \rangle + \langle\widehat{P}^d(\k_1\rangle \langle\widehat{P}^d(\k_2) \, \dng^d\rangle-\langle\widehat{P}^d(\k_1)^2\rangle \langle\widehat{P}^d(\k_2)\rangle \langle(\dng^d)^2\rangle \bigg] \\ & \ \ 
 +\textbf{C}_{\ell_1,\ell_2}^\textup{T}(k_1,k_2)
\end{split}
\end{equation}
The ingredients for the terms in the square brackets have been already calculated. We then need to calculate the Gaussian and trispectrum contributions including SN. A naive approximation to the Gaussian covariance in the discrete case can be obtained by replacing in the continuous covariance ($\textbf{C}_{\ell_1,\ell_2}^\textup{G}$ in Eq.~(\ref{eq:C_GaussFinal})), the power spectrum multipoles in the discrete case as

\begin{equation}
P_\ell (k) \rightarrow P_\ell (k)+\frac{\I_{12}}{\I_{22}}\delta^K_{\ell,0}\, ,
\label{eq:SNgaussPowerSubstitution}\end{equation}
where $\delta^K$ is the kronecker delta function and thus the SN contributes only to the monopole power. However this approximation to $\textbf{C}_{\ell_1,\ell_2}^\textup{G}$ reduces the accuracy of our results by 20\% as compared to the full calculation presented in Appendix \ref{apx:SN_exact_gaussian}. The final result is given by appending the continuous covariance in Eq.~(\ref{eq:C_GaussFinal}) by an additional term as:

\begin{equation}
\begin{split}
\textbf{C}^\textup{G}_{\ell_1\ell_2} (k_1,k_2)= \textbf{C}^\textup{G(cont.)}_{\ell_1\ell_2} (k_1,k_2)+\Big[ \sum_{\ell'} P_{\ell'}(k_1)\mathcal{W}^{(2)}_{\ell_1,\ell_2,\ell'} (k_1,k_2) +(k_1 \leftrightarrow k_2)\Big]+\mathcal{W}^{(3)}_{\ell_1,\ell_2}(k_1,k_2)
\end{split}\label{eq:covaSN-G}\end{equation}
where the explicit expressions for the $\mathcal{W}^{(2)}$ and $\mathcal{W}^{(3)}$ kernels are written in Eq.~(\ref{eq:W_kernel}) and we calculate these kernels in a similar manner $\mathcal{W}^{(1)}$ in Eq.~(\ref{eq:C_GaussFinal}) by using different FFTs constructed from the survey random catalog. Finally, the only term in the covariance remaining to re-calculate is the trispectrum contribution including SN,

\begin{equation}\begin{split}
\textbf{C}&^\textup{T}_{\ell_1 \ell_2}(k_1,k_2)
=\frac{1}{\I_{22}^2} \int_{\hat{\k}_{\ell_1},\hat{\k}_{\ell_2}}\langle \widehat{P}^d(\k_1) \widehat{P}^d(\k_2) \rangle_c\\
=&\frac{1}{\I_{22}^2} \int_{\hat{\k}_{\ell_1},\hat{\k}_{\ell_2}} \bigg\langle \bigg( \sum_{i}^{\ng} -\alpha \sum_{i}^{\nr} \bigg)\bigg( \sum_{i'(\neq i)}^{\ng} -\alpha \sum_{i'(\neq i)}^{\nr} \bigg) w_i w_{i'}e^{-i \k_1\cdot(\x_i-\x_{i'})}\bigg( \sum_{j}^{\ng}-\alpha \sum_{j}^{\nr} \bigg)\bigg( \sum_{j'(\neq j)}^{\ng} -\alpha \sum_{j'(\neq j)}^{\nr} \bigg) w_j w_{j'}e^{-i \k_2\cdot(\x_j-\x_{j'})} \bigg\rangle_c.
\end{split}\label{eq:TrispSN}\end{equation}

The six possible SN contributions here are diagrammatically shown in Fig.~\ref{fig:TrispSNcontributions}. If one naively counts the number of possible pairs, one would expect there to be 14 SN terms \cite{MarVerHea9710}. However, some of these vanish because self-pairs are already removed in the estimator in Eq.~(\ref{eq:4.1}) (see also \cite{OcoEisVar1611,Lac1806}). Let us now consider $i=j\neq i' \neq j'$ for example; its contribution to the covariance becomes
\begin{figure}
\includegraphics[scale=0.6,keepaspectratio=true]{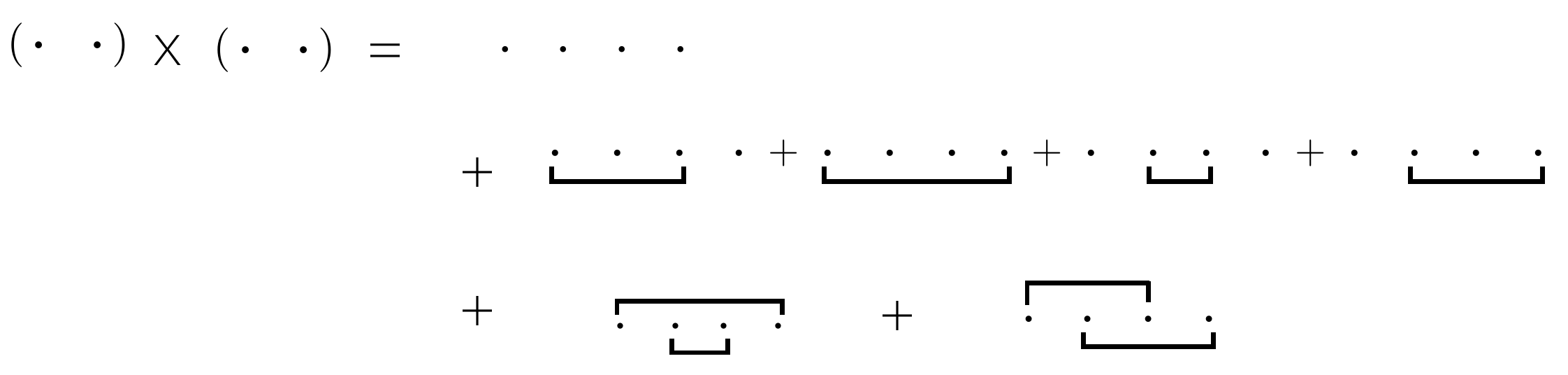}
\caption{ SN contributions to the power spectrum covariance involving the trispectrum, see Eq.~(\ref{eq:TrispSN}).}
\label{fig:TrispSNcontributions}
\end{figure}

\begin{equation}
\begin{split}
\frac{1}{\I_{22}^2}& \int_{\hat{\k}_{\ell_1},\hat{\k}_{\ell_2}} \bigg\langle \bigg( \sum_{i}^{\ng} -\alpha \sum_{i}^{\nr} \bigg)\bigg( \sum_{j(\neq i)}^{\ng} +\alpha^2 \sum_{j(\neq i)}^{\nr} \bigg)\bigg( \sum_{j'(\neq j)(\neq i)}^{\ng} -\alpha \sum_{j'(\neq j)(\neq i)}^{\nr} \bigg)  w_i w^2_j w_{j'}e^{-i \k_1\cdot \x_i}e^{i (\k_1-\k_2)\cdot \x_j} e^{i \k_2\cdot \x_{j'}}\bigg\rangle_c\\
\simeq&\frac{1}{\I_{22}^2} \int_{\hat{\k}_{\ell_1},\hat{\k}_{\ell_2},\x_1,\x'_1,\x_2} W_{11}(\x_1)W_{12}(\x'_1)W_{11}(\x_2)\langle \delta(\x_1)(1+\bar{\alpha}+\delta(\x'_1))\delta(\x_2)\rangle_c\ e^{-i \k_1\cdot \x_1+i (\k_1-\k_2)\cdot \x'_1+i \k_2\cdot \x_2}\, .
\label{CTstep1}
\end{split}\end{equation}
We only include one of the contributions in the connected term in the above equation as per $\langle \delta(\x_1)(1+\bar{\alpha}+\delta(\x'_1))\delta(\x_2)\rangle_c = \langle \delta(\x_1)\delta(\x'_1)\delta(\x_2)\rangle_c$, because the remaining term $(1+\bar{\alpha})\langle \delta(\x_1)\delta(\x_2)\rangle$ corresponds to a  term  already taken into account in the Gaussian contribution, see Eq.~(\ref{eq:cova_SN-G}). Therefore, Eq.~(\ref{CTstep1}) becomes,
\begin{equation}\begin{split}
& \frac{1}{\I_{22}^2} \int_{\hat{\k}_{\ell_1},\hat{\k}_{\ell_2}} \langle \delta_W(\k_1) \delta_{W_{12}}(\k_2-\k_1)  \delta_W(-\k_2) \rangle_c =\frac{\I_{34}}{\I_{22}^2} \int_{\hat{\k}_{\ell_1},\hat{\k}_{\ell_2}} B(\k_1,\k_2)\\
&= 2\frac{\I_{34}}{\I_{22}^2} \int_{\hat{\k}_{\ell_1},\hat{\k}_{\ell_2}} \bigg( P_{\rm L}(k_1) P_{\rm L}(k_2) Z_1 (\k_1) Z_1 (\k_2)Z_2(\k_1,\k_2)\\
&\qquad \qquad+P_{\rm L}(\k_1+\k_2)Z_1(\k_1+\k_2)[P_{\rm L}(\k_1)Z_1(\k_1) Z_2(-\k_1,\k_1+\k_2)+ (\k_1\leftrightarrow \k_2)]\bigg).
\label{CTstep2}
\end{split}
\end{equation}
We get a similar expression in the remaining three cases ($j'=i,i=j,i'=j'$). The second type of SN contribution arises when there are two sets of particles at the same position, which are shown as terms with both an underbracket and an overbracket in Fig. \ref{fig:TrispSNcontributions}. This corresponds to either ($i=j)'\neq (i'=j$) or ($i=j)\neq (i'=j'$) and their combined contribution to the covariance becomes
\begin{equation}
\begin{split}
\frac{2}{\I_{22}^2} &\int_{\hat{\k}_{\ell_1},\hat{\k}_{\ell_2}} \bigg\langle \bigg( \sum_{i}^{\ng} +\alpha^2 \sum_{i}^{\nr} \bigg)\bigg( \sum_{j(\neq i)}^{\ng} +\alpha^2 \sum_{j(\neq i)}^{\nr} \bigg) w^2_i w^2_{j}
e^{-i (\k_1-\k_2)\cdot (\x_i-\x_j)} \bigg\rangle_c\\
=&  \frac{2}{\I_{22}^2} \int_{\hat{\k}_{\ell_1},\hat{\k}_{\ell_2}} \int_{\x_1, \x_2} W_{12}(\x_1) W_{12}(\x_2) \langle (1+\bar{\alpha}+\delta(\x_1))(1+\bar{\alpha}+\delta(\x_2)) \rangle_c\, e^{-i (\k_1-\k_2)\cdot (\x_1-\x_2)}\\
\simeq&\, 2\,\frac{\I_{24}}{\I_{22}^2} \int_{\hat{\k}_{\ell_1},\hat{\k}_{\ell_2}} P_{\rm L}(\k_1-\k_2) Z_1^2(\k_1-\k_2)\, ,
\end{split}\label{eq:SN3}\end{equation}
\end{widetext}
where we have again used that $\langle (1+\bar{\alpha}+\delta(\x_1))(1+\bar{\alpha}+\delta(\x_2)) \rangle_c = \langle \delta(\x_1)\delta(\x_2) \rangle_c$, because the remaining term,
given by $(1+\bar{\alpha})^2$, has already been considered in the disconnected covariance case in Eq.~(\ref{eq:cova_SN-G}).

\subsection{Shot Noise Estimator: True vs FKP Shot Noise}
\label{sec:TrueSN}

As mentioned above, in our estimator~\citep{Sco1510}, we subtract the {\em true} SN~\cite{Ham0002} using simply the self-pairs in the data and random catalogs respectively. This is unlike the FKP SN~\cite{FelKaiPea9405} which subtracts the {\em expected} SN and is purely calculated by the random catalog and then rescaled. The FKP SN is obtained by replacing in the power spectrum estimator in Eq.~(\ref{eq:4.1}) by

\begin{figure}[h!]
\centering
\includegraphics[scale=0.7,keepaspectratio=true]{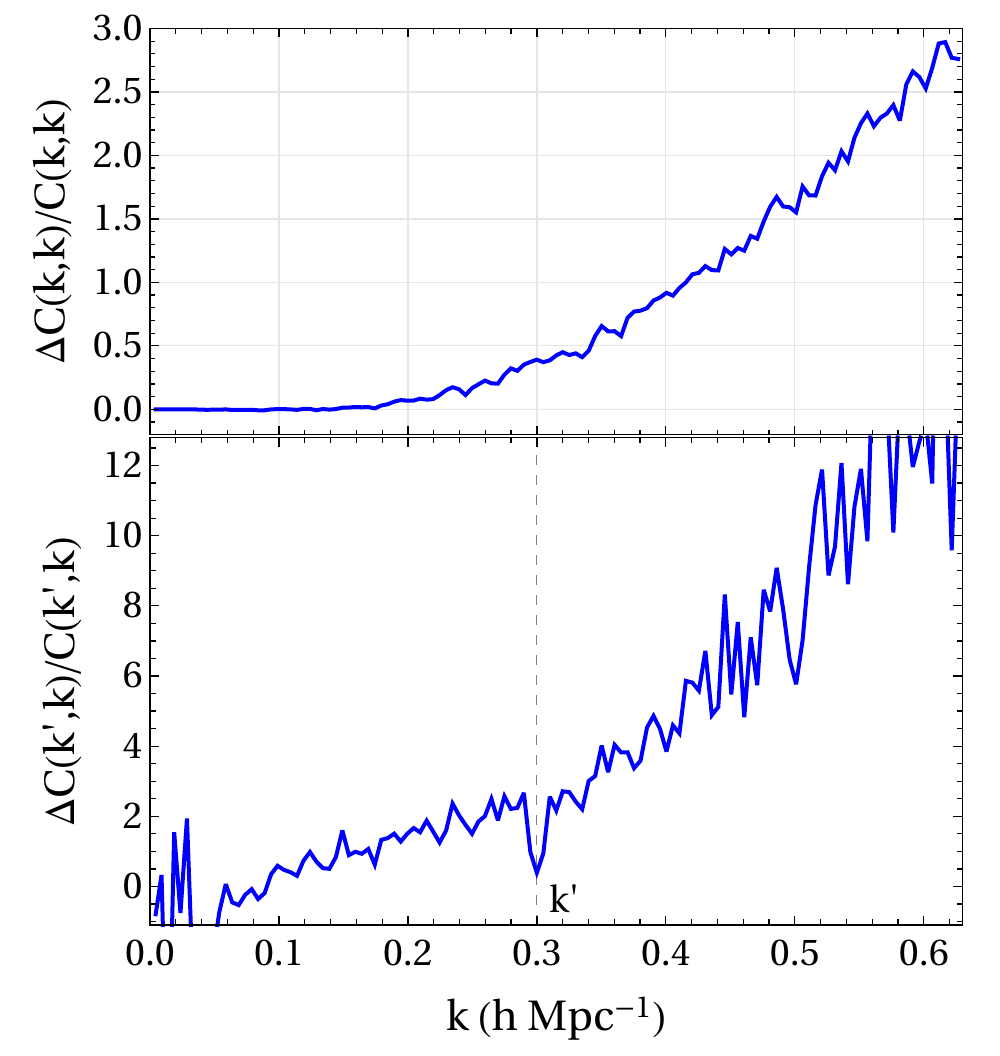}
\caption{Enhancement in the diagonal elements (top panel) of the power spectrum monopole auto-covariance due to using the FKP shot noise estimator as compared to the true shot noise estimator. The bottom panel shows the enhancement of the non-diagonal elements for $k'=0.3 \kMpc$.}
\label{fig:TrueSN}
\end{figure}
\begin{figure}[h!]
\centering
\includegraphics[scale=0.6,keepaspectratio=true]{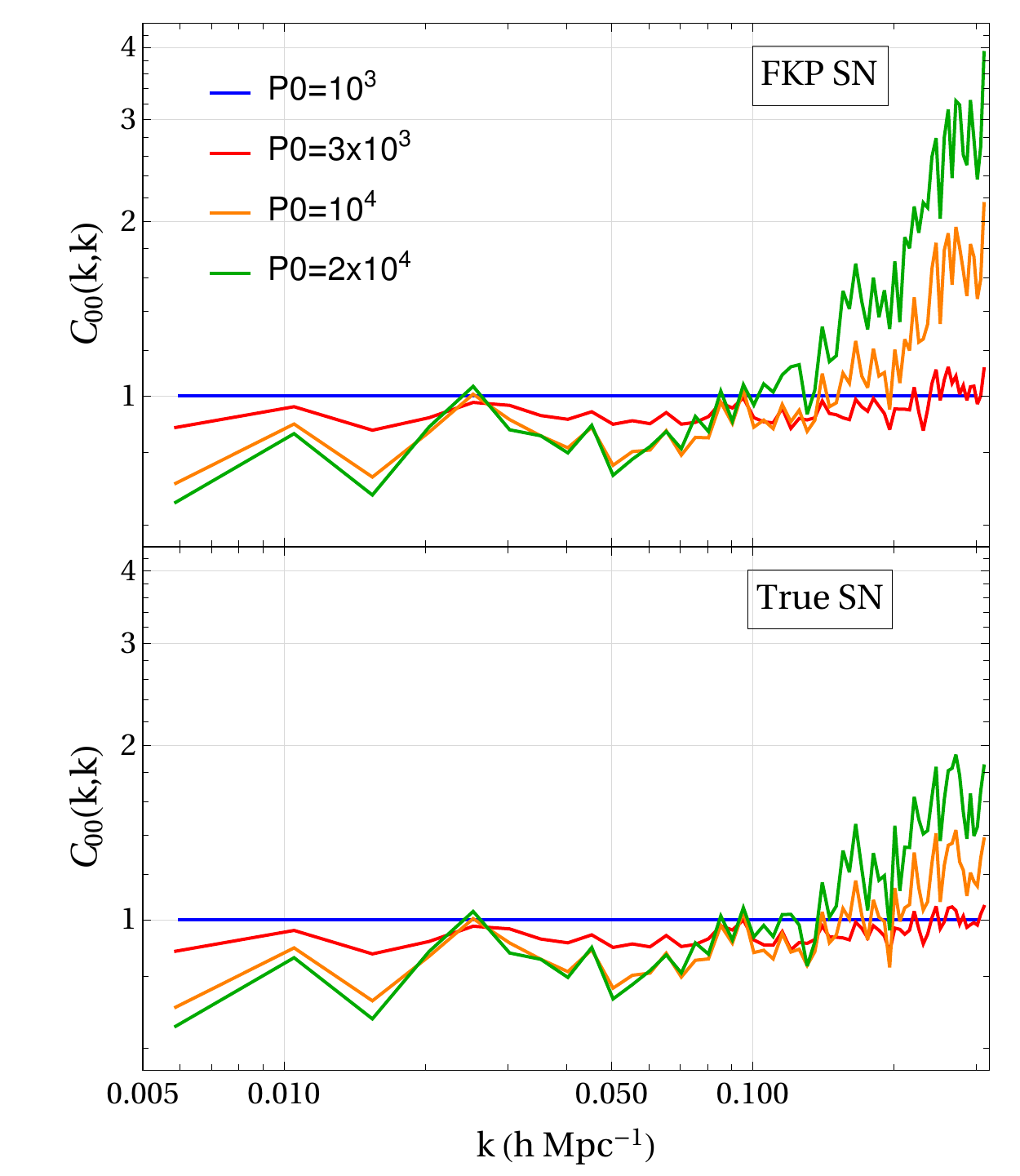}
\caption{Impact of the SN estimator (top: FKP shot noise, bottom: true shot noise) on varying the $P_0$ parameter in the FKP weights (Eq.~(\ref{eq:weight_FKP})), relative to the case with $P_0=10^3$~(Mpc/$h$)$^3$ in blue. It is more optimal to use a higher $P_0$ value at low-$k$ and vice versa in the FKP weights as the variance is lower.}
\label{fig:TrueSN_P0}
\end{figure}

\beq
\bigg( \sum_{j=1}^{\ng} +\alpha^2 \sum_{j=1}^{\nr}\bigg) w^2_j \longrightarrow \alpha(1+\alpha)\sum_{j=1}^{\nr} w^2_j
\label{TrueToFKPsn}
\eeq
This difference can have important consequences at small scales where the covariance is shot-noise dominated. To see this, Fig.~\ref{fig:TrueSN} shows the excess diagonal monopole covariance (top panel) as a function of $k$ that results from using the FKP SN as opposed to the true SN. The increased covariance is quite significant for $k\ga 0.2 \kMpc$, reaching 100\% enhancement by $k=0.4 \kMpc$. An even stronger effect is seen in the off-diagonals elements (bottom panel in Fig.~\ref{fig:TrueSN}). It is not difficult to understand why the covariance increases when using the expected SN, since there is an extra contribution from the SN fluctuations that now adds to the covariance. In Appendix~\ref{apx:FKP_SN_analytic} we discuss what these contributions are and present an evaluation of the extra covariance due to SN fluctuations that show good agreement with Fig.~\ref{fig:TrueSN}. We also explore the impact of the SN estimator on the cumulative signal-to-noise of the monopole power spectrum. 

There is one more interesting consequence of using the true SN. In the data analyses of most of the redshift surveys, the reference power spectrum $P_0$ parameter in the FKP weight is usually assumed constant  (Eq.~(\ref{eq:weight_FKP})) i.e independent of $k$. Following the BOSS analysis in~\cite{BeuSeoSai1704}, we have also adopted $P_0=10^4$ (Mpc/$h$)$^3$, independent of $k$, for all the results in this paper, including Fig.~\ref{fig:TrueSN}. Upon studying the effect of varying the $P_0$ parameter on the covariance, we find that a constant $P_0$ leads to a far from ideal FKP weight in the shot-noise dominated regime. Fig.~\ref{fig:TrueSN_P0} shows the change in the monopole covariance (relative to the case with $P_0=10^3$ (Mpc/$h$)$^3$) when using the FKP SN (top panel) and using the true SN (bottom panel). Overall, we see that the FKP estimator is optimal when one uses $P_0$ closer to the value of the power spectrum $P(k)$ at the $k$ being measured, as expected; in other words, a higher $P_0$ value at low-$k$ and vice versa. We also see that using the FKP SN degrades the signal (i.e. enhances the covariance) significantly at high-$k$ unless one tunes the choice of $P_0$ appropriately (something that in practice is never done). We plan to use different values of $P_0$ to analyze different $k$-bins of the power spectrum and probe the corresponding effect on cosmological parameter constraints in an upcoming work \cite{WadIvaSco20}.

\section{Understanding the Full Covariance}
\label{sec:SNres}

\begin{figure*}
\includegraphics[width=0.45\textwidth]{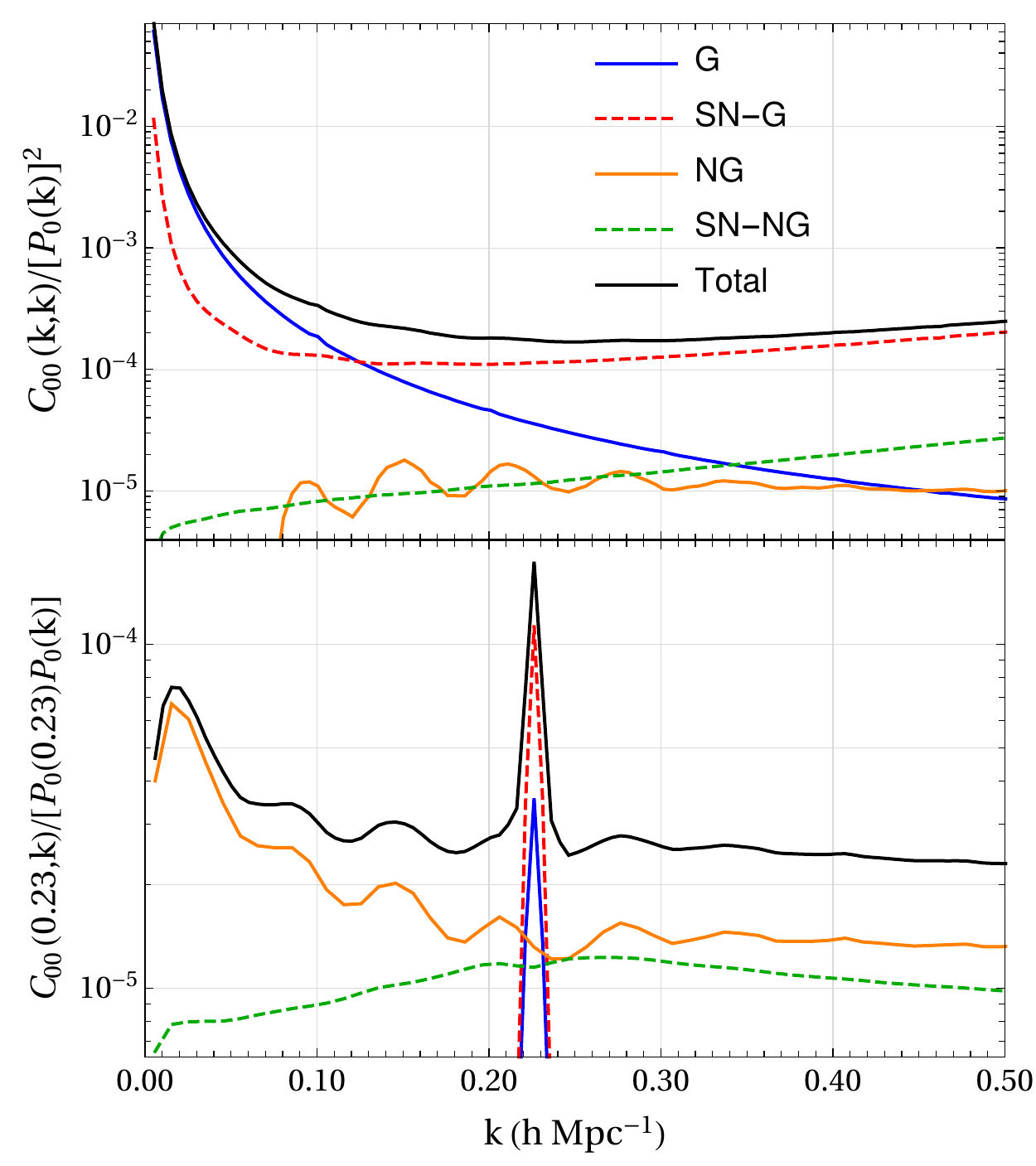}
\includegraphics[width=0.45\textwidth]{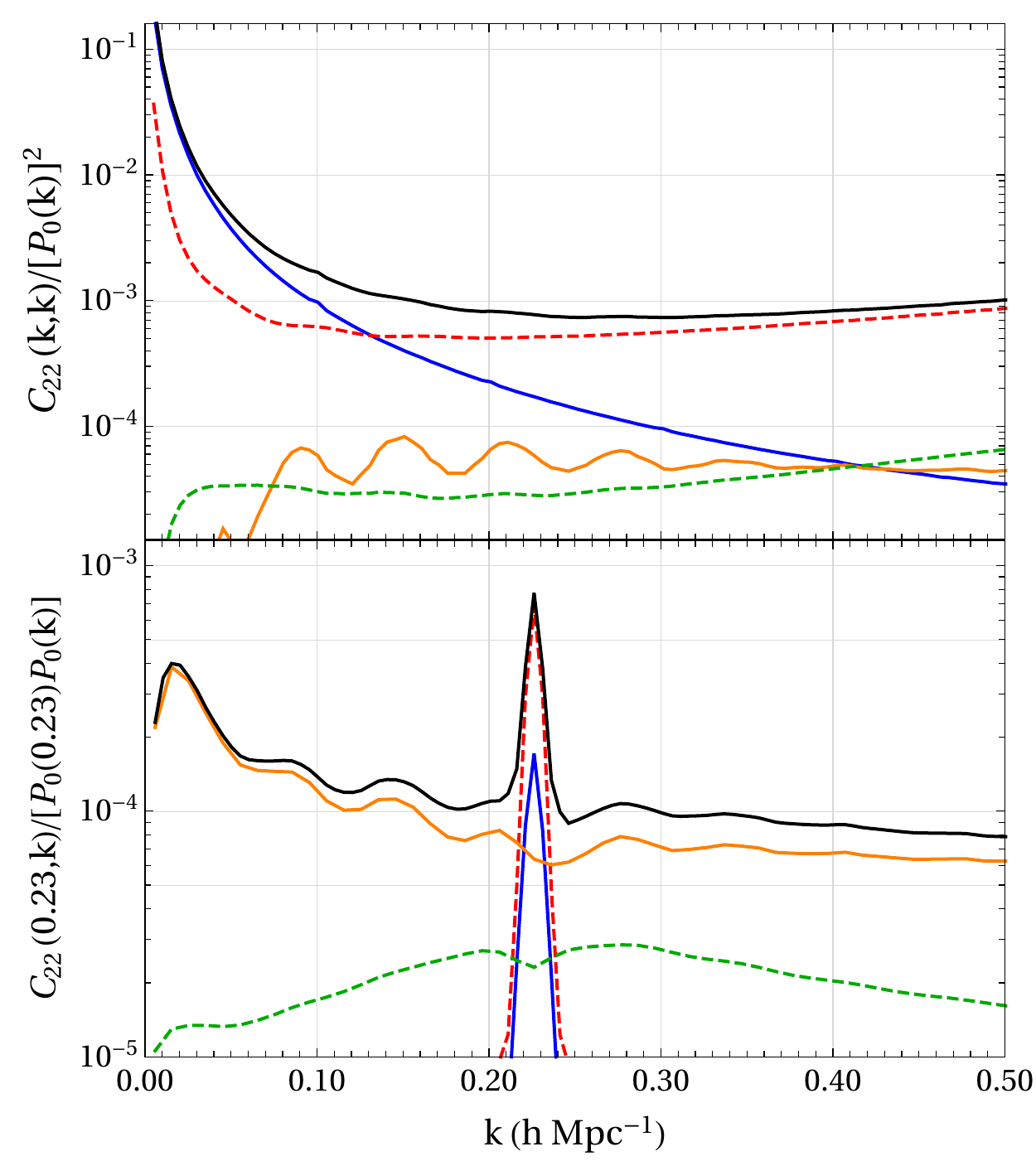}
\includegraphics[width=0.45\textwidth]{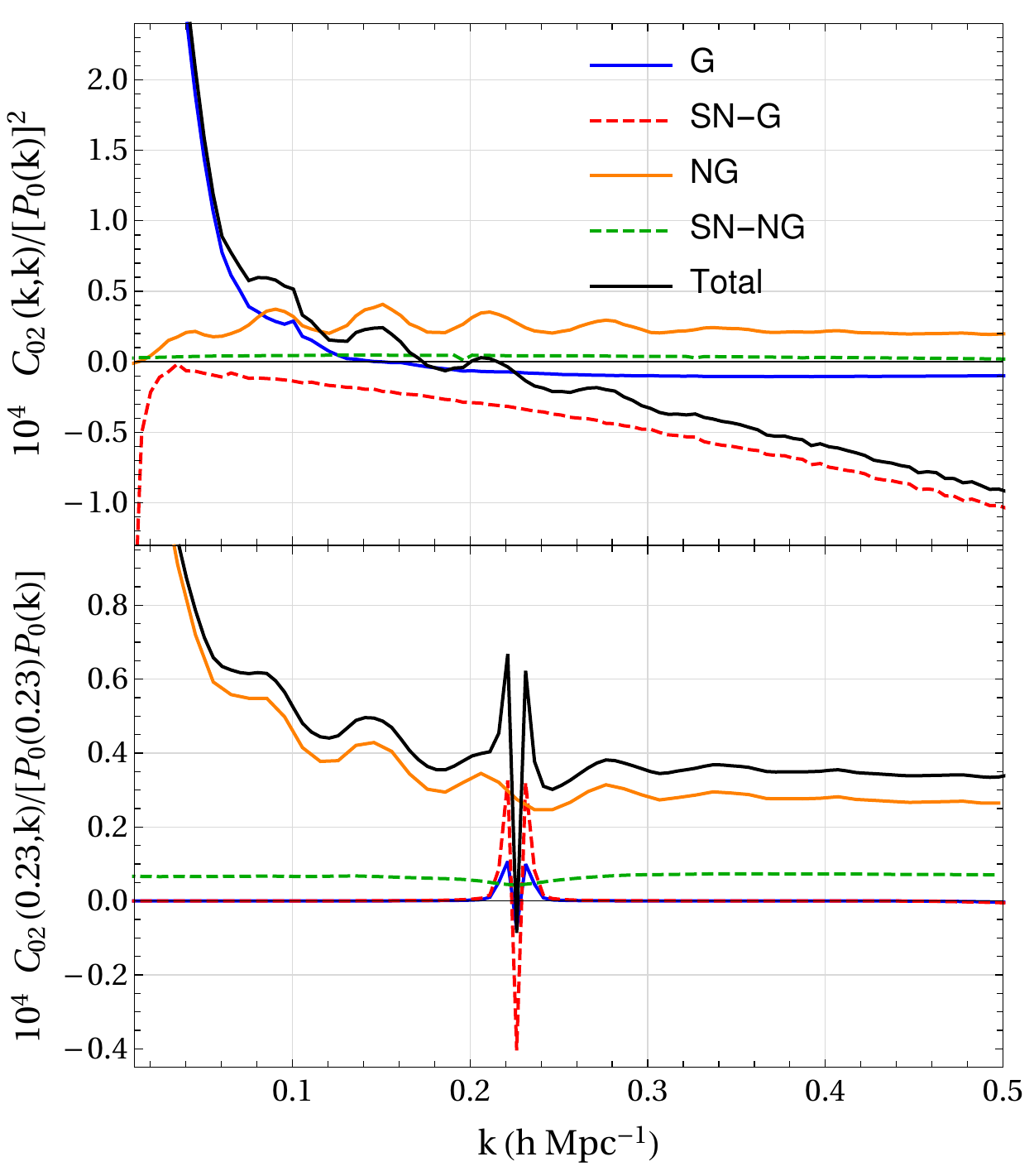}
\includegraphics[width=0.45\textwidth]{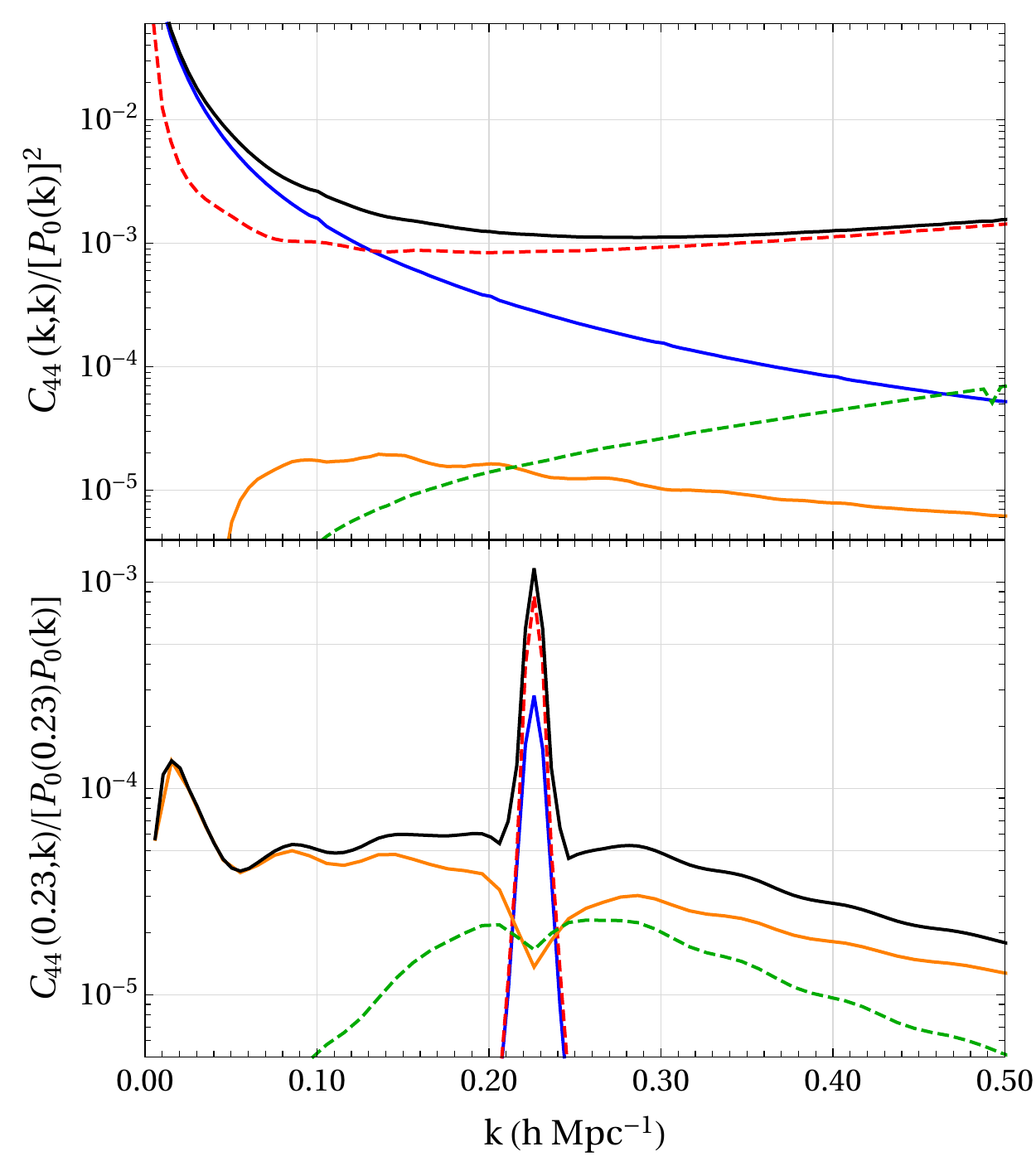}
\caption{Individual contributions to the multipole covariance matrix elements. Top (bottom) sub-panel shows diagonals (a particular row). In the case when the fields are continuous (i.e no shot noise (SN)), we show the gaussian part (G) of covariance in blue and the non-gaussian part (NG) in orange. The dashed lines represent the additional contribution of SN to the corresponding terms coming from discreteness of galaxies. The black line is the sum total of all the components. Before reaching the nonlinear regime, the diagonal covariance is already dominated by shot noise and therefore can be accurately modeled analytically.}
 \label{fig:SN_contrib}
\end{figure*}

Now that we have calculated all the relevant effects that enter in the prediction of the  galaxy power spectrum multipoles covariance in redshift surveys, we are ready to derive some insights from the relative size of different components. Fig.~\ref{fig:SN_contrib} shows  the contribution of each physical effect to the $\ell=0,2,4$ auto-covariances and the monopole-quadrupole cross-covariance. 

We break the contributions in terms of the continuous Gaussian  (G) and non-Gaussian (NG) covariances, and similarly the discreteness contributions (SN-G and SN-NG). Overall, we see from Fig.~\ref{fig:SN_contrib} the following behavior: at low-$k$, diagonal auto-covariances are dominated by their continuous Gaussian contribution, while at high $k$ by their Gaussian SN contributions (with a subdominant, although rising, non-Gaussian SN component).  
On the
other hand, the monopole-quadrupole cross-covariance shows a more non-trivial interplay between different effects, with the continuous non-Gaussian (NG) contribution dominating at intermediate scales along the diagonal.
The matrix elements, which are very close the diagonal, are mostly dominated by Gaussian components (the continuous and the discrete ones due to the width of the survey window as shown in Fig.~\ref{fig:SN_contrib} for the particular case of $k'=0.23 \kMpc$). On the other hand, the elements far from the diagonal are dominated by the NG continuous component (that includes regular trispectrum $T_0$, beat-coupling and local average effects) with a subdominant but not entirely negligible contribution from non-Gaussian SN. Looking at the breakdown of the NG continuous covariance in Fig.~\ref{fig:ED_components}, we see that the super survey mode coupling dominates the covariance at high-$k$. 

The main lesson from these results is that a perturbative treatment of the covariance is very well justified since before reaching the nonlinear regime (where PT breaks down); the covariance at high-$k$ is dominated by shot noise (which can be predicted fairly well from knowledge of the selection function) and by super survey modes (which are in the linear regime and can therefore be accurately modeled). In addition, complicated effects like loop corrections and velocity dispersion are subdominant for the same reason. More precisely, the situation is controlled by the quantity $\bar{n} P$, i.e. the importance of shot noise compared to the clustering signal. Typically, redshift surveys are designed so that $\bar{n} P \sim{\rm few}$ at the BAO scale \cite{FonMcDMos1405}, which implies that, at scales where non-linearities and velocity dispersion become strong, the covariance will generally be shot-noise dominated. This makes a perturbative approach to the covariance very compelling. For reference, the high-$z$ bin of the BOSS DR12 sample (which we use throughout the main text) has $\bar{n} P=1$ at $k=0.2 \kMpc$ and the low-$z$ bin (results presented in Appendix~\ref{apx:LowZ_comparison}) has $\bar{n} P=1$ at $k=0.3 \kMpc$. 

Finally it is worth estimating, very roughly, the impact that loop corrections might have on these results. Loop corrections for the matter power spectrum covariance have been computed in  in~\cite{BerSchSol1606,BerSchSol1606b,MohSelVla1704,TarNisJeo20} and start dominating at about $k\gtrsim 0.15 \kMpc$. We can see from Fig.~\ref{fig:ED_components} that, for the case of $C_{00}(0.23,\, k>0.15 \kMpc)$, the regular trispectrum ($T_0$) contribution to the continuous NG covariance is $\lesssim 20$\%. On the other hand, Fig.~5 in~Ref. \cite{MohSelVla1704} shows that loop corrections are about twice the regular trispectrum contribution at similar scales. If we also consider the contribution of shot noise to the non-diagonal elements as given in Fig.~\ref{fig:SN_contrib}, we can roughly infer that the loop corrections would have $\lesssim$ 20 \% contribution to the non-diagonal $C_{00}$ elements.
This is of course only a simple estimate, since the loop correction value is taken from the real-space matter covariance, and furthermore one expects significant competing effects from velocity dispersion at these scales. 

As there are many intermediate results in this work, we provide the reader with summary of the equations used in our final calculation of the covariance. We use Eq.~(\ref{eq:C_GaussFinal}) for the continuous Gaussian covariance and Eq.~(\ref{eq:covaSN-G}) for the shot noise contribution to it. The non-Gaussian part is made up of two components:
 the regular trispectrum in Eq.~(\ref{eq:CovaT0}) and the contribution from super survey modes which itself is a combination of Eq.~(\ref{CBCl1l2k1k2}) for beat-coupling and Eq.~(\ref{CLAredshift}) for local average effects (with the respective substitutions in  Eqs.~(\ref{eq:LOS_BC}, \ref{eq:LOS_BC2}) to account for the varying LOS effect).
Finally, there are various shot noise contributions to the non-Gaussian part which are discussed in Sec.~\ref{sec:SNcova} (Eqs.~(\ref{CTstep2}, \ref{eq:SN3}) for the regular trispectrum and Eqs.~(\ref{eq:4.2}, \ref{eq:SNresponse}) for the local average terms).

\section{Comparison with previous results in the Literature}
\label{sec:Literature_comparison}
Let us start by outlining a major difference in our approach and the extensive previous literature on how the super-survey modes affect the power spectrum and its covariance, in particular as it entails to galaxy redshift surveys. In this paper, we adopt the FKP estimator which is universally used to calculate the power spectrum from galaxy surveys,  Eq.~(\ref{eq:delta_FKP}), repeating it here for convenience
\begin{equation}
\hat{\delta}^\textup{FKP}(\x) \simeq \frac{1}{\sqrt{\I_{22}}} \frac{\delta_W (\x)}{(1+\dng)^{1/2}}\, ,
\label{eq:delta_FKP2}
\end{equation}
However, all the previous literature dealing with the effect of super-survey modes on galaxy surveys (eg. \cite{PutWagMen1204,TakHu1306, LiHuTak1404, LiHuTak1411,BalSelSen1609,MohSelVla1704,ChaDizNor1802,AkiTakLi1704,LiSchSel1802,AkiSugShi1907, KlyPra1910})
uses different overdensity and power estimators, which are
\begin{equation}
\hat{\delta} \equiv \frac{\delta_W}{1+\delta_b}\, ,\ \ \ \ \ \ \ \ \ \ \widehat{P} \equiv \frac{P_W}{(1+\delta_b)^2}\, ,
\label{eq:delta_OtherLit}\end{equation}
where $\delta_b$ is the super-survey {\em matter in real space} fluctuation, as opposed to the {\em redshift-space galaxy }  fluctuation $\dng$ in Eq.~(\ref{eq:delta_FKP2}).
The estimators in Eq.~(\ref{eq:delta_OtherLit}) were first used in~\cite{PutWagMen1204}, referred to as the deP12 estimators hereafter. The deP12 estimator is motivated by the fact that one typically normalizes the density fluctuations {\em in simulations} by the mean density of the simulated box. One of the differences between the FKP and deP12 estimators is immediately obvious: the normalization factor is relatively weaker in the FKP estimator. Physically, this arises because, for a non-trivial (space-dependent) selection function, the normalization is done at the level of the power spectrum, not the overdensity. 

One way to quickly gauge the effect on the covariance is to calculate the response of power spectrum to a super-survey mode for the two estimators, assuming a top-hat survey window, as is commonly done. Let us for simplicity take the case of no bias and no redshift distortions. For the deP12 estimators, we can rescale the response of the power spectrum to a super-survey mode in the presence of the normalization factors  as
\beq
 \frac{\partial P(k)}{\partial \delta_b} \rightarrow \frac{\partial P(k)}{\partial \delta_b} -2\, P(k)\, ,
\eeq
which gives \cite{LiHuTak1411}
 \begin{equation}
\frac{\partial P(k)}{\partial \delta_b} = P(k) \left(\frac{26}{21}- \frac{1}{3} \frac{d\ln k^3P}{d\ln k}\right)\, .
\label{eq:response1}\end{equation}
 
However, the FKP estimator leads instead to the following rescaling
\beq
\frac{\partial P(k)}{\partial \dng} \rightarrow \frac{\partial P(k)}{\partial \dng} -P(k)\, ,
\eeq
which gives
 \begin{equation}
\frac{\partial P(k)}{\partial \dng} = P(k) \left(\frac{47}{21}- \frac{1}{3} \frac{d\ln k^3P}{d\ln k}\right)\, ,
\label{eq:response2}\end{equation}
which is the same as our earlier Eq.~(\ref{meanPKFPreal}) but for the particular case of unbiased tracers ($b_1=1$, $b_2^{\rm sph}=0$) and a top-hat window.
Comparing Eqs.~(\ref{eq:response1}) and (\ref{eq:response2}), we see that the final response of the power spectrum is about $ 2.3$ times larger at $k=0.1 \kMpc$ for the FKP estimator as compared to the deP12 estimator. To see the effect of the estimator on the covariance is straightforward because the contribution of super-survey modes to the covariance is given by the widely used super survey covariance result~\cite{TakHu1306}
\begin{equation}
\textbf{C}^\textup{SSC}(k_1,k_2)= \sigma^2_{\rm TH} \ \frac{\partial P(k_1)}{\partial \delta_b}\, \frac{\partial P(k_2)}{\partial \delta_b}\, .
\label{eq:SSC}\end{equation}
We thus find that the real-space SSC matter covariance obtained using the FKP estimator is about {\em five times larger} at $k=0.1 \kMpc$ than that obtained using the deP12 estimator.

As we stressed already, there is another important difference between the estimators in Eqs.~(\ref{eq:delta_FKP2}) and~(\ref{eq:delta_OtherLit}) as $\delta_b \neq \dng$. The SSC approach calculates responses with respect to the real-space matter mode $\delta_b$, and to include the effect of redshift-space distortions, calculates in addition responses with respect to the tidal field. To include shot noise and non-local bias up to cubic order, one presumably needs to calculate additional responses. In our approach, we bypass all this: 1) we write the estimator of galaxy fluctuations, 2) we calculate its power spectrum covariance. The FKP estimator in Eq.~(\ref{eq:delta_FKP2}) has $\dng$ which is in equal footing with $\delta_W$, i.e. $\dng$ corresponds to the galaxy fluctuations in redshift space, and has in it nonlinear evolution, bias, redshift distortions and shot noise as discussed in the previous sections.

We have checked that for the case of matter with a top-hat survey window and the deP12 estimator, our results agree with the SSC approach results. Beyond this, we are not aware of results with nonlinear and non-local bias and arbitrary survey windows in redshift space. It is worth emphasizing that for a realistic window function one can no longer write the covariance due to super-survey modes (BC+LA in our nomenclature) as a product of responses times the variance of the super-survey modes as seen in Eq.~(\ref{eq:SSC}). Instead the result contains a number of terms weighted by different variance measures, e.g. see Eq.~(\ref{eq:CovaNOrsdSimplified}) where  $\sigma_{22}^2=\sigma_{22 \times 10}^2=\sigma^2_{10}$ (defined in Eqs.~\ref{sigma22}, ~\ref{sigma10x22} and~\ref{sigma10}) follows only for a top-hat survey window. For a realistic survey window such as  BOSS DR12, the quantities $\sigma_{22}^2, \sigma_{22 \times 10}^2$ and $\sigma^2_{10}$ differ up to $ 35$\% from each other.  Physically, the appearance of different variance measures is due to the difference in origins of the effects: trispectrum, bispectrum and power spectrum, respectively. 
Another major difference between our calculations for the super-survey modes and previous work on super survey modes is that we implement the radial redshift distortions (taking into account the changing LOS across the survey) instead of assuming that the LOS is fixed throughout the survey volume.

\begin{figure*}
        \includegraphics[scale=0.5,keepaspectratio=true]{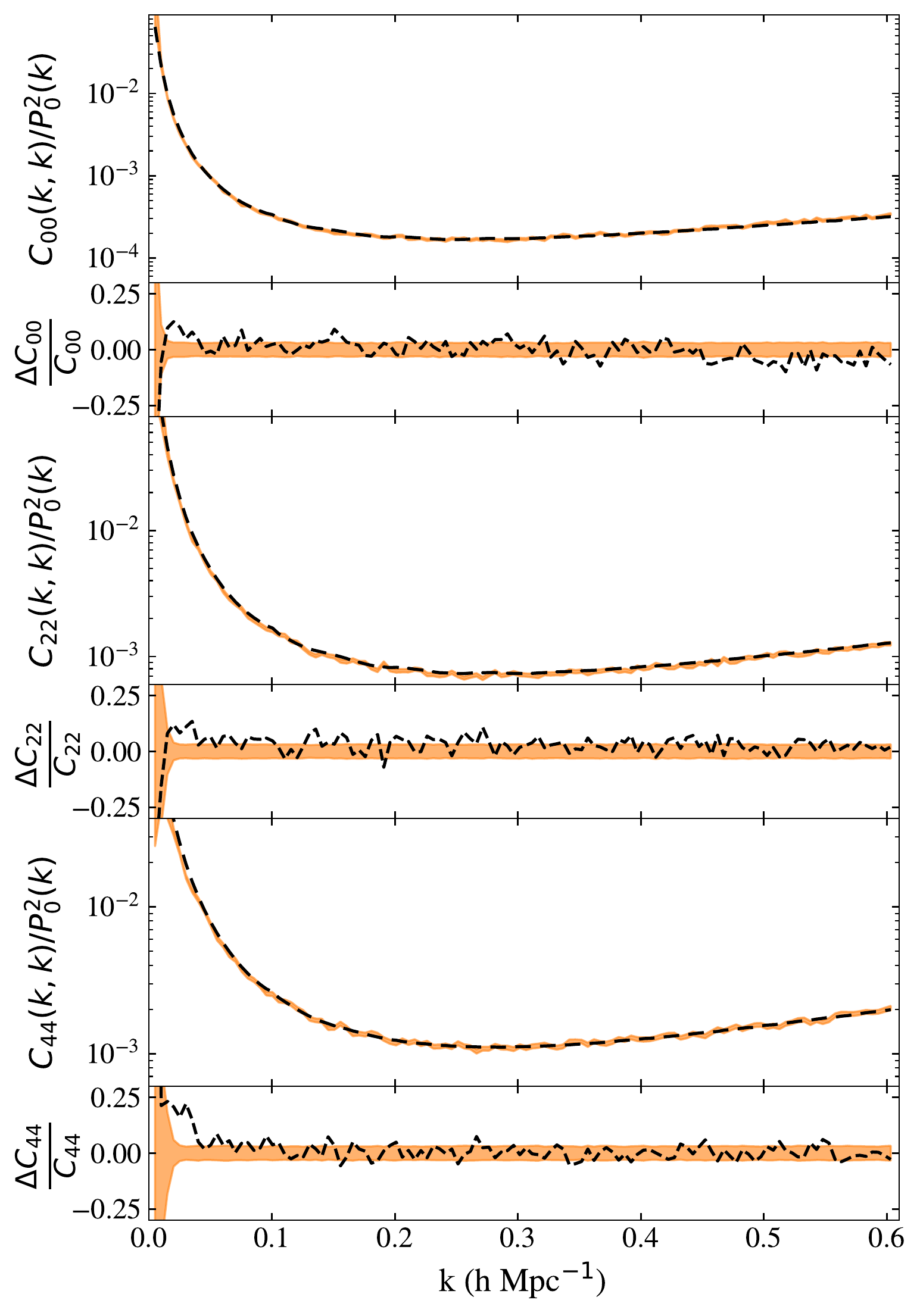}
        \includegraphics[scale=0.5,keepaspectratio=true]{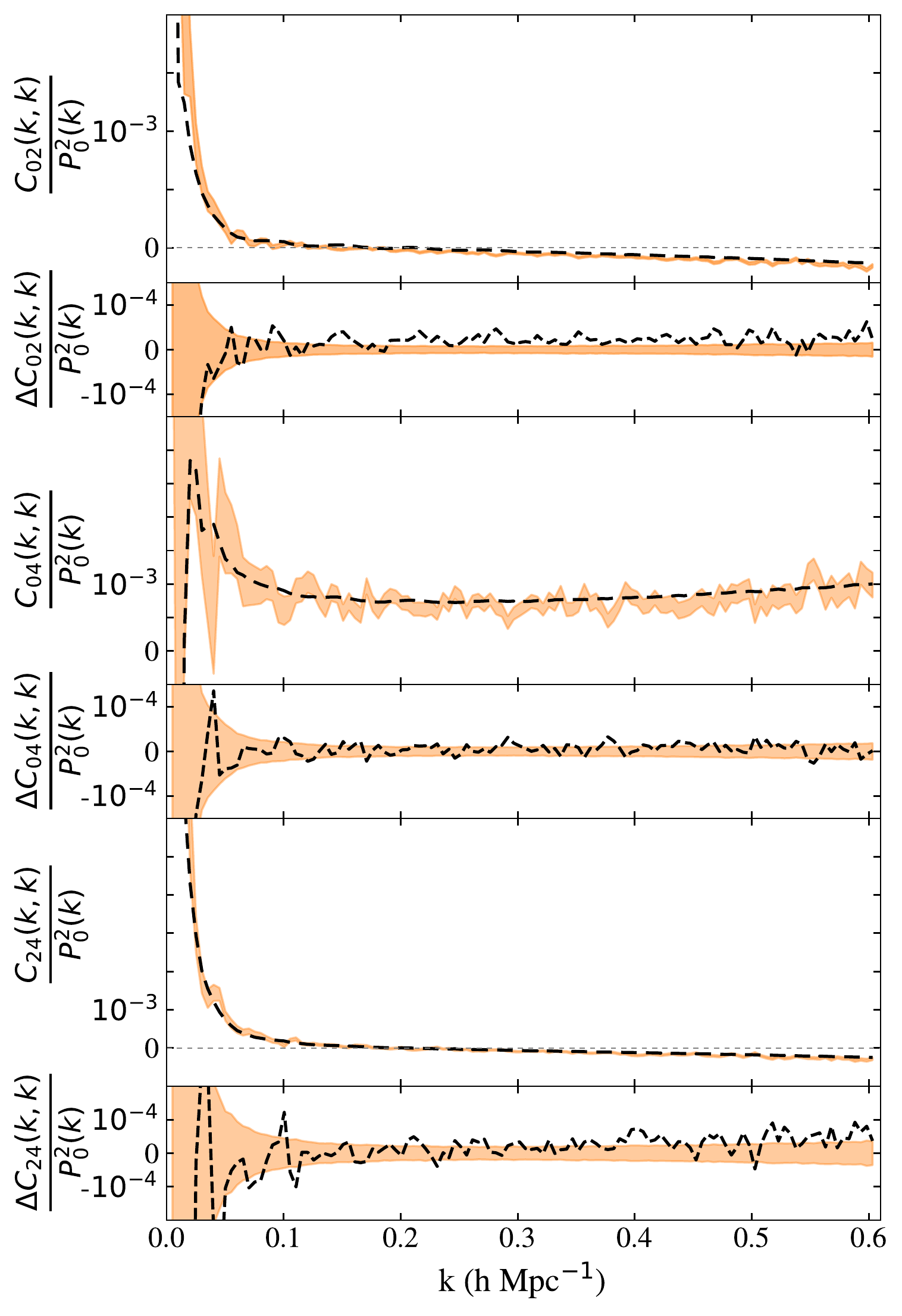}
    \caption{Diagonal elements of the auto-covariance (left panels) and cross-covariance (right panels) matrix of power spectrum multipoles obtained from the BOSS DR12 Patchy mocks (orange band) and the corresponding predictions from our perturbative calculation (dashed black). We find an excellent agreement even though no parameter fitting to the Patchy mocks was performed in our analysis.}
    \label{fig:Ckk}
\end{figure*}

Let us now discuss the comparison of our Gaussian covariance with the previous literature. The work in~\cite{GriSanSal1604} provides a simple analytic estimate of the gaussian multipole covariance but does not include the effect of changing LOS and the non-trivial survey geometry. Ref.~\cite{BlaCarKod1810} includes the effect of changing LOS but their expressions are relatively more complicated and need Monte-Carlo integration to evaluate; we use simplified expressions using well-motivated assumptions --- the major simplification in our case coming from neglecting the convolution of the power spectrum with the survey window as given in our Eq.~(\ref{eqn:convolved_power}).
Ref.~\cite{LiSinYu1811} also includes the effect of changing LOS but briefly argues that their results in the plane-parallel approximation can be generalized to a varying LOS by replacing $\hat{\textbf{n}} \to \hat{\x}$. However, as we derive rigorously here, it is not as simple, since even the expansion of the power spectrum in multipoles does not make sense with a varying LOS. We therefore use the local power spectrum $P_{\rm local}(\k;\x)$ which does admit such an expansion. Ignoring this point, our Gaussian covariance including shot noise and that in~\cite{LiSinYu1811} are similar, up to the replacement 
\beq\begin{split}
&\langle F_{\ell_1}(\k_1) F_{\ell_2}(-\k_2)\rangle \langle F_{0}(\k_2) F_{0}(-\k_1) \rangle\\
&\rightarrow \langle F_{\ell_1}(\k_1) F_{0}(-\k_2)\rangle \langle F_{\ell_2}(\k_2) F_{0}(-\k_1) \rangle
\end{split}\label{eq:LiSin19}\eeq
in our Eq.~(\ref{eq:3.2}). This makes a difference only for the covariance of higher-order multipoles. When we compare our results to the measurements of the covariance from mock catalogs (discussed in Sec.~\ref{sec:Patchy_compare}),
we find the replacement in Eq.~(\ref{eq:LiSin19}) introduces some amount of inaccuracy. For the particular case of diagonals of the following three matrices, our results become less accurate by: $\sim$5\% for $\C_{22}$(quadrupole auto-covariance), $\sim$10\% for $\C_{44}$ and $\sim$20\% for $\C_{24}$.

\section{Comparison with Mock Catalogs}
\label{sec:Patchy_compare}

To show how well our  approach for the covariance works, we compare our results with the V6C MultiDark-Patchy mock galaxy catalogs~\cite{KitRodChu1603} (hereafter referred to as Patchy mocks), which were used in SDSS-BOSS parameter estimation analysis~\cite{AlaAtaBai1709}. These catalogs were generated by using the \textsc{patchy} code \cite{KitYepPra14} and calibrated using the BigMultiDark $N$-body simulation \cite{KlyYep1604, RodChuPra1610}. BOSS DR12 uses two main non-overlapping redshift bins in their analysis. We compare our method to the mocks in the North Galactic Cap (NGC) region; in this section we show  results corresponding to the the high redshift bin ($0.5<z<0.75$) while  results for the low redshift bin ($0.2<z<0.5$) are shown in Appendix~\ref{apx:LowZ_comparison}. The high-$z$ bin used here has a mean redshift of $z=0.58$ with a corresponding  linear growth factor of $D_+=0.82$ and growth rate  $f=0.78$, derived from the fiducial cosmology adopted in the Patchy mocks:  $\Omega_m=0.307$, $h=0.678$ and $\sigma_8=0.829$. To minimize the shot noise contribution from the random catalog, we use a random catalog with $100$ times the number density of the mock catalogs (i.e. $\bar{\alpha}=0.01$).  We measure the redshift space power spectrum in 2048 realizations of the mocks using the estimator in~\cite{Sco1510} implemented numerically following~\cite{SefCroSco1512}.

To make our predictions, we also need to specify the galaxy bias parameters. We obtain the linear bias
 parameter by fitting the Patchy power spectrum monopole at low-$k$ predicted using linear PT, giving $b_1=2.01$.  The nonlinear bias parameters at quadratic and cubic order are not obtained by fitting to the Patchy data, instead we obtain the local bias parameters by using the fitting functions to N-body simulations for $b_n(b_1)$ of~\cite{LazWagBal1602} where we get\footnote{Note that \cite{LazWagBal1602} uses a different basis of biased operators and their values have been appropriately converted (using Eqs.~C28 and C31 of \cite{EggScoSmi1906}) to match our bias scheme.} $b_2 = -0.47$ and $b_3= -2.47$. For the non-local bias parameters we use the bias basis of~\cite{ChaScoShe1204,EggScoSmi1906} and assume the 
local Lagrangian approximation,

\begin{figure*}
    \centering
        \includegraphics[scale=0.5,keepaspectratio=true]{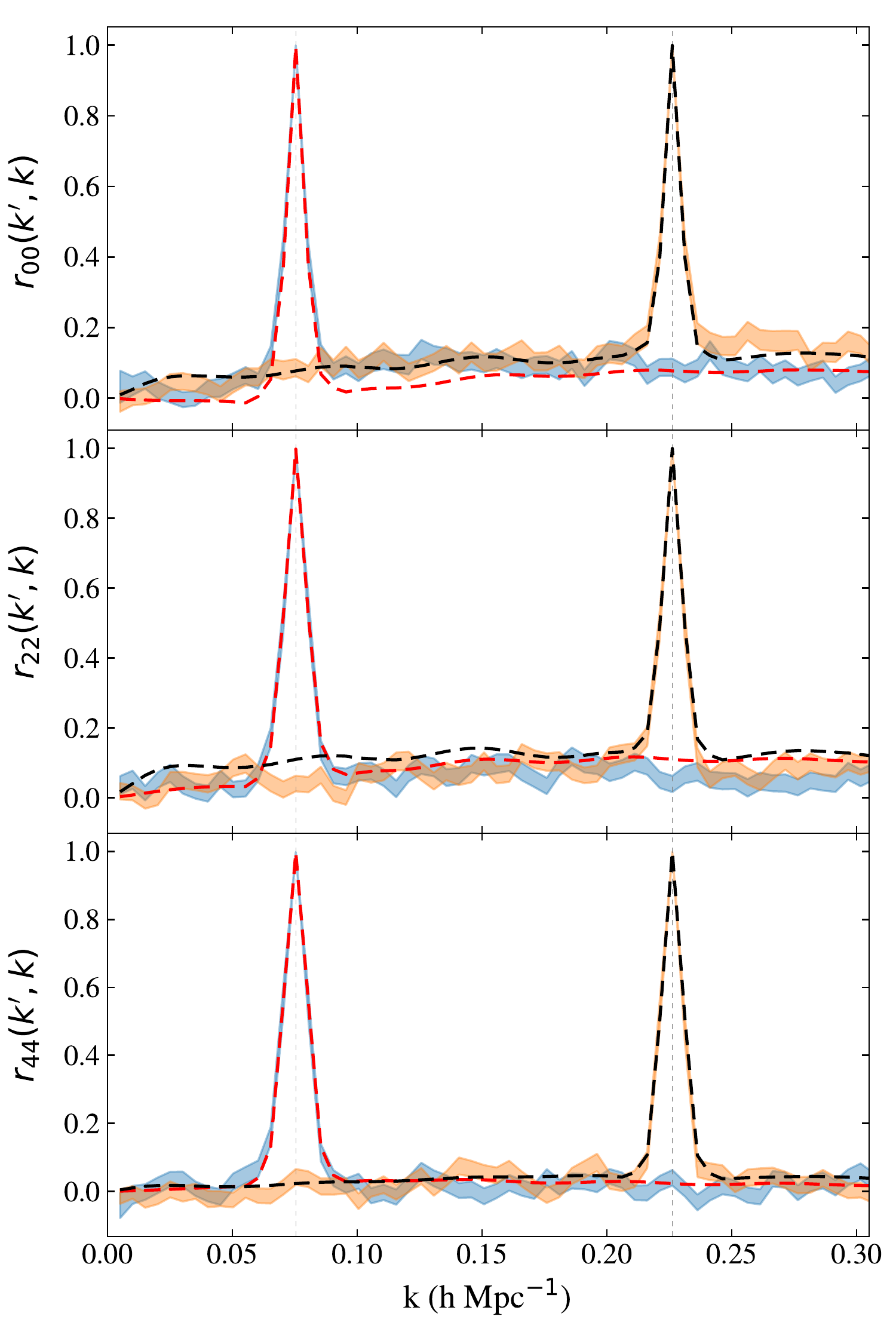}
        \includegraphics[scale=0.5,keepaspectratio=true]{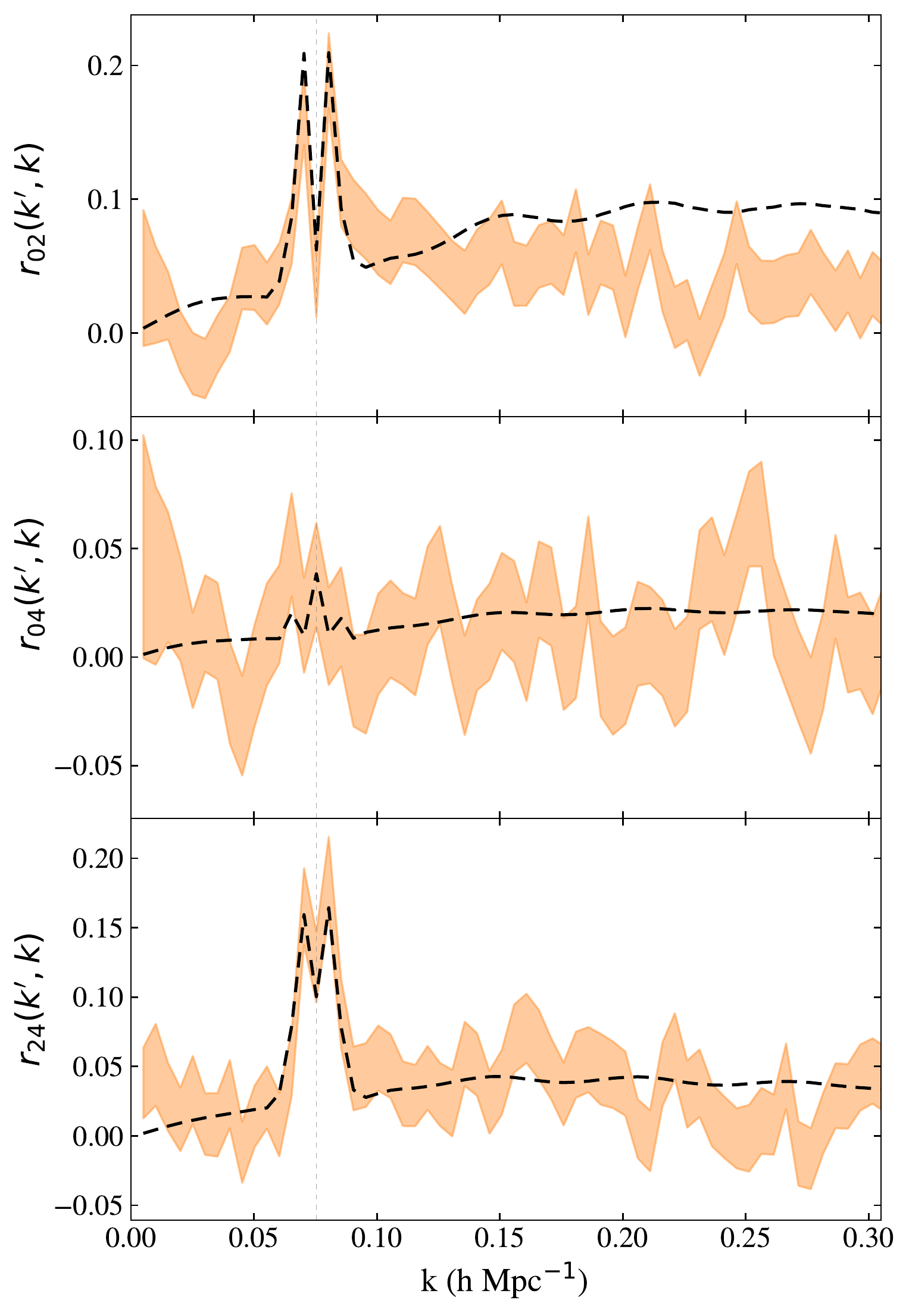}
        \includegraphics[scale=0.5,keepaspectratio=true]{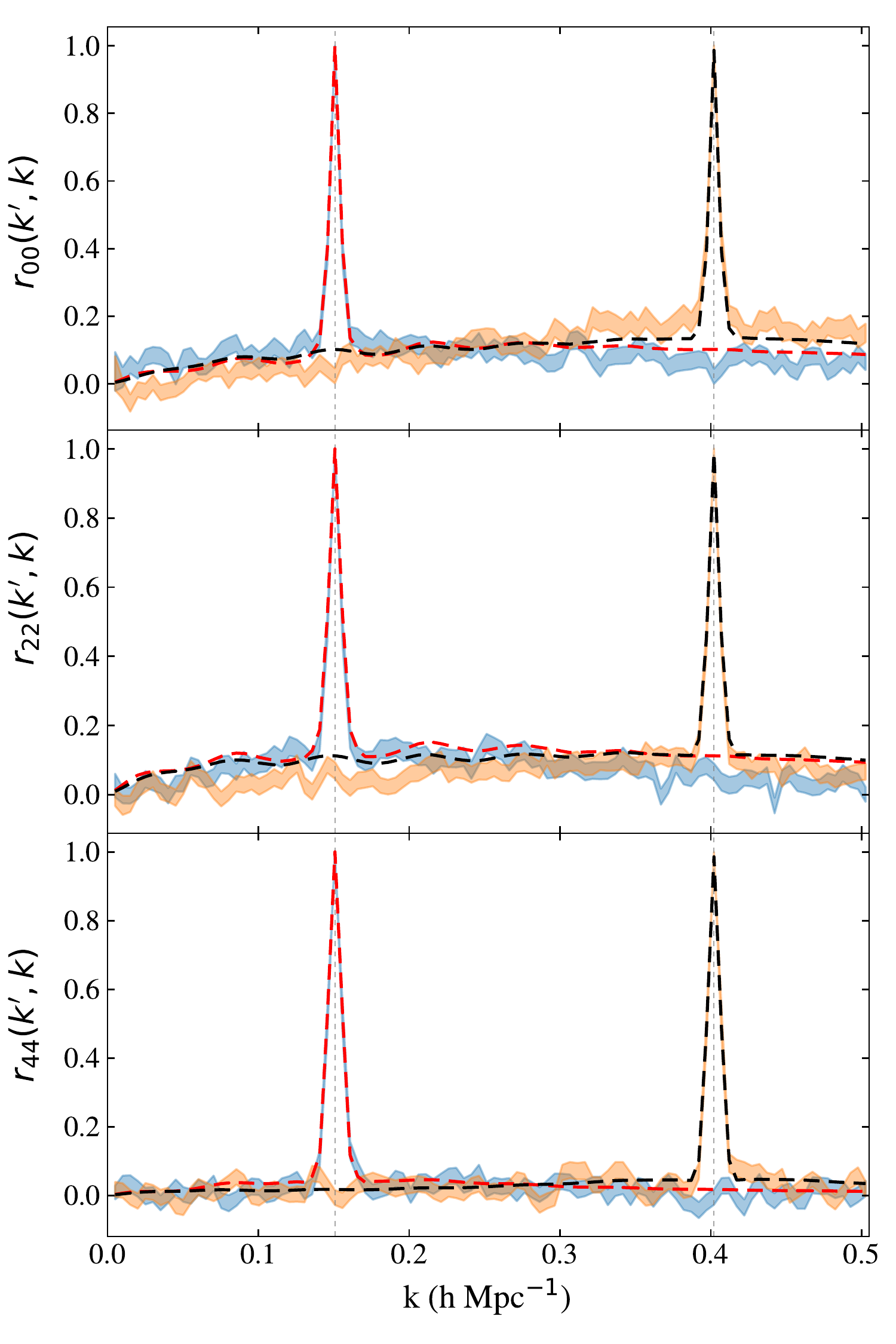}
        \includegraphics[scale=0.5,keepaspectratio=true]{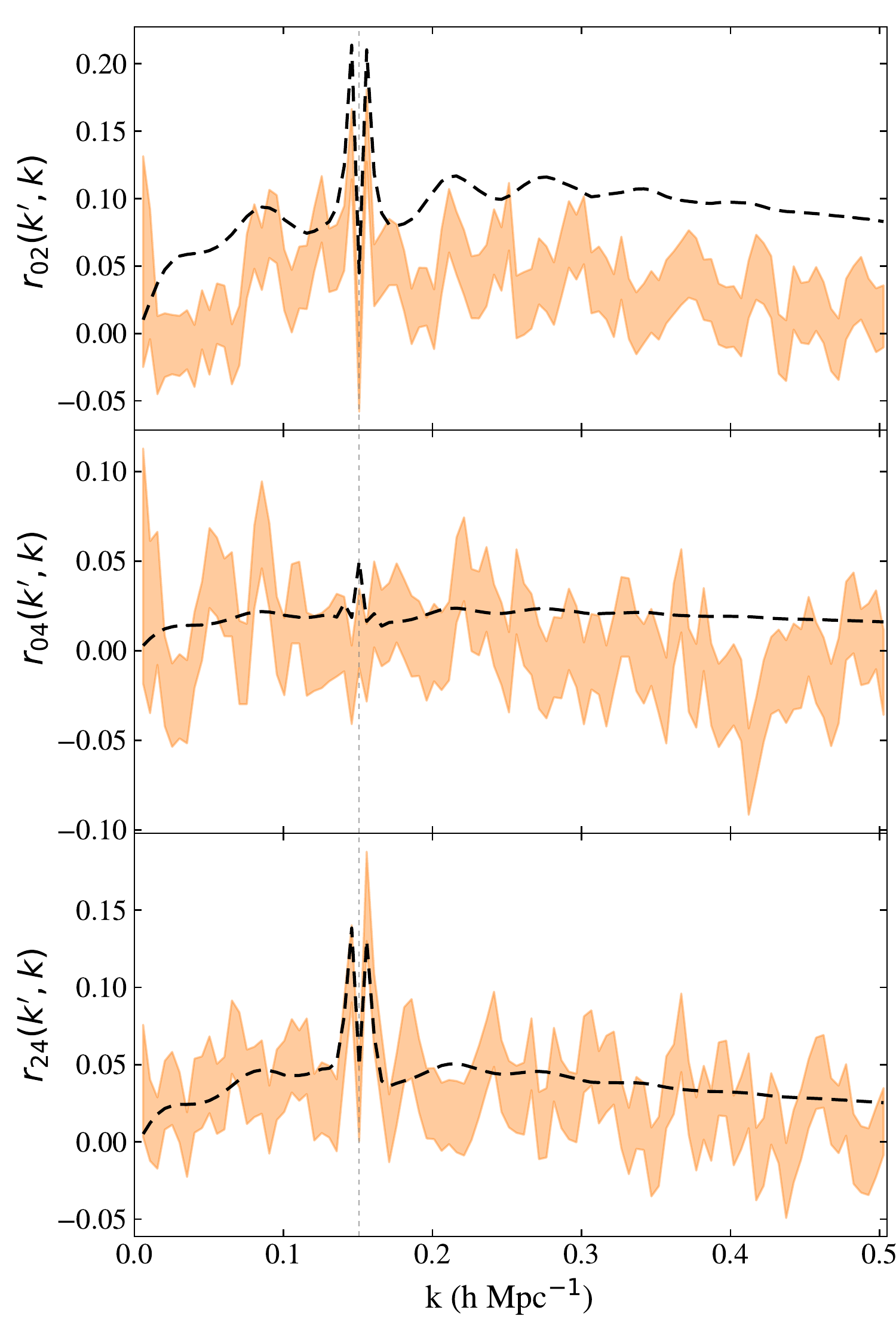}
    \caption{Rows of the reduced auto and cross covariance matrix of power spectrum multipoles obtained from the Patchy mocks (orange and blue) and the corresponding row from our perturbative calculation (dashed red and dashed black). Even though 2048 Patchy mocks were generated for the BOSS DR12 analysis, there is still a large numerical noise in the Patchy cross-covariance.}
    \label{fig:rij}
\end{figure*}

\beqa 
&\gamma_2&= -{2\over7} (b_1-1), \ \ \ \ \ \ \ \ \ \ \gamma_3={11\over 63}(b_1 - 1), \nonumber \\
& \gamma_2^\times& = -{2\over 7} \, b_2, \ \ \ \ \ \ \ \ \ \ \ \ \ \ \  \gamma_{21} = -{22\over 147} (b_1-1),
\label{LLA}
\eeqa
which has been shown to be quite accurate when compared to numerical simulations in the literature~\cite{ChaScoShe1204,SheChaSco1304,SaiBalVla1405,LazWagBal1602,AbiBal1807} and has also been shown to be consistent with the BOSS data (see Fig.~4 of \cite{EggScoCro20} and also \cite{IvaSimZal19}). 
In principle, a more accurate estimate of the nonlinear bias parameters in the mock catalogs can be obtained by fitting the bispectrum and trispectrum to the Patchy mocks. However, since the Patchy method itself is  approximate (using 2LPT combined with spherical collapse for the nonlinear dynamics, plus prescriptions for bias and velocity dispersion), our simpler approach seems a very reasonable first step.

The Patchy mocks are constructed by reshaping simulation boxes of  (2.5 Gpc/$h$)$^3$ volume, which while larger than the survey volume, is still missing larger scale modes and this might cause the  beat coupling and local average effects to not be entirely correct.  In fact, a recent paper~\cite{SugSaiBeu1908} argues that such effects cannot be trusted in Patchy mocks. We disagree with this assessment for the following reasons. 
First, because the large-scale power spectrum below the turnover is so blue,  the integrals describing the amplitude of BC and LA effects are dominated
 by their high-$k$ cutoff set by the width of the survey window, not the largest super survey mode.  Second, one can actually compare how well the variance of  $N_g$ fluctuations measured in the Patchy mock realizations agrees with theoretical expectations, and they do fairly well: while the measurements give $\langle \dng^2 \rangle = 3.095\times 10^{-5}$, Eq.~(\ref{eq:4.2}) gives $3.286\times 10^{-5}$ integrating over all super survey modes
, a difference of less than 6\%. This shows that effects coming from super survey modes can be trusted in Patchy mocks to a good accuracy.  

As a result of this, in our predictions, we simply integrate over all super survey modes. 
This point highlights an another advantage of a perturbative calculation, where it is straightforward to account for the super survey modes (which are in the linear regime). On the other hand, such modes increase the expense of using numerical simulation methods (because larger simulation volumes are required). 
 
To estimate the power spectrum covariance from a sample of $\textup{N}_\textup{m}$ mocks, we use 
\begin{equation}\begin{split}
&\hat{\textbf{C}}(k_i,k_j)\\
&\equiv \frac{1}{\textup{N}_\textup{m}-1} \left[\sum_n^{\textup{N}_\textup{m}}\, [P^{(n)}(k_i)- \bar{P}(k_i)] [P^{(n)}(k_j)- \bar{P}(k_{j})] \right]\, ,
\end{split}\end{equation}
where the sample mean power spectrum is given by $\bar{P}(k_i)=\sum_n^{\textup{N}_\textup{m}} P^{(n)}(k_i)$. To estimate the error on the measured covariance, we use bootstrapping to generate new sets of $\textup{N}_\textup{m}$ mocks and compute the error bars from the scatter in the resulting covariances. In Appendix~\ref{apx:CovaError}, we estimate analytically the expected covariance error and find it to be in good agreement with the bootstrap technique.

\begin{figure*}
    \centering
        \includegraphics[scale=0.6,keepaspectratio=true]{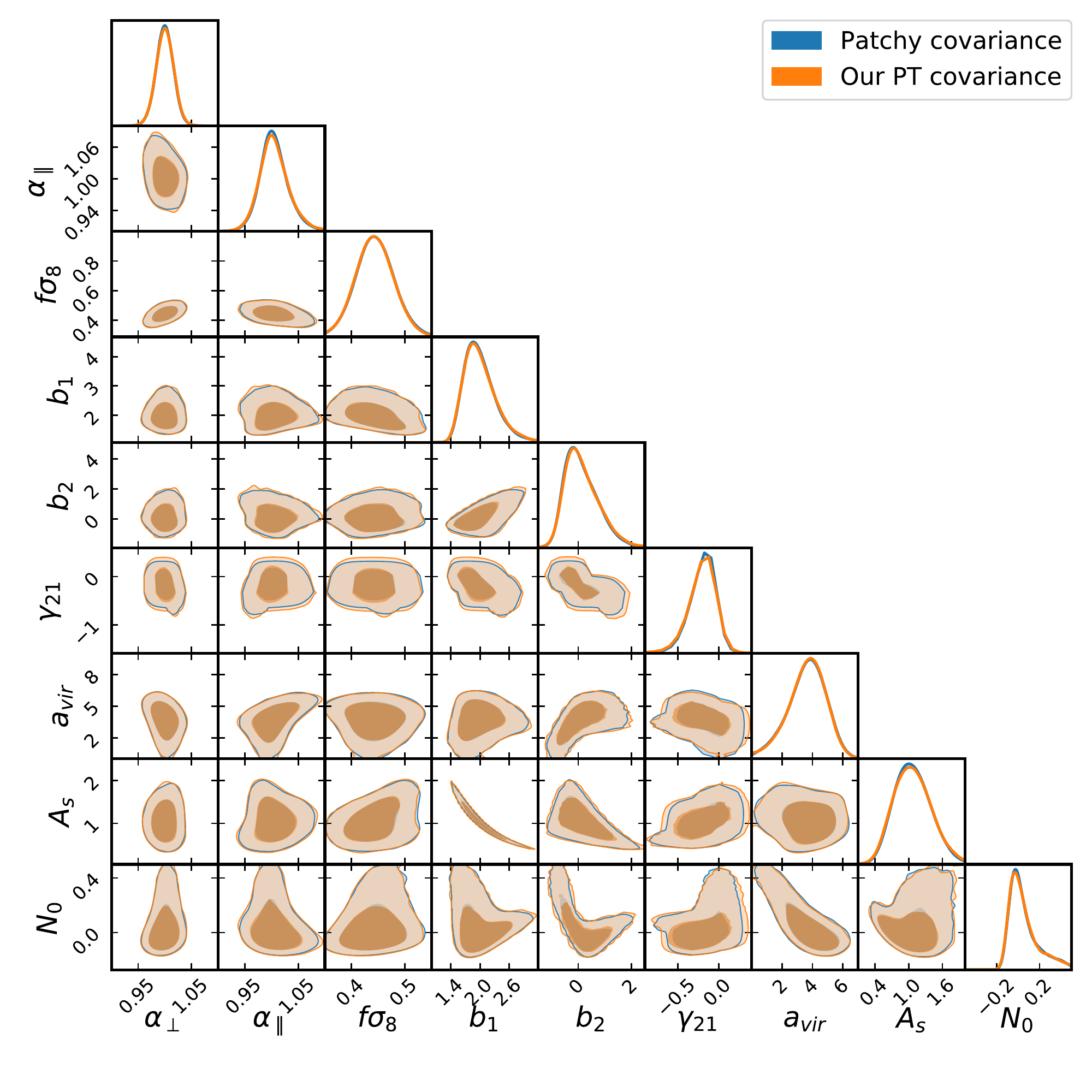}
    \caption{Constraints on cosmological parameters (for fixed linear spectrum shape) when changing the galaxy power spectrum multipoles covariance matrix: covariance from the BOSS DR12 Patchy mocks (blue) and from our perturbative calculation (orange). The data vector in this analysis is taken as the theoretical model computed at the fiducial cosmological parameters. The results for the full BOSS parameter analysis will be presented in an upcoming work~\cite{WadIvaSco20}.}
    \label{fig:param_constraints}
\end{figure*}
Let us now compare the diagonal elements of the Patchy covariance matrix to our analytic calculations. Fig.~\ref{fig:Ckk} shows the auto-covariances on the left panels, and the cross-covariances on the right panels for all possible multipole combinations that include $\ell=0,2,4$. We see that the PT predictions (dashed lines) are in excellent agreement with the measurements from the Patchy mock catalogs (orange bands quantifying the error bars). In each case we also show the relative error, note in this case the dashed line becomes noisy due to the scatter in the Patchy covariance matrix. We again emphasize that no fitting at all has been done to make this comparison;  as explained above, we only fit the linear bias parameter $b_1$ to the low-$k$ power spectrum (not to the covariance matrix) and the remaining bias parameters are not fit to the data (see Eq.~\ref{LLA}). Appendix~\ref{apx:LowZ_comparison} extends the comparison in Fig.~\ref{fig:Ckk} to the low-$z$ bin ($0.25<z<0.5$) with similar conclusions.

In Fig.~\ref{fig:rij} we compare the extradiagonal elements of the auto (left panels) and cross (right panels) covariances. We plot the following reduced covariances, 
\begin{equation}
r_{\ell,\ell'} (k,k') = \frac{C_{\ell,\ell'} (k,k')}{\sqrt{C_{\ell,\ell} (k,k) C_{\ell',\ell'} (k',k')}}\, .
\end{equation}
For the auto-covariances (left panels, $\ell=\ell'$) this corresponds to the standard cross-correlation coefficient, whereas for the cross-covariances the definition of the denominator (which always involves the diagonals of the {\em auto}-covariances) guarantees that the reduced covariance is bounded, avoiding the problem of having zeros in the denominator if one were to use e.g. $C_{\ell,\ell'} (k,k)$.  

Overall the level of agreement seen in Fig.~\ref{fig:rij} is very good, especially given that we have not fitted the nonlinear bias parameters which have more impact for extradiagonal matrix elements. In addition, recall that unlike the case of the Gaussian covariance, for the non-Gaussian covariance which dominates along the extradiagonal elements, we do not include any velocity dispersion effects, thus there is certainly room for improvement.
This is particularly important in the monopole-quadrupole cross-covariance, which as we saw in Sec.~\ref{sec:SNres} is most subject to nonlinear corrections not included in our treatment (being the least shot-noise dominated).

Until now we have directly compared the individual elements of covariance matrix, let us now compare the impact of the covariance matrix on cosmological parameter constraints. We use the theoretical framework presented in \cite{SanScoCro1701} to determine errors on cosmological parameters and show the results in Fig.~\ref{fig:param_constraints}. The shape of the linear spectrum is fixed for simplicity, and we take as data vector the theoretical model computed at the fiducial cosmological parameters (we will present results for the full BOSS parameter analysis in an upcoming work~\cite{WadIvaSco20}). We performed the current analysis for $k_{\textup{max}}=0.3$ h/Mpc and used the following parameters: the bias parameters as discussed earlier; the standard anisotropic Alcock-Paczynski parameters $\alpha_\perp$ and $\alpha_\parallel$;  a parameter related to the small-scale galaxy velocity dispersion $a_\textup{vir}$; the normalization factor in the power spectrum $A_\textup{s}$; and a constant shot noise term in the power spectrum estimator $N_0$. The Markov chain Monte Carlo (MCMC) chains were run using the COMPASS code. The one-loop theoretical calculation  for the galaxy power spectrum multipoles follows \cite{SanScoCro1701}, where it was shown to robustly extract cosmological information at scales larger than 20 Mpc/h by testing it with QPM, Patchy mocks, and  on the Minerva simulations~\cite{GriSanSal1604}. As in~\cite{SanScoCro1701}, all one-loop bias parameters were free except for $\gamma_2$ which, being degenerate with $\gamma_{21}$, is fixed to the local Lagrangian value in Eq.~(\ref{LLA}). The results of Fig.~\ref{fig:param_constraints} show that the constraints in this simple test are essentially indistinguishable.

\begin{figure}
\includegraphics[scale=0.55,keepaspectratio=true]{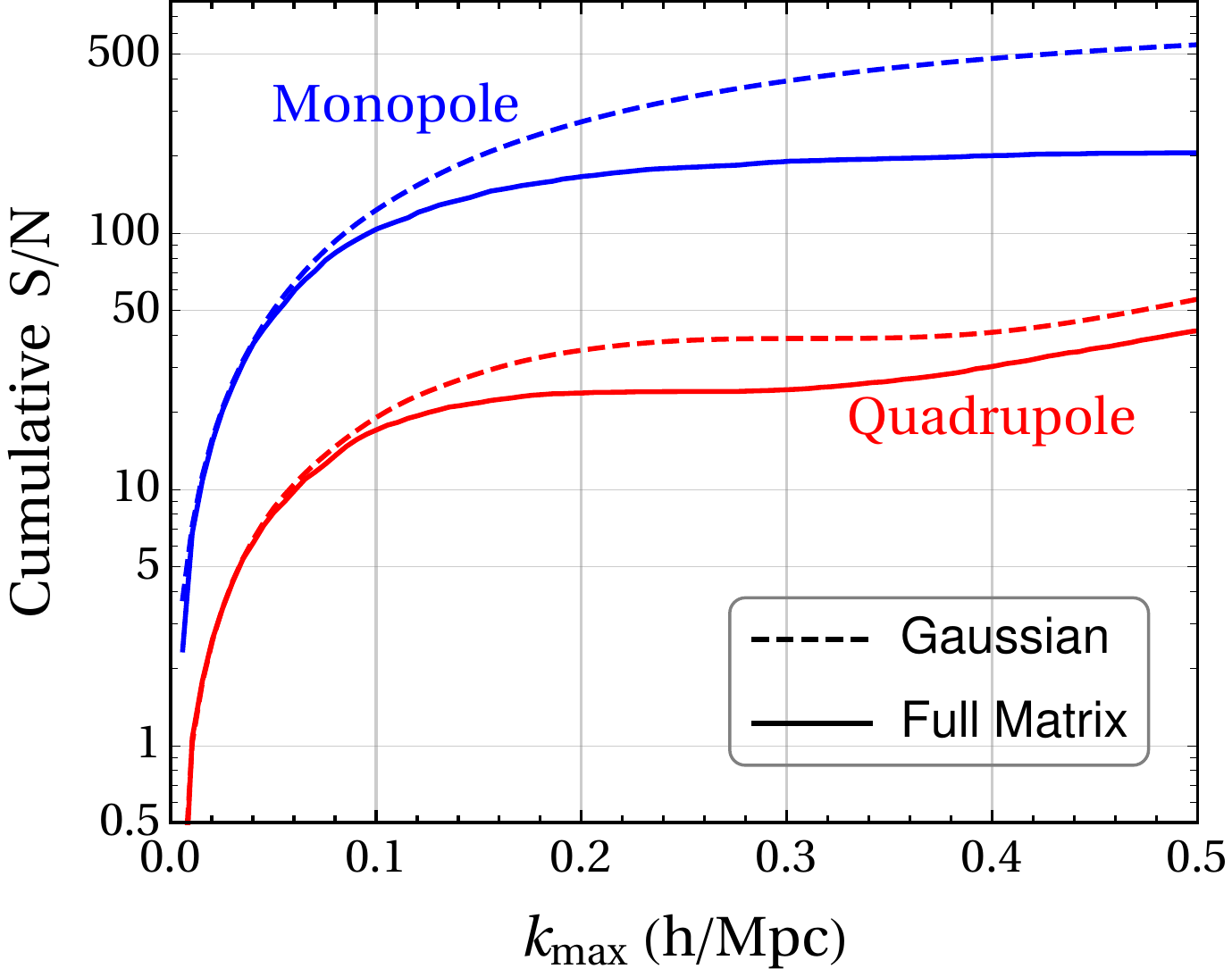}
\caption{Cumulative signal-to-noise ratio S/N given by Eq. (\ref{eq:S_N}) for the monopole and the quadrupole power spectra. The dashed line for each case corresponds to using the covariance matrix containing only the Gaussian contribution and the solid line corresponds to using the full matrix which includes the non-Gaussian contribution (see Fig.~\ref{fig:SN_contrib} for the decomposition of the covariance). Even though non-diagonal elements sourced by non-Gaussian effects are small, they do have an overall impact on the S/N for $k > 0.1 \kMpc$ and should be appropriately modeled.}
\label{fig:S_N}
\end{figure}

\section{Conclusions}
\label{sec:Conclusions}


In this paper we presented a perturbation theory (PT) approach to the full covariance matrix of the galaxy power spectrum multipoles, including the effects of non-linear evolution, nonlinear and nonlocal bias, redshift-space distortions beyond the plane-parallel approximation, non-trivial survey window functions and shot noise. We found an excellent agreement between our analytic covariance and that estimated from Patchy mock catalogs used for BOSS DR12 parameter estimation in the whole range we tested, up to $k=0.6 \kMpc$ (see Figs.~\ref{fig:Ckk},~\ref{fig:rij}, ~\ref{fig:Ckk_LowZ} and~\ref{fig:param_constraints}).

We discussed all the ingredients that enter into the calculation of the power spectrum covariance, and the approximations made. In particular, we used PT for non-linear evolution, bias and redshift-space distortions, and included shot noise in the Poisson limit. We worked at scales smaller than the survey window, where uncertainties are small and the most cosmological information is present; this helps in decoupling the covariance into factors that depend on survey geometry alone (which can be precomputed for a given survey using FFTs) and the factors that depend on the physics of clustering. This also make our results particularly efficient for numerical implementation in data analysis.

We found that the nature of the FKP estimator, which is widely used in the literature for measurements on galaxy redshift surveys, is somewhat different than the estimator assumed in most of the literature on the effects of super-survey modes, leading to a stronger super-sample covariance (Sec.~\ref{sec:Literature_comparison}). We also showed that the regular trispectrum contribution sourced by small-scale modes is not necessarily negligible compared to SSC (Fig.~\ref{fig:ED_components}), and both effects must be included to properly treat extra-diagonal elements. In connection to this, it is worth showing the impact of going beyond a Gaussian covariance matrix, which is typically assumed when making fisher forecasts in the literature. For this purpose we calculate the cumulative signal-to-noise (S/N) for the monopole and quadrupole power spectrum. We show in Fig.~\ref{fig:S_N} the calculation for $\ell=0,2$ of 
\beq
\left(\frac{\textup{S}}{\textup{N}}\right)^2 = \sum_{i,j}^{k_\textup{max}} P_\ell (k_i)\ \textbf{C}^{-1}_{k_\textup{max}} [P_\ell(k_i),P_\ell(k_j)]\ P_\ell(k_j)\, 
\label{eq:S_N}
\eeq
where we invert the covariance matrix in two different cases.
 From these results we see that even though extra-diagonal elements are small, as there are many of them, they can affect the S/N appreciably, i.e. by $\sim$ 40 \% (30 \%)  at $k_\textup{max}=0.2 \kMpc$ for the monopole (quadrupole) case. A more relevant question in practice is how would these different cases affect cosmological parameter constraints (rather than overall S/N), but this will be addressed in upcoming work \cite{WadIvaSco20}. 

Breaking up the different effects that contribute to the full covariance (see Fig.~\ref{fig:SN_contrib}), we showed that the perturbative approach is particularly well-suited for galaxy redshift surveys which are typically shot-noise dominated in the regime where PT breaks down, making the need for non-perturbative solutions (i.e. numerical simulations) less compelling given the significant computational time they require to reliably measure covariance matrices, and their inability to easily recompute it for a different cosmology or bias parameter set. 

One can estimate from the value of $\bar{n}P$ for a given biased tracer that this state of affairs is likely to remain in upcoming surveys (see Fig.~2 of \cite{FonMcDMos1405}). For example, $\bar{n}P(k = 0.14\, \kMpc) \lesssim 4$ for BOSS while considering all the upcoming redshift surveys in the near future such as Euclid, DESI, WFIRST, HETDEX, the least shot noise dominated case would be  the bright galaxy sample in DESI  and only by a factor of $\lesssim 2.5$ compared to BOSS. From Fig.~\ref{fig:SN_contrib} we can see that even after damping the shot noise strength by a significant factor, it  would still dominate the covariance at weakly non-linear scales. Therefore we expect the hierarchical trends among the various components of the covariance presented here to hold for such upcoming surveys. The equations used to derive our final results are summarized in Sec.~\ref{sec:SNres} and encoded in a code \textsc{CovaPT} which will be made publicly available\footnote{\textsc{CovaPT} is available for download at \href{https://github.com/JayWadekar/CovaPT}{\textcolor{blue}{https://github.com/JayWadekar/CovaPT}}}.

On the more technical side, there are a number of improvements that can be done to our treatment. In the large-scale limit one can improve our treatment of radial distortions by going beyond leading order in $(kd)^{-1}$ (see Eq.~\ref{PlocalLeg})  and redshift selection function effects, which should improve the treatment  at large scales. In addition, including a better treatment of window convolution at low-$k$ would be desirable to have robust constraints at scales comparable to survey size, i.e. for models with primordial non-Gaussianity of local type. At small scales, improvements to include would be a treatment of velocity dispersion in the trispectrum and loop corrections from nonlinear evolution and bias and also non-Poissonian shot noise terms.  Having said all this, however, it is not clear at this point how much these improvements matter in practice. We plan to address this by comparing our analytic results with more robust mock catalogs built with full $N$-body simulations~\cite{VilHahMas1909} in the near future.  In addition, we plan to use the approach presented here for computing the covariance of the bispectrum, which is computationally prohibitive using numerical simulations.

\section*{Acknowledgements}

We thank Misha Ivanov, Alex Barreira, Linda Blot, Kwan Chan, Albert Chuang, Mart\'\i n Crocce, Atsushi Taruya, Marcel Schmittfull, Elizabeth Krause, Chang Hahn, Yin Li, Shirley Ho,  Shun Saito, Ariel S\'anchez, Emiliano Sefusatti, David Spergel, Masahiro Takada, Mat\'{\i}as Zaldarriaga, Martin White and Uros Seljak  for useful discussions. We also thank Alex Eggemeier for his prodigious help with the COMPASS code.
This research was supported by the Munich Institute for Astro- and Particle Physics (MIAPP) which is funded by the Deutsche Forschungsgemeinschaft (DFG, German Research Foundation) under Germany's Excellence Strategy --- EXC-2094 --- 390783311.

\appendix
\onecolumngrid
\section{Redshift-Space Distortions Kernels}
\label{apx:RSDkernels}


To compute the trispectrum in tree-level PT, we use the redshift-space kernels in standard Eulerian PT up to cubic order. Here we write all the kernels corresponding to a fixed LOS $\hat{\textbf{n}}$ following~\cite{ScoCouFri9906} extended for non-local bias in the bias basis of~\cite{ChaScoShe1204,EggScoSmi1906}. The redshift-space PT kernels are defined as,
\beq
\delta (\k) = \sum_{n=1}^\infty \int_{\q_1..\q_n} \delta_D(\k - \q_1...-\q_n) \ Z_n(\q_1,..,\q_n)\ \delta_L(\q_1)...\delta_L(\q_n)
\label{eq:ZnDef}
\eeq
They are,
\begin{equation}
Z_1(\k_1)= b_1+f \mu_1^2\, ,
\end{equation}
\begin{equation}
\begin{split}
Z_2(\k_1,\k_2) &= b_1F_2(\k_1,\k_2)+f\mu^2G_2(\k_1,\k_2)+\frac{f\mu k}{2}\left[\frac{\mu_1}{k_1}(b_1+f\mu_2^2)+\frac{\mu_2}{k_2}(b_1+f\mu_1^2)\right]+\frac{b_2}{2} + \gamma_2 K\\
=& (b_1+f\mu^2)F_2(\k_1,\k_2)+\left(\frac{2}{7}f\mu^2+\gamma_2\right)K(\k_1,\k_2)+\frac{f\mu k}{2}\left[\frac{\mu_1}{k_1}(b_1+f\mu_2^2)+\frac{\mu_2}{k_2}(b_1+f\mu_1^2)\right]+\frac{b_2}{2}\, ,
\end{split}
\end{equation}

\begin{equation}
\begin{split}
Z_3(\k_1,\k_2,\k_3) &= b_1F^{(s)}_3(\k_1,\k_2,\k_3)+f\mu^2G^{(s)}_3(\k_1,\k_2,\k_3)+f\mu k\left[b_1F^{(s)}_2(\k_1,\k_2)+f\mu_{12}^2 G^{(s)}_2(\k_1,\k_2)\right] \frac{\mu_3}{k_3}\\
&\ + f\mu k (b_1+f\mu_1^2) \frac{\mu_{23}}{k_{23}}G^{(s)}_2(\k_2,\k_3)+\frac{(f\mu k)^2}{2}(b_1+f\mu_1^2)\frac{\mu_2}{k_2} \frac{\mu_3}{k_3}+3b_2 F^{(s)}_2(\k_1,\k_2)+ \frac{b_3}{6}\\
&\ + \gamma_3 L(\k_1,\k_2,\k_3)+\frac{\gamma^x_2}{3}(K(\k_1,\k_2)+K(\k_2,\k_3)+K(\k_1,\k_3))\\
&\ +\frac{\gamma_{21}}{3} (K(\k_1,\k_2)K(\k_1+\k_2,\k_3)+K(\k_2,\k_3)K(\k_2+\k_3,\k_1)+K(\k_1,\k_3)K(\k_1+\k_3,\k_2))\, ,
\end{split}
\end{equation}
where we denote $\mu \equiv \k \cdot \hat{\textbf{n}}/|\k|$ where $\k=\k_1+...+\k_n$ and $\mu_i \equiv \k_i\cdot\hat{\textbf{n}}/|\k_i|$ and $\gamma_2$ is the non-local bias
and the embedded kernels are taken from \cite{BerColGaz0209}; we show some of the most useful kernels: 
\begin{equation}
\begin{split}
F_2(\k,\textbf{q}) &= \frac{5}{7} + \frac{(\k.\textbf{q})}{2} \left( \frac{1}{k^2} + \frac{1}{q^2}\right) + \frac{2}{7} \frac{(\k.\textbf{q})^2}{k^2 q^2}\, ,\\
G_2(\k,\textbf{q}) &= \frac{3}{7} + \frac{(\k.\textbf{q})}{2} \left( \frac{1}{k^2} + \frac{1}{q^2}\right) + \frac{4}{7} \frac{(\k.\textbf{q})^2}{k^2 q^2}\, ,\\
K(\k,\textbf{q}) &= (\hat{\k}\cdot\hat{\textbf{q}})^2-1\, ,\\
L(\k_1,\k_2,\k_3) &= 2\frac{(\hat{\k}_1\cdot \hat{\k}_2)(\hat{\k}_2\cdot \hat{\k}_3)(\hat{\k}_1\cdot \hat{\k}_3)}{k^2_1\, k^2_2\, k^2_3}-\frac{(\hat{\k}_1\cdot \hat{\k}_2)^2}{k_1^2\, k_2^2}-\frac{(\hat{\k}_2\cdot \hat{\k}_3)^2}{k_2^2\, k_3^2}-\frac{(\hat{\k}_1\cdot \hat{\k}_3)^2}{k_1^2\, k_3^2}+1\, .
\end{split}\label{eq:ModeCouplingKernels}\end{equation}
We now outline the procedure to derive the explicit form of the $\mathcal{Z}_{21}$ kernel which has been frequently used in our calculations. We start with the definition of the $\mathcal{Z}_{21}$ kernel presented in Eq.~(\ref{eq:Z12_definition})
\begin{equation}
\bigg\{4\int_{\hat{\k}_{\ell_1},\p_1}W_{11}(\p_1) W_{11}(\e-\p_1) P(\k_1-\p_1) Z_1(\k_1-\p_1) Z_2(\k_1-\p_1,\e)\bigg\} \equiv P(k_1) W_{22}(\e)\mathcal{Z}_{12} (k_1,\ell_1,\hat{\e}\cdot\hat{\textbf{n}})\, ,
\label{eq:Z12_def2}
\end{equation}
where the short mode $\k_1$ is perturbed by long modes like $\e, \p_1$. We can therefore expand the terms on the LHS using the limits $|\p_1|\ll |\k_1|$ and $|\e| \ll |\k_1|$.
One of the identities most useful for our calculation is
\begin{equation}
\mu_{\k-\e}^n = \left(\frac{(\k-\e)\cdot \hat{\textbf{n}}}{|\k-\e|}\right)^n = \mu_{\k}^n\frac{\left(1-\frac{\e\cdot\hat{\textbf{n}}}{\mu_{\k}k}\right)^n}{\left(1-\frac{\hat{\k}\cdot\e}{k^2}\right)^n} = \mu_{\k}^n\left(1+n\frac{\hat{\k}\cdot\e}{k}\right)-n \mu_{\k}^{n-1}\frac{\hat{\textbf{n}}\cdot\e}{k}+\mathcal{O} \left(\frac{\epsilon^2}{k^2}\right)\, .
\label{eq:mu_k}\end{equation}
We substitute the following expansions in Eq.~(\ref{eq:Z12_def2}) 
\begin{subequations}
\begin{equation}
P(\k_1-\p_1) = P(\k)\left[1-\frac{\p_1\cdot\k_1}{k^2_1}\frac{d \ln P}{d \ln k}\right]+\mathcal{O} \left(\frac{p^2_1}{k^2_1}\right)
\end{equation}
\begin{equation}
Z_1(\k_1-\p_1) = b_1+f\mu_{\k_1-\p_1}^2=b_1+f (\hat{\k}_1\cdot\hat{\textbf{n}})^2 \left( 1+2\frac{\p_1\cdot\k_1}{k^2_1}-2\frac{\p_1\cdot\hat{\textbf{n}}}{\hat{\k}_1\cdot\hat{\textbf{n}}}\right)+\mathcal{O} \left(\frac{p^2_1}{k^2_1}\right)
\end{equation}
\begin{equation}
\begin{split}
Z_2(\k&_1-\p_1,\e)
=(b_1+f\mu_{\k_1-\p_1+\e}^2)F_2(\k_1-\p_1,\e)+\left(\frac{2}{7}f\frac{(\k_1\cdot\hat{\textbf{n}})}{k_1^2}+\gamma_2\right)\left(\frac{(\k_1\cdot\e)^2}{\epsilon^2k_1^2}-1\right)\\
&+\frac{f (\k_1-\p_1+\e)\cdot \hat{\textbf{n}}}{2} \left[\frac{\k_1\cdot\hat{\textbf{n}}}{k^2_1}(b_1+f\mu_{\e}^2)+\frac{(\e\cdot\hat{\textbf{n}})}{\epsilon^2}(b_1+f\mu_{\k_1-\p_1}^2)\right]+\frac{b_2}{2} +\mathcal{O} \left(\frac{p_1}{k_1},\frac{\epsilon}{k_1}\right)\, ,
\end{split}
\end{equation}
\label{eq:P,Z1,Z2_kernels}
\end{subequations}

and then we use the window identities in Eq.~(\ref{eq:WindowIdentities}) to calculate the integral over $\p_1$ space and get

\begin{equation}
\begin{split}
4\int_{\hat{\k}_{\ell_1},\p_1}&W_{11}(\p_1) W_{11}(\e-\p_1) P(\k_1-\p_1) Z_1(\k_1-\p_1) Z_2(\k_1-\p_1,\e)\\
\simeq&\, 4\, W_{22}(\e) P(k) \bigg\{ \int_{\hat{\k}_{\ell_1}}\left[1-\frac{\e\cdot\k_1}{2k^2_1}\frac{d \ln P}{d \ln k}\right]\left[b_1+f (\hat{\k}_1\cdot\hat{\textbf{n}})^2 \left( 1+\frac{\e\cdot\k_1}{k^2_1}-\frac{\e\cdot\hat{\textbf{n}}}{\hat{\k}_1\cdot\hat{\textbf{n}}}\right)\right]\\
&\times \bigg[ (b_1+f\mu_{\k_1+\e/2}^2)F_2(\k_1-\e/2,\e)+\left(\frac{2}{7}f\frac{(\k_1\cdot\hat{\textbf{n}})}{k_1^2}+\gamma_2\right)\left(\frac{(\k_1\cdot\e)^2}{\epsilon^2k_1^2}-1\right)\\
&\ \ \ \ \ \ +\frac{f (\k_1+\e/2)\cdot \hat{\textbf{n}}}{2} \left[\frac{\k_1\cdot\hat{\textbf{n}}}{k^2_1}(b_1+f\mu_{\e}^2)+\frac{(\e\cdot\hat{\textbf{n}})}{\epsilon^2}(b_1+f\mu_{\k_1-\e/2}^2)\right]+\frac{b_2}{2}\bigg] \bigg\} 
\end{split}\label{eq:Z12_expansion}
\end{equation}
The integral in the curly bracket in the above expression gives us the explicit form of $\mathcal{Z}_{21} (k_1,\ell_1,\hat{\e}\cdot\hat{\textbf{n}})$.


\section{Shot Noise contribution to the Gaussian covariance}
\label{apx:SN_exact_gaussian}
We generalize the results of Sec. \ref{sec:GauCova} for the case of discrete point particles. We will again label all the discrete terms with a superscript `$d$' in this section. We can write a discrete analogue of the overdensity in Eq.~(\ref{eq:fl_estimator}) as: $F^d_{\ell}(\k)=\left(\sum_{i}^{\ng} -\alpha \sum_{i}^{\nr}\right) w_i e^{-i \k\cdot \x} \L(\hat{\k}\cdot\hat{\x})$. We write the two point correlations as

\begin{equation}\begin{split}
\langle F^d_{\ell_1}(\k_1) F^d_{\ell_2}(\k_2) \rangle =& \bigg\langle\bigg( \sum_{i}^{\ng} -\alpha \sum_{i}^{\nr} \bigg) \bigg( \sum_{j}^{\ng} -\alpha \sum_{j}^{\nr} \bigg) w_i w_j e^{-i \k_1\cdot\x_i-i\k_2\cdot\x_j} \L_{\ell_1}(\hat{\k}_1\cdot\hat{\x}_i)\L_{\ell_2}(\hat{\k}_2\cdot\hat{\x}_j)\bigg\rangle\\
=& \bigg\langle \bigg( \sum_{i}^{\ng} -\alpha \sum_{i}^{\nr} \bigg) \bigg( \sum_{j\neq i}^{\ng} -\alpha \sum_{j \neq i}^{\nr} \bigg) w_i w_j e^{-i \k_1\cdot\x_i-i\k_2\cdot\x_j} \L_{\ell_1}(\hat{\k}_1\cdot\hat{\x}_i)\L_{\ell_2}(\hat{\k}_2\cdot\hat{\x}_j)\bigg\rangle\\
&+\bigg\langle\bigg( \sum_{i}^{\ng} +\alpha^2 \sum_{i}^{\nr} \bigg) w^2_i e^{-i (\k_1+\k_2)\cdot\x_i} \L_{\ell_1}(\hat{\k}_1\cdot\hat{\x}_i)\L_{\ell_2}(\hat{\k}_2\cdot\hat{\x}_i)\bigg\rangle\\
\simeq & \int_{\x_1,\x_2}W_{11}(\x_1)W_{11} (\x_2)\langle \delta(\x_1)\delta (\x_2) \rangle e^{-i \k_1\cdot\x_i-i\k_2\cdot\x_j} \L_{\ell_1}(\hat{\x}_1\cdot\hat{\k}_1)\L_{\ell_2}(\hat{\x}_2\cdot\hat{\k}_2)\\
&+ (1+\bar{\alpha})\int_{\x} W_{12}(\x) e^{-i (\k_1+\k_2)\cdot\x} \L_{\ell_1}(\hat{\k}_1\cdot\hat{\x})\L_{\ell_2}(\hat{\k}_2\cdot\hat{\x})\\
=& \langle F_{\ell_1}(\k_1) F_{\ell_2}(\k_2)\rangle + \mathcal{I}_{12}^{\ell_1,\ell_2} (\k_1,\k_2)\, ,
\end{split}\label{eq:b.7}\end{equation}
where we have decomposed the correlations into two components each of which has continuous integrals instead of discrete sums over particle indices. We have also introduced the following notation in Eq.~(\ref{eq:b.7}) for the shot noise component:
\begin{equation}
\mathcal{I}_{12}^{\ell_1,\ell_2} (\k_1,\k_2)\equiv(1+\bar{\alpha})\int_{\x}W_{12}(\x)e^{-i(\k_1+\k_2)\cdot\x} \L_{\ell_1}(\hat{\k}_1\cdot\hat{\x}) \L_{\ell_2}(\hat{\k}_2\cdot\hat{\x})\, .
\end{equation}
Gaussian covariance in the discrete case can be written analogous to Eq.~(\ref{eq:3.2}) as
\begin{equation}
\begin{split}
\textbf{C}^{\textup{G}}_{\ell_1\ell_2} (k_1,k_2)
=\frac{(2\ell_1+1)(2\ell_2+1)}{\I_{22}^2}\int_{\hat{\k}_1,\hat{\k}_2}& \bigg[ \langle F^d_{\ell_1}(\k_1) F^d_{0}(-\k_2)\rangle \langle F^d_{\ell_2}(\k_2) F^d_{0}(-\k_1) \rangle+ \langle F^d_{\ell_1}(\k_1) F^d_{\ell_2}(-\k_2)\rangle \langle F^d_{0}(\k_2) F^d_{0}(-\k_1) \rangle\bigg]
\end{split}
\end{equation}
Upon substituting Eq.~(\ref{eq:b.7}), the Gaussian covariance can be broken down into a continuous and a shot noise component ($\textbf{C}^{\textup{G}}_{\ell_1\ell_2}=\textbf{C}^\textup{G(cont.)}_{\ell_1\ell_2} + \textbf{C}^\textup{SN-G}_{\ell_1\ell_2}$). We dealt with the continuous component in Sec. \ref{sec:GauCova} and now we focus on calculating the shot noise component  
\begin{equation}
\begin{split}
&C^\textup{SN-G}_{\ell_1\ell_2} (k_1,k_2)=\frac{(2\ell_1+1)(2\ell_2+2)}{\I_{22}^2}\int_{\hat{\k}_1,\hat{\k}_2}\\
&\Big[\langle F_{\ell_1}(\k_1) F_{\ell_2}(-\k_2)\rangle \mathcal{I}_{12}^{0 ,0} (-\k_1,\k_2)+\langle F_0(\k_1) F_0(-\k_2)\rangle \mathcal{I}_{12}^{\ell_1, \ell_2} (-\k_1,\k_2)+\langle F_{\ell_1}(\k_1) F_{0}(-\k_2)\rangle \mathcal{I}_{12}^{0,\ell_2} (-\k_1,\k_2)\\
&+\langle F_0(\k_1) F_{\ell_2}(-\k_2)\rangle \mathcal{I}_{12}^{\ell_1,0} (-\k_1,\k_2) +\mathcal{I}_{12}^{\ell_1,0} (\k_1,-\k_2)\mathcal{I}_{12}^{0,\ell_2} (-\k_1,\k_2)+\mathcal{I}_{12}^{\ell_1,\ell_2} (\k_1,-\k_2)\mathcal{I}_{12}^{0,0} (-\k_1,\k_2)\Big]\, ,
\end{split}\label{eq:cova_SN-G}\end{equation}
which has two types of SN terms: $\mathcal{O}[(\mathcal{I}_{12})^2]$ and $\mathcal{O}[\mathcal{I}_{12}]$.
Let us first calculate the $\mathcal{O}[(\mathcal{I}_{12})^2]$ type terms as 
\begin{equation}
\begin{split}
&\int_{\hat{\k}_1,\hat{\k}_2}\mathcal{I}_{12}^{\ell_1,\ell_2} (\k_1,-\k_2)\mathcal{I}_{12}^{0,0} (-\k_1,\k_2)+ \mathcal{I}_{12}^{\ell_1,0} (\k_1,-\k_2)\mathcal{I}_{12}^{0,\ell_2} (-\k_1,\k_2)\\
&=(1+\bar{\alpha})^2\int_{\hat{\k}_1,\hat{\k}_2,\x_1,\x_2} W_{12}(\x_1)W_{12}(\x_2) e^{-i(\k_1-\k_2)\cdot(\x_1-\x_2)}\L_{\ell_1}(\hat{\x}_1\cdot\hat{\k}_1)\Big[\L_{\ell_2}(\hat{\x}_1\cdot\hat{\k}_2)+\L_{\ell_2}(\hat{\x}_2\cdot\hat{\k}_2)\Big]\, ,
\end{split}\label{eq:b6}
\end{equation}

while the remaining four terms in Eq.~(\ref{eq:cova_SN-G}) are $\mathcal{O}[\mathcal{I}_{12}]$ type. The derivation of all the four terms is similar and we show the most general one below

\begin{equation}\begin{split}
\langle F_{\ell_1}(\k_1) F_{\ell_2}(-\k_2)\rangle=&\int_{\x_1,\x'_1} e^{-i\k_1 \cdot \x_1+ i\k_2 \cdot \x'_1} \langle \delta(\x_1)\delta(\x'_1) \rangle W_{11}(\x_1)W_{11} (\x'_1) \mathcal{L}_{\ell_1}(\hat{\x}_1\cdot\hat{\k}_1)\mathcal{L}_{\ell_2}(\hat{\x}_1 \cdot \hat{\k}_2 )\, .
\end{split}\end{equation}
which can be evaluated using the steps similar to Sec.~\ref{sec:GauCova}, as we now show.
We first substitute the relative coordinate $\x'_1 \rightarrow \x_1-\textbf{s}_1$

\begin{equation}
\begin{split}
\langle F_{\ell_1}(\k_1) F_{\ell_2}(-\k_2)\rangle=& \int_{\x_1,\textbf{s}_1} e^{-i (\k_1-\k_2)\cdot \x_1} e^{ -i \k_2 \cdot \textbf{s}_1} \xi(\textbf{s}_1; \x_1) W_{11}(\x_1)W_{11}(\x_1-\textbf{s}_1) \mathcal{L}_{\ell_1}(\hat{\x}_1 \cdot \hat{\k}_1 ) \mathcal{L}_{\ell_2}(\hat{\x}_1\cdot\hat{\k}_2)\, .
\end{split}\label{eq:b7}
\end{equation}

We then use Eq.~(\ref{eq:xi(s)_simplify}) to simplify the integral over the $\textbf{s}_1$ and then use the multipole expansion to leading order  in  $(kd)^{-1}$ as discussed following Eq.~(\ref{meanPell}) to get

\begin{equation}
\begin{split}
\langle F_{\ell_1}(\k_1) F_{\ell_2}(-\k_2)\rangle\simeq & \int_{\x_1} e^{-i (\k_1-\k_2)\cdot \x_1} P_{\rm local}(\k_2;\x_1) W_{22}(\x_1) \mathcal{L}_{\ell_1}(\hat{\x}_1 \cdot \hat{\k}_1 ) \mathcal{L}_{\ell_2}(\hat{\x}_1 \cdot \hat{\k}_2 )\\
=& \sum_{\ell'}P_{\ell'}(k_2) \int_{\x_1} e^{-i (\k_1-\k_2)\cdot \x_1}  W_{22}(\x_1)
\mathcal{L}_{\ell'}(\hat{\x}_1 \cdot \hat{\k}_2 )
\mathcal{L}_{\ell_1}(\hat{\x}_1 \cdot \hat{\k}_1 )  \mathcal{L}_{\ell_2}(\hat{\x}_1 \cdot \hat{\k}_2 )\, .
\end{split}\label{eq:b9}
\end{equation}

Note that the symmetry $\{\k_1,\ell_1\} \leftrightarrow \{\k_2,\ell_2\}$ is broken as a consequence of choosing the LOS in Eq.~(\ref{eq:b7}) to be along one of the galaxies (i.e. $\x_1$ instead of $\x'_1$). In order to restore this symmetry, we can rewrite Eq.~(\ref{eq:b9}) as
\begin{equation}
\begin{split}
\langle F_{\ell_1}(\k_1) F_{\ell_2}(-\k_2)\rangle\simeq \int_{\x_1} e^{-i (\k_1-\k_2)\cdot \x_1}  W_{22}(\x_1) \mathcal{L}_{\ell_1}(\hat{\x}_1 \cdot \hat{\k}_1 ) \mathcal{L}_{\ell_2}(\hat{\x}_1 \cdot \hat{\k}_2 ) \sum_{\ell'}\frac{1}{2}\Big[P_{\ell'}(k_1) \mathcal{L}_{\ell'}(\hat{\x}_1 \cdot \hat{\k}_1 )+P_{\ell'}(k_2) \mathcal{L}_{\ell'}(\hat{\x}_1 \cdot \hat{\k}_2 )\Big]\, .
\end{split}\label{eq:b10}
\end{equation}

Using Eqs.~(\ref{eq:b6}) and (\ref{eq:b10}), the SN contribution to Gaussian covariance in Eq.~(\ref{eq:cova_SN-G}) simplifies to

\begin{equation}
\begin{split}
C^\textup{SN-G}_{\ell_1\ell_2}& (k_1,k_2)=\frac{(2\ell_1+1)(2\ell_2+1)}{\I_{22}^2}\int_{\hat{\k}_1,\hat{\k}_2,\x_1,\x_2}\\
\times \bigg \{& (1+\bar{\alpha})^2 W_{12}(\x_1)W_{12}(\x_2) e^{-i(\k_1-\k_2)\cdot(\x_1-\x_2)}\L_{\ell_1}(\hat{\x}_1\cdot\hat{\k}_1)\Big[\L_{\ell_2}(\hat{\x}_1\cdot\hat{\k}_2)+\L_{\ell_2}(\hat{\x}_2\cdot\hat{\k}_2)\Big]\\
&+(1+\bar{\alpha})
e^{-i(\k_1-\k_2)\cdot (\x_1-\x_2)} W_{22}(\x_1)W_{12}(\x_2) \Big[ \frac{1}{2}\sum_{\ell'}P_{\ell'}(k_1) \L_{\ell'}(\hat{\x}_1\cdot\hat{\k}_1) +\frac{1}{2}\sum_{\ell'}P_{\ell'}(k_2) \L_{\ell'}(\hat{\x}_1\cdot\hat{\k}_2)\Big]\\
&\ \ \times \Big[
\L_{\ell_1}(\hat{\x}_1\cdot\hat{\k}_1)
\L_{\ell_2}(\hat{\x}_2\cdot\hat{\k}_2)+\L_{\ell_1}(\hat{\x}_2\cdot\hat{\k}_1)
\L_{\ell_2}(\hat{\x}_1\cdot\hat{\k}_2)+\L_{\ell_1}(\hat{\x}_1\cdot\hat{\k}_1)
\L_{\ell_2}(\hat{\x}_1\cdot\hat{\k}_2)+\L_{\ell_1}(\hat{\x}_2\cdot\hat{\k}_1)
\L_{\ell_2}(\hat{\x}_2\cdot\hat{\k}_2)\Big] \bigg \}
\end{split}\end{equation}
and can be more compactly rewritten as 
\begin{equation}
\begin{split}
C^\textup{SN-G}_{\ell_1\ell_2}& (k_1,k_2) =\Big[ \sum_{\ell'} P_{\ell'}(k_1)\mathcal{W}^{(2)}_{\ell_1,\ell_2,\ell'} (k_1,k_2) +(k_1 \leftrightarrow k_2)\Big]+\mathcal{W}^{(3)}_{\ell_1,\ell_2}(k_1,k_2)
\end{split}\end{equation}
which was presented in the main text as Eq.~(\ref{eq:covaSN-G}) and the kernels are defined as
\beq\begin{split}
\mathcal{W}^{(2)}_{\ell_1,\ell_2,\ell'} (k_1,k_2)\equiv&\frac{(1+\bar{\alpha})}{2} \frac{(2\ell_1+1)(2\ell_2+1)}{\I_{22}^2}\int_{\hat{\k}_1,\hat{\k}_2,\x_1,\x_2}
e^{-i(\k_1-\k_2)\cdot (\x_1-\x_2)} W_{22}(\x_1)W_{12}(\x_2) \L_{\ell'}(\hat{\x}_1\cdot\hat{\k}_1)\\
 \times \Big[
\L_{\ell_1}(\hat{\x}_1\cdot\hat{\k}_1)&
\L_{\ell_2}(\hat{\x}_2\cdot\hat{\k}_2)+\L_{\ell_1}(\hat{\x}_2\cdot\hat{\k}_1)
\L_{\ell_2}(\hat{\x}_1\cdot\hat{\k}_2)+\L_{\ell_1}(\hat{\x}_1\cdot\hat{\k}_1)
\L_{\ell_2}(\hat{\x}_1\cdot\hat{\k}_2)+\L_{\ell_1}(\hat{\x}_2\cdot\hat{\k}_1)
\L_{\ell_2}(\hat{\x}_2\cdot\hat{\k}_2)\Big] \bigg \}\\
\mathcal{W}^{(3)}_{\ell_1,\ell_2}(k_1,k_2)\equiv&(1+\bar{\alpha})^2 \frac{(2\ell_1+1)(2\ell_2+1)}{\I_{22}^2}\int_{\hat{\k}_1,\hat{\k}_2,\x_1,\x_2} W_{12}(\x_1)W_{12}(\x_2) e^{-i(\k_1-\k_2)\cdot(\x_1-\x_2)}\\
&\qquad \qquad \qquad \times \L_{\ell_1}(\hat{\x}_1\cdot\hat{\k}_1)\Big[\L_{\ell_2}(\hat{\x}_1\cdot\hat{\k}_2)+\L_{\ell_2}(\hat{\x}_2\cdot\hat{\k}_2)\Big]\\
\end{split}\label{eq:W_kernel}\eeq

\section{Line-of-sight (LOS) variation corresponding to super-survey modes} 
\label{apx:LOS_BeatTerms}
The LOS changes appreciably for modes whose size is comparable to that of the survey. Here we present an approach to account for the changing LOS when dealing with such modes. Neglecting selection effects, one can show that to leading order in $(kd)^{-1}$ where $d$ is the distance to galaxies, the PT kernels with radial RSDs can be written as those in the plane-parallel approximation  but with a local LOS that follows the radial direction, i.e. we have,
\beq
\delta (\x) \simeq \sum_{n=1}^\infty \int_{\q_1..\q_n} e^{i(\q_1+...+\q_n)\cdot\x} \ Z_n^{\hat{\x}}(\q_1,..,\q_n)\ \delta_L(\q_1)...\delta_L(\q_n)
\label{eq:radialRSDdelta}\eeq
where $Z_n^{\hat{\x}}$ is the usual plane-parallel kernel presented in Appendix~\ref{apx:RSDkernels} but with a fixed LOS $\hat{\textbf{n}}$ replaced by the varying LOS $\hat{\textbf{x}}$.  In the plane-parallel limit, when $Z_n^{\hat{\x}} \to Z_n^{\hat{\textbf{n}}}$, Fourier transform leads to the usual Dirac delta function from translation invariance, leading to Eq.~(\ref{eq:ZnDef}) above. We now re-derive the terms presented in Sec.~\ref{sec:RSD_Trispectrum} for the case of radial RSDs and varying LOS.

\subsection{Beat-Coupling using radial RSD kernels}
\label{apx:BC_radialRSD}

\begin{figure}
\centering
\includegraphics[scale=0.45,keepaspectratio=true]{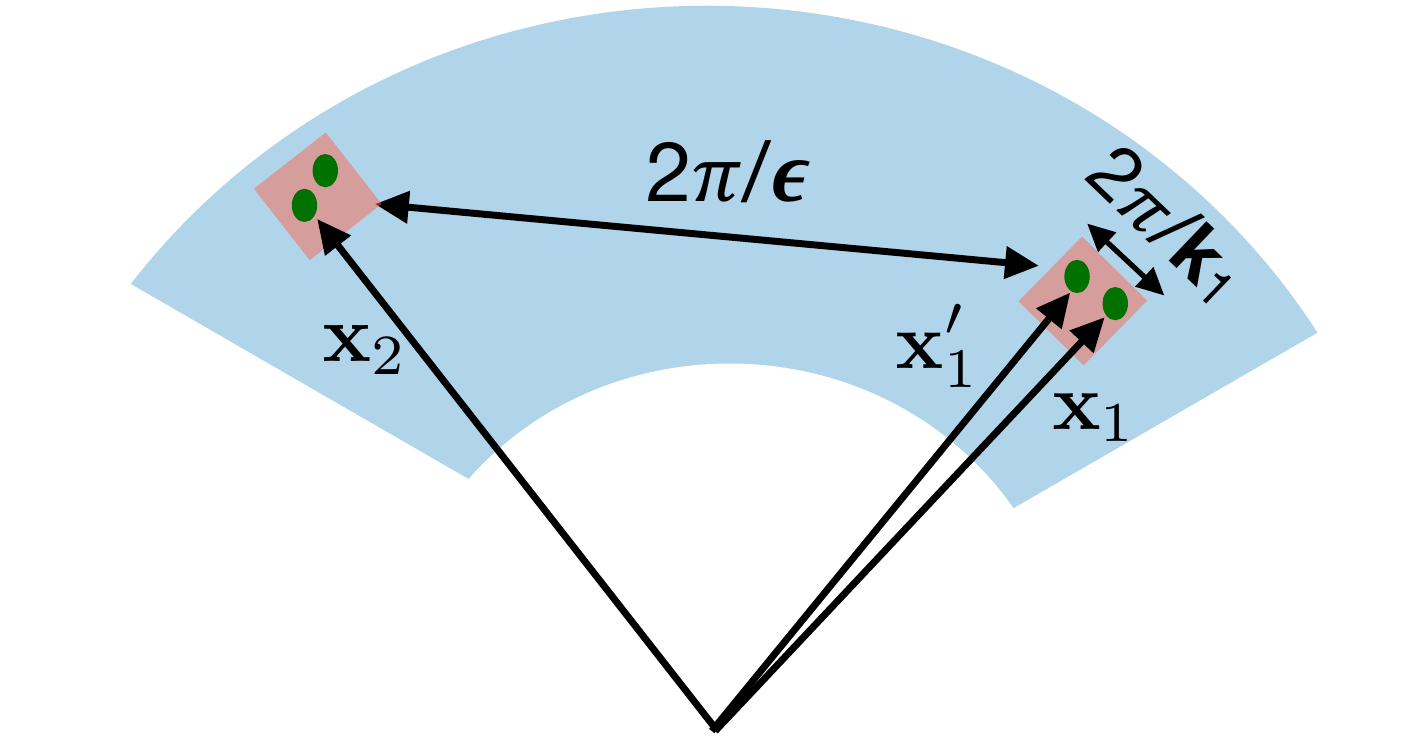}
\caption{A schematic diagram corresponding to Eq. (\ref{eq:d5}) showing a wide-angle survey where the LOS changes significantly along the survey. 
This configuration space diagram is shown to emphasize that keeping the LOS fixed is a good approximation for the small-scale modes ($\k_1,\k_2$) but not a good approximation when calculating integrals involving the beat mode ($\e$).} 
\label{fig:VaryingLOS}
\end{figure}

Let us start with the beat-coupling contribution to the covariance
\beq\begin{split}
&\textbf{C}^\textup{BC}_{\ell_1\ell_2} (k_1,k_2)\frac{\I_{22}^2}{(2\ell_1+1)(2\ell_2+1)}\\
&=4\int_{\hat{\k}_1,\hat{\k}_2,\x'_1,..,\x'_2} e^{-i\k_1\cdot(\x'_1-\x_1)} e^{-i\k_2\cdot(\x'_2-\x_2)} \langle \delta(\x'_1) \delta(\x_1)\delta(\x_2)\delta(\x'_2)\rangle_c\\
&\times W_{11}(\x'_1)W_{11}(\x_1)W_{11}(\x_2)W_{11}(\x'_2)\L_{\ell_1}(\hat{\x}_1\cdot\hat{\k}_1)\L_{\ell_2}(\hat{\x}_2\cdot\hat{\k}_2)\\
&= 4\int_{\hat{\k}_1,\hat{\k}_2,\x'_1,..,\x'_2,\q_1,..,\q_2} e^{-i\k_1\cdot(\x'_1-\x_1)} e^{-i\k_2\cdot(\x'_2-\x_2)} e^{i(\q_1\cdot\x'_1+\q_2\cdot\x'_2)}e^{i[(\q'_1+\q''_1)\cdot\x_1+(\q'_2+\q''_2)\cdot\x_2]} W_{11}(\x'_1)W_{11}(\x_1)W_{11}(\x_2)W_{11}(\x'_2)\\
&\times  \langle \delta_L(\q_1) \delta_L(\q'_1)\delta_L(\q''_1)\delta_L(\q''_2)\delta_L(\q'_2) \delta_L(\q_2)\rangle_c Z_1^{\hat{\x}'_1}(\q_1) Z_2^{\hat{\x}_1}(\q'_1,\q''_1)Z_2^{\hat{\x}_2}(\q'_2,\q''_2)Z_1^{\hat{\x}'_2}(\q_2) \L_{\ell_1}(\hat{\x}_1\cdot\hat{\k}_1)\L_{\ell_2}(\hat{\x}_2\cdot\hat{\k}_2)
\end{split}\eeq
We use Gaussian pairings in the connected six-point function and substitute a variable $\e$ for the beat mode
($\q'_1=-\q_1$, $\q'_2=-\q_2$, $\q''_2=-\q''_1 \equiv \e$) to get

\beq\begin{split}
16& \int_{\hat{\k}_1,\hat{\k}_2,\x,\e,\q_1,\q_2,\p_1,..,\p_2} e^{-i\k_1\cdot(\x'_1-\x_1)} e^{-i\k_2\cdot(\x'_2-\x_2)} e^{i(\q_1\cdot\x'_1+\q_2\cdot\x'_2)}e^{i[(-\e-\q_1)\cdot\x_1+(\e-\q_2)\cdot\x_2]} e^{i(\p_1\cdot\x'_1+\p'_1\cdot\x_1+\p_2\cdot\x'_2+\p_2\cdot\x_2)}\\
&\times W_{11}(\p_1)W_{11}(\p'_1)W_{11}(\p'_2)W_{11}(\p_2) P_{\rm L}(\e) P_{\rm L}(\q_1) P_{\rm L}(\q_2) Z_1^{\hat{\x}'_1}(\q_1) Z_2^{\hat{\x}_1}(\q_1,\e)Z_2^{\hat{\x}_2}(\q_2,-\e)Z_1^{\hat{\x}'_2}(\q_2) \L_{\ell_1}(\hat{\x}_1\cdot\hat{\k}_1)\L_{\ell_2}(\hat{\x}_2\cdot\hat{\k}_2)
\end{split}\label{eq:d5}\eeq
Because the LOS does not change over scales of modes $k_1, k_2$ that we are interested in, $Z_i^{\hat{\x}'_1}\rightarrow Z_i^{\hat{\x}_1}$ and $Z_i^{\hat{\x}'_2}\rightarrow Z_i^{\hat{\x}_2}$ (see Fig.~\ref{fig:VaryingLOS} for a diagram). We therefore integrate over the $\x'_1, \x'_2$ space and use the resulting delta functions $\Ddel(-\k_1+\p_1+\q_1)$ and $\Ddel(-\k_2+\p_2+\q_2)$ to integrate out the $\q_1, \q_2$ space as

\beq\begin{split}
&\textbf{C}^\textup{BC}_{\ell_1\ell_2} (k_1,k_2)\I_{22}^2=16(2\ell_1+1)(2\ell_2+1)\int_{\hat{\k}_1,\hat{\k}_2,\x_1,\x_2,\e,\p_1,\p'_1,\p_2,\p'_2} e^{i(\p_1+\p'_1-\e)\cdot\x_1}e^{i(\p_2+\p'_2+\e)\cdot\x_2} W_{11}(\p_1)W_{11}(\p'_1)W_{11}(\p_2)W_{11}(\p'_2)\\
&\times P_{\rm L}(\e) P_{\rm L}(\k_1-\p_1) P_{\rm L}(\k_2-\p_2) Z_1^{\hat{\x}_1}(\k_1-\p_1) Z_2^{\hat{\x}_1}(\k_1-\p_1,\e)Z_2^{\hat{\x}_2}(\k_2-\p_2,-\e)Z_1^{\hat{\x}_2}(\k_2-\p_2)\L_{\ell_1}(\hat{\x}_1\cdot\hat{\k}_1)\L_{\ell_2}(\hat{\x}_2\cdot\hat{\k}_2)
\end{split}\label{eq:d4}\eeq

If we use the plane parallel approximation ($\hat{\x}_1, \hat{\x}_2 \rightarrow \hat{\textbf{n}}$), we recover the expression given in Eq.~(\ref{CBCl1l2}) as
\beq\begin{split}
&16\, (2\ell_1+1)(2\ell_2+1)\int_{\hat{\k}_1,\hat{\k}_2,\e,\p_1,\p_2} W_{11}(\p_1)W_{11}(\e-\p_1)W_{11}(\p_2)W_{11}(-\e-\p_2)\\
&\times P_{\rm L}(\e) P_{\rm L}(\k_1-\p_1) P_{\rm L}(\k_2-\p_2) Z_1(\k_1-\p_1) Z_2(\k_1-\p_1,\e)Z_2(\k_2-\p_2,-\e)Z_1(\k_2-\p_2)\L_{\ell_1}(\hat{\k}_1\cdot\hat{\textbf{n}})\L_{\ell_2}(\hat{\k}_2\cdot\hat{\textbf{n}})
\end{split}\eeq

However, we want to derive expressions for a wide-angle survey where we cannot use the plane parallel approximation. To do this, we rearrange Eq.~(\ref{eq:d4}) into two similar integrals
\beq\begin{split}
&\int_{\e} P_{\rm L}(\epsilon)\bigg\{4(2\ell_1+1)\int_{\hat{\k}_1,\x_1,\p_1,\p'_1} e^{i(\p_1+\p'_1-\e)\cdot\x_1} W_{11}(\p_1)W_{11}(\p'_1) P_{\rm L}(\k_1-\p_1) Z_1^{\hat{\x}_1}(\k_1-\p_1) Z_2^{\hat{\x}_1}(\k_1-\p_1,\e)\L_{\ell_1}(\hat{\x}_1\cdot\hat{\k}_1)\bigg\}\\
&\times \bigg\{ 4(2\ell_2+1)\int_{\hat{\k}_2,\x_2,\p_2,\p'_2}e^{i(\p_2+\p'_2+\e)\cdot\x_2} W_{11}(\p_2)W_{11}(\p'_2) P_{\rm L}(\k_2-\p_2) Z_1^{\hat{\x}_2}(\k_2-\p_2)  Z_2^{\hat{\x}_2}(\k_2-\p_2,-\e)\L_{\ell_2}(\hat{\x}_2\cdot\hat{\k}_2)\bigg\}
\end{split}\eeq

Let us focus on the first term in curly brackets 

\beq\begin{split}
&\bigg\{4(2\ell_1+1)\int_{\hat{\k}_1,\x_1,\p_1,\p'_1} e^{i(\p_1+\p'_1-\e)\cdot\x_1}W_{11}(\p'_1)W_{11}(\p_1)P_{\rm L}(\k_1-\p_1) Z_1^{\hat{\x}_1}(\k_1-\p_1) Z_2^{\hat{\x}_1}(\k_1-\p_1,\e)\L_{\ell_1}(\hat{\x}_1\cdot\hat{\k}_1) \bigg\}\\
&=4(2\ell_1+1)\int_{\x_1,\p}e^{i(\p-\e)\cdot\x_1}\int_{\hat{\k}_1,\p_1} W_{11}(\p-\p_1) W_{11}(\p_1)P_{\rm L}(\k_1-\p_1) Z_1^{\hat{\x}_1}(\k_1-\p_1) Z_2^{\hat{\x}_1}(\k_1-\p_1,\e) \L_{\ell_1}(\hat{\x}_1\cdot\hat{\k}_1)\, ,
\end{split}\label{eq:d7}\eeq
where we have changed variables such that $\p'_1\rightarrow\p-\p_1$.
Using the expansions in Eq. (\ref{eq:P,Z1,Z2_kernels}) to the lowest order we can write our expressions in a compact way by making the following kernel substitutions 
\beq
\int_{\hat{\k}_1}P_{\rm L}(\k_1-\p_1) Z_1^{\hat{\x}_1}(\k_1-\p_1) Z_2^{\hat{\x}_1}(\k_1-\p_1,\e)\L_{\ell_1}(\hat{\x}_1\cdot\hat{\k}_1)=A(\hat{\e}\cdot\hat{\x}_1)+(\p_1\cdot\hat{\x}_1) B(\hat{\e}\cdot\hat{\x}_1)+(\p_1\cdot \hat{\e})C(\hat{\e}\cdot\hat{\x}_1)\, ,
\label{eq:d8}\eeq

where we have explicitly shown the dependence of the $A, B, C$ kernels only on $\hat{\x}_1$ and $\hat{\e}$ to make it clear that it is non-trivial to integrate out the $\x_1$ or $\e$ spaces at this point. We move ahead with using the window identities in Eq.~(\ref{eq:WindowIdentities}) to calculate the integral over the $\p_1$ space and Eq.~(\ref{eq:d7}) becomes

\beq\begin{split}
&4(2\ell_1+1)\int_{\x_1,\p}e^{i(\p-\e)\cdot\x_1}\int_{\p_1} W_{11}(\p-\p_1) W_{11}(\p_1) [A(\hat{\e}\cdot\hat{\x}_1)+(\p_1\cdot\hat{\x}_1) B(\hat{\e}\cdot\hat{\x}_1)+(\p_1\cdot \hat{\e})C(\hat{\e}\cdot\hat{\x}_1)]\\
&=4(2\ell_1+1)\int_{\x_1,\p}e^{i(\p-\e)\cdot\x_1} W_{22}(\p) \Big[A(\hat{\e}\cdot\hat{\x}_1)+ \frac{\p}{2}\cdot\Big(\hat{\x}_1 B(\hat{\e}\cdot\hat{\x}_1)+\hat{\e}\, C(\hat{\e}\cdot\hat{\x}_1)\Big)\Big]\\
&=4(2\ell_1+1)\bigg[\int_{\x_1}e^{-i\e\cdot\x_1} W_{22}(\x_1) A(\hat{\e}\cdot\hat{\x}_1)+\frac{1}{2i}\int_{\x_1}e^{-i\e\cdot\x_1} \nabla_{\x_1} (W_{22}(\x_1))\cdot \Big(\hat{\x}_1 B(\hat{\e}\cdot\hat{\x}_1)+\hat{\e}\, C(\hat{\e}\cdot\hat{\x}_1)\Big)\bigg]
\end{split}\label{eq:d9}\eeq
where the second term on the RHS after the substitution: $e^{-i\e\cdot\x_1} \nabla_{\x_1} (W_{22}(\x_1))= \nabla_{\x_1} (e^{-i\e\cdot\x_1} W_{22}(\x_1))- W_{22}(\x_1) \nabla_{\x_1} (e^{-i\e\cdot\x_1})= i\e\, e^{-i\e\cdot\x_1} W_{22}(\x_1)+\nabla_{\x_1} (e^{-i\e\cdot\x_1} W_{22}(\x_1))$ can be written as
\beq\begin{split}
&\frac{1}{2}\int_{\x_1} e^{-i\e\cdot\x_1}W_{22}(\x_1)\, \e\cdot\Big(\hat{\x}_1 B(\hat{\e}\cdot\hat{\x}_1)+\hat{\e}\, C(\hat{\e}\cdot\hat{\x}_1)\Big)+\frac{1}{2i}\int_{\x_1}\nabla_{\x_1} (e^{-i\e\cdot\x_1} W_{22}(\x_1))\cdot \Big(\hat{\x}_1 B(\hat{\e}\cdot\hat{\x}_1)+\hat{\e}\, C(\hat{\e}\cdot\hat{\x}_1)\Big)\, .
\end{split}\eeq
The term involving the integral of dot products of the gradient $\nabla_{\x_1} (e^{-i\e\cdot\x_1} W_{22}(\x_1))$ can be shown to be zero using the fact that the survey window has finite extent in every direction: $W_{22}(\x_1)\rightarrow 0$ as $|\x_1|\rightarrow\infty$. Eq.~(\ref{eq:d7}) can therefore be finally written as

\beq\begin{split}
&\bigg\{4(2\ell_1+1)\int_{\hat{\k}_1,\x_1,\p_1,\p'_1} e^{i(\p_1+\p'_1-\e)\cdot\x_1}W_{11}(\p'_1)W_{11}(\p_1)P_{\rm L}(\k_1-\p_1) Z_1^{\hat{\x}_1}(\k_1-\p_1) Z_2^{\hat{\x}_1}(\k_1-\p_1,\e)\L_{\ell_1}(\hat{\x}_1\cdot\hat{\k}_1) \bigg\}\\
&=4(2\ell_1+1)\int_{\x_1}e^{-i\e\cdot\x_1} W_{22}(\x_1) \Big[A(\hat{\e}\cdot\hat{\x}_1)+\frac{\e}{2}\cdot\Big(\hat{\x}_1 B(\hat{\e}\cdot\hat{\x}_1)+\hat{\e}\, C(\hat{\e}\cdot\hat{\x}_1)\Big)\Big]\\
&=\int_{\x_1}e^{-i\e\cdot\x_1} W_{22}(\x_1) 4(2\ell_1+1)\int_{\hat{\k}_1} P_{\rm L}\Big(\k_1-\frac{\e}{2}\Big) Z_1^{\hat{\x}_1}\Big(\k_1-\frac{\e}{2}\Big) Z_2^{\hat{\x}_1}\Big(\k_1-\frac{\e}{2},\e\Big)\L_{\ell_1}(\hat{\x}_1\cdot\hat{\k}_1)\\
&=\int_{\x_1}e^{-i\e\cdot\x_1} W_{22}(\x_1) P_{\rm L}(k_1) \mathcal{Z}_{21}(k_1,\ell_1,\e\cdot\x_1)\, ,
\end{split}\label{eq:d11}\eeq
where we have used the Eq.~(\ref{eq:d8}) for substituting the $A, B, C$ kernels and also used the definition of the $\mathcal{Z}_{21}$ kernel from Eq.~(\ref{eq:Z12_definition}). Using the radial RSD kernels, the final expression for the beat-coupling contribution to covariance becomes

\begin{equation}
\begin{split}
\textbf{C}&^\textup{BC}_{\ell_1\ell_2} (k_1,k_2)= \frac{1}{\I_{22}^2}P_{\rm L}(k_1)P_{\rm L}(k_2) \int_{\e,\x_1,\x_2} P_{\rm L}(\epsilon) W_{22}(\x_1) W_{22}(\x_2) e^{-i\e\cdot(\x_1-\x_2)}  \mathcal{Z}_{21}(k_1,\ell_1,\hat{\e}\cdot\hat{\x}_1)\mathcal{Z}_{21}(k_2,\ell_2,-\hat{\e}\cdot\hat{\x}_2)\, .
\end{split}
\end{equation}
which results in Eq. (\ref{eq:LOS_BC}).
\subsection{Local Average effect with radial RSD}
\label{apx:LA_radialRSD}
Let us now derive the results of Sec.~\ref{sec:RSD_LA} for the case of radial RSD. We first start with writing the variance of $\ng$ fluctuations as
\begin{equation}
\begin{split}
\langle \dng^2 \rangle =& \frac{1}{\I_{10}^2}\bigg \langle \int_{\x_1 ,\x_2} W_{10}(\x_1) \delta(\x_1) W_{10}(\x_2) \delta(\x_2)\bigg \rangle\\
=&\frac{1}{\I_{10}^2}\bigg \langle\int_{\x_1 ,\e_1}W_{10}(\x_1)\delta_L(\e_1) Z_1^{\hat{\x}_1}(\e_1) e^{i\e_1\cdot\x_1} \int_{\x_2 ,\e_2}W_{10}(\x_2)\delta_L(\e_2) Z_1^{\hat{\x}_2}(\e_2) e^{i\e_2\cdot\x_2}\bigg \rangle\\
=&\frac{1}{\I_{10}^2} \int_{\e} P_{\rm L}(\epsilon) \bigg \{ \int_{\x_1,\x_2} W_{10}(\x_1)W_{10}(\x_2) e^{-i\e\cdot(\x_1-\x_2)}Z_1^{\hat{\x}_1}(\e_1)Z_1^{\hat{\x}_2}(\e_2)\bigg \}\,.
\end{split}\label{eq:VaryingLOS}\end{equation}
which results in Eq.~(\ref{eq:dngRMS}).
We now rewrite the three-point contribution in Eq.~(\ref{3ptTermRSD}) for the case of radial RSD
\begin{equation}
\begin{split}
\frac{1}{\I_{22}}& \int_{\hat{\k}_{\ell_1}} \langle|\delta_W(\k_1)|^2 \dng\rangle = \frac{2 (2\ell_1+1)}{\I_{22} \I_{10}} \int_{\hat{\k}_1,\x}e^{-i\k_1\cdot(\x'_1-\x_1)}\langle\delta(\x'_1)\delta(\x_1)\delta(\x_2) \rangle W_{11}(\x'_1) W_{11}(\x_1)W_{10}(\x_2)\L_{\ell_1}(\hat{\x}_1\cdot\hat{\k}_1)\\
=&\frac{2(2\ell_1+1)}{\I_{22} \I_{10}} \int_{\hat{\k}_1,\x_1,\x'_1,\x_2,\q_1,..,\q_2}e^{-i\k_1\cdot(\x'_1-\x_1)} e^{i[\q_1\cdot\x'_1+\q_2\cdot\x_2+(\q'_1+\q''_1)\cdot\x_1]} W_{11}(\x'_1) W_{11}(\x_1)W_{10}(\x_2)\\
&\times \langle\delta_L(\q_1)\delta_L(\q'_1)\delta_L(\q''_1)\delta_L(\q_2) \rangle\L_{\ell_1}(\hat{\x}_1\cdot\hat{\k}_1)
\end{split}
\end{equation}
We use Gaussian pairings in the connected six-point function and substitute a variable $\e$ for the beat mode
($\q'_1=-\q_1$, $\q_2=-\q''_1 \equiv \e$) to get 

\beq\begin{split}
4(2\ell_1+1)& \int_{\hat{\k}_1,\hat{\k}_2,\x,\p_1,\p'_1,\e,\q_1,\q_2} e^{-i\k_1\cdot(\x'_1-\x_1)} e^{i[\p_1\cdot\x'_1+\p'_1\cdot\x_1]} e^{i(\q_1\cdot\x'_1+\e\cdot\x_2)}e^{-i(\e+\q_1)\cdot\x_1} W_{11}(\p_1) W_{11}(\p'_1)W_{10}(\x_2)\\
&\times P_{\rm L}(\e) P_{\rm L}(\q_1) Z_1^{\hat{\x}'_1}(\q_1) Z_2^{\hat{\x}_1}(\q_1,\e) \L_{\ell_1}(\hat{\x}_1\cdot\hat{\k}_1)Z_1^{\hat{\x}_2}(\e)\, ,
\end{split}\eeq
Because the LOS does not change over scales of mode $k_1$ that we are interested in, $Z_i^{\hat{\x}'_1}\rightarrow Z_i^{\hat{\x}_1}$ (c.f Fig.~\ref{fig:VaryingLOS}), we can integrate over the $\x'_1$ space and use the resulting $\Ddel(-\k_1+\p_1+\q_1)$ function to integrate over the $\q_1$ space as
\beq\begin{split}
&\int_{\e,\x_2} P_{\rm L}(\epsilon)e^{i\e\cdot\x_2} W_{10}(\x_2)  Z_1^{\hat{\x}_2}(\e)\\
&\times \bigg\{4(2\ell_1+1)\int_{\hat{\k}_1,\x_1,\p_1,\p'_1} e^{i(\p_1+\p'_1-\e)\cdot\x_1} W_{11}(\p_1)W_{11}(\p'_1) P_{\rm L}(\k_1-\p_1) Z_1^{\hat{\x}_1}(\k_1-\p_1) Z_2^{\hat{\x}_1}(\k_1-\p_1,\e)\L_{\ell_1}(\hat{\x}_1\cdot\hat{\k}_1)\bigg\}\, .
\end{split}\eeq
We have already simplified the term present in the curly brackets in Eq.~(\ref{eq:d11}) and we get
\begin{equation}
\begin{split}
&\frac{1}{\I_{22}} \int_{\hat{\k}_{\ell_1}} \langle|\delta_W(\k_1)|^2 \dng\rangle =\frac{P_{\rm L}(k_1)}{\I_{10}\I_{22}} \int_{\e,\x_1,\x_2}P_{\rm L}(\epsilon) W_{22}(\x_1)W_{10}(\x_2)e^{-i\e\cdot(\x_1-\x_2)}Z^{\hat{\x}_2}_1(\e) \mathcal{Z}_{21}(k_1,\ell_1,\hat{\e}\cdot\hat{\x_1})
\end{split}
\end{equation}
which results in Eq.~(\ref{eq:LOS_BC2}).
\section{Analytic analysis of true and FKP shot noise contributions to the covariance}
\label{apx:FKP_SN_analytic}
\begin{figure}
\centering
\includegraphics[scale=0.5,keepaspectratio=true]{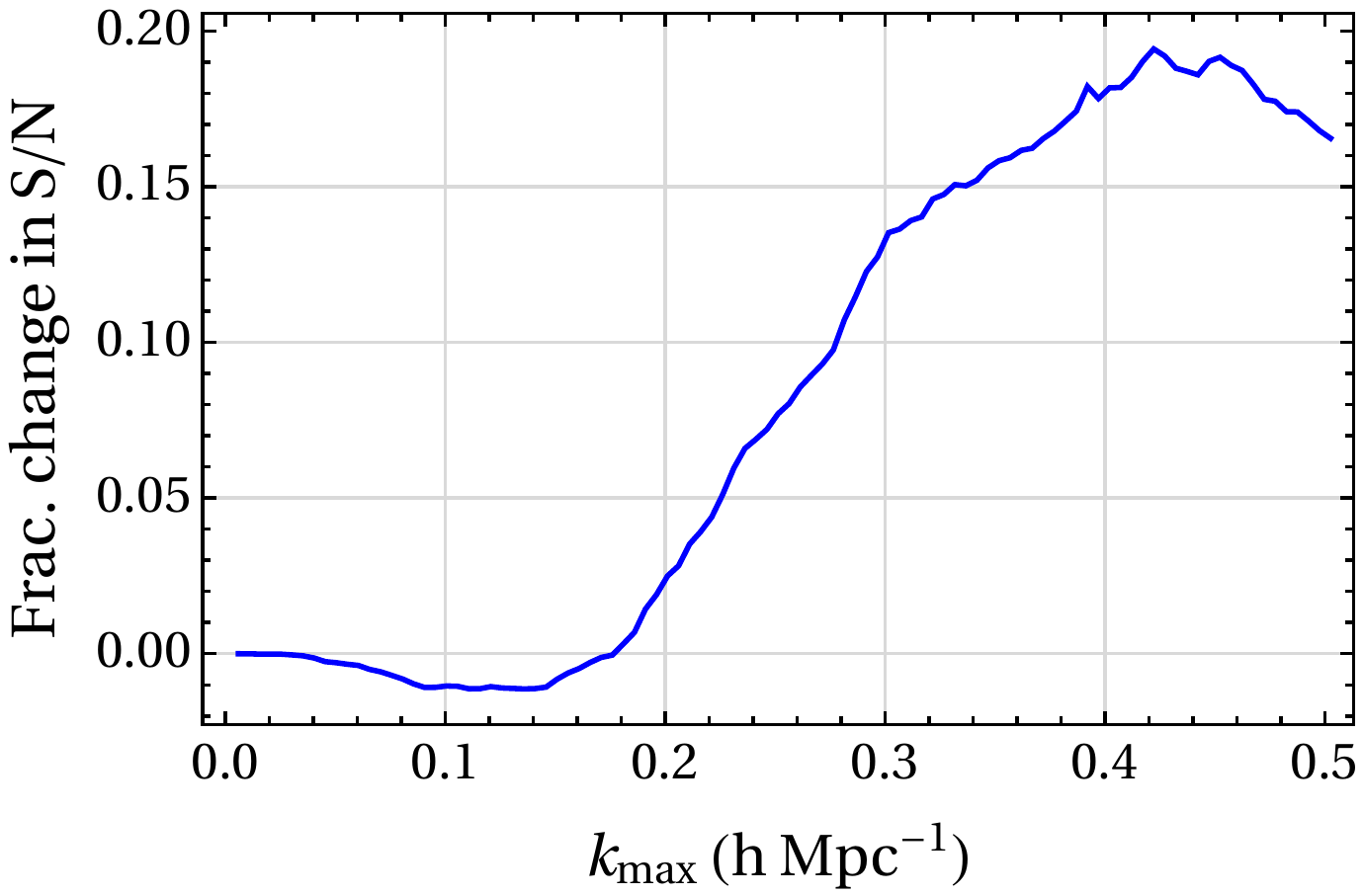}
\caption{Fractional change in the cumulative signal-to-noise ratio S/N of the monopole power on using the FKP SN estimator relative to using the True SN estimator (see  Eq.~\ref{TrueToFKPsn}).}
\label{fig:TrueSN_S_N}
\end{figure}

Sec.~\ref{sec:TrueSN} presented the comparison of the covariance for the cases of using True SN and FKP SN in the power spectrum estimator. In this section, we derive the difference in the analytic covariance on using FKP and True SN in the power estimator. If we use the FKP SN instead of true SN, the power spectrum estimator changes from Eq. (\ref{eq:4.1}) to

\begin{equation}
\begin{split}
\hat{P}^\textup{FKP-SN}(\k) \equiv& \frac{1}{\alpha\, \I^r_{22}} \bigg[\bigg( \sum_{i}^{\ng} -\alpha \sum_{i}^{\nr} \bigg)\bigg( \sum_{i'}^{\ng}-\alpha \sum_{i'}^{\nr} \bigg) w_i w_{i'}e^{-i \k\cdot(\x_i-\x_{i'})}- \bigg( \alpha\sum_{j}^{\nr} +\alpha^2 \sum_{j}^{\nr} \bigg) w^2_j\bigg]\\
 =&\frac{1}{\I_{22}(1+\dng^d)} \bigg[\bigg( \sum_{i}^{\ng} -\alpha \sum_{i}^{\nr} \bigg)\bigg( \sum_{i'(\neq i)}^{\ng} -\alpha \sum_{i'(\neq i)}^{\nr} \bigg) w_i w_{i'}e^{-i \k\cdot(\x_i-\x_{i'})}+\bigg(\sum_{j}^{\ng}- \alpha\sum_{j}^{\nr} \bigg) w^2_j\bigg]\\
  =&\frac{1}{\I_{22}(1+\dng^d)} \bigg[\bigg( \sum_{i}^{\ng} -\alpha \sum_{i}^{\nr} \bigg)\bigg( \sum_{i'(\neq i)}^{\ng} -\alpha \sum_{i'(\neq i)}^{\nr} \bigg) w_i w_{i'}e^{-i \k\cdot(\x_i-\x_{i'})}+\bigg(\sum_{j}^{\ng}- \bar{\alpha}\sum_{j}^{\nr} \bigg) w^2_j - (\alpha-\bar{\alpha})\I^r_{12}\bigg]\\
 \equiv& \frac{\hat{P}^d(\k)}{\I_{22}(1+\dng^d)}+\frac{\I_{12}}{\I_{22}} \frac{\d12^d-\dng^d}{1+\dng^d}\, ,
\end{split}\end{equation}
where $\d12^d \equiv (\sum_i^{\ng}-\bar{\alpha} \sum_{i}^{\nr})w^2_i/\I_{12}$ (which is a discrete version of $\d12 \equiv \frac{1}{\I_{12}}\int_{\x} W_{12}(\x) \delta(\x)$) and the rest of the notations are the same as in Sec. \ref{sec:TrueSN}.
The expectation value of the estimator becomes 
\beq
\langle\hat{P}^\textup{FKP-SN}(\k)\rangle \simeq  \bigg\langle\frac{\hat{P}^d(\k)}{\I_{22}(1+\dng^d)} \bigg\rangle - \frac{\I_{12}}{\I_{22}} (\langle \d12^d\dng^d\rangle-\langle(\dng^d)^2\rangle)\, .
\eeq
The corresponding covariance is given by 
\beq\begin{split}
\textbf{C}^\textup{FKP-SN}(\k_1,\k_2)=&\bigg\langle\bigg(\frac{\hat{P}^d(\k_1)}{\I_{22}(1+\dng^d)}+\frac{\I_{12}}{\I_{22}} \frac{\d12^d-\dng^d}{1+\dng^d}\bigg)\bigg(\frac{\hat{P}^d(\k_2)}{\I_{22}(1+\dng^d)}+\frac{\I_{12}}{\I_{22}} \frac{\d12^d-\dng^d}{1+\dng^d}\bigg)\bigg\rangle\\
&-\bigg\langle\bigg(\frac{\hat{P}^d(\k_1)}{\I_{22}(1+\dng^d)}+\frac{\I_{12}}{\I_{22}} \frac{\d12^d-\dng^d}{1+\dng^d}\bigg)\bigg\rangle\bigg\langle\bigg(\frac{\hat{P}^d(\k_2)}{\I_{22}(1+\dng^d)}+\frac{\I_{12}}{\I_{22}} \frac{\d12^d-\dng^d}{1+\dng^d}\bigg)\bigg\rangle\\
\simeq&\bigg\langle\frac{\hat{P}^d(\k_1)}{\I_{22}(1+\dng^d)}\frac{\hat{P}^d(\k_2)}{\I_{22}(1+\dng^d)}\bigg\rangle-\bigg\langle\frac{\hat{P}^d(\k_1)}{\I_{22}(1+\dng^d)}\bigg\rangle\bigg\langle\frac{\hat{P}^d(\k_2)}{\I_{22}(1+\dng^d)}\bigg\rangle\\
&+ \frac{\I_{12}}{\I^2_{22}}\bigg\langle\frac{(\d12^d-\dng^d)(\hat{P}^d(\k_1)+\hat{P}^d(\k_2))}{(1+\dng^d)^2}\bigg\rangle+\frac{\I_{12}^2}{\I_{22}^2}\bigg\langle \frac{(\d12^d-\dng^d)^2}{(1+\dng^d)^2}\bigg\rangle\\
&-\frac{\I_{12}}{\I^2_{22}}\bigg\langle\frac{\d12^d-\dng^d}{1+\dng^d}\bigg\rangle\bigg\langle\frac{\hat{P}^d(\k_1)+\hat{P}^d(\k_2)}{1+\dng^d}\bigg\rangle-\frac{\I_{12}^2}{\I_{22}^2} (\langle \d12^d\dng^d\rangle-\langle(\dng^d)^2\rangle)^2
\end{split}\eeq
The difference between the covariance on using the true SN versus the FKP SN in the power estimator becomes
\beq\begin{split}
&\textbf{C}^\textup{FKP-SN}(\k_1,\k_2)-\textbf{C}^\textup{True-SN}(\k_1,\k_2)\\
&\simeq \frac{\I_{12}}{\I^2_{22}}\bigg\langle(\d12^d-\dng^d)(\hat{P}^d(\k_1)+\hat{P}^d(\k_2))(1-2\dng^d+3(\dng^d)^2)\bigg\rangle+\frac{\I_{12}^2}{\I_{22}^2}\langle (\d12^d-\dng^d)^2\rangle+\frac{\I_{12}}{\I^2_{22}}\langle\hat{P}^d(\k_1)+\hat{P}^d(\k_2)\rangle\langle \d12^d\dng^d-(\dng^d)^2\rangle\\
&\simeq \frac{\I_{12}}{\I_{22}}\bigg(\frac{1}{\I_{22}}\langle(\d12^d-\dng^d)(\hat{P}^d(\k_1)+\hat{P}^d(\k_2))\rangle 
-(Z_1^2(\k_1) P(k_1)+Z_1^2(\k_2) P(k_2))\langle\d12^d\dng^d -(\dng^d)^2\rangle\bigg)\\
&\quad+\frac{\I_{12}^2}{\I_{22}^2}\langle (\d12^d)^2+(\d12^d)^2- 2( \d12^d \dng^d)\rangle\, ,
\end{split}\label{eq:C_FKPvsTrueSN}\eeq
where there are two point and three point terms in the final expression. The calculations of all these terms are similar to those performed earlier in Sec.~\ref{sec:TrueSN}. The two point terms are the
various background fluctuation terms which can be calculated similar to Eq. (\ref{eq:4.2}) as

\begin{equation}
\begin{split}
\langle (\d12^d)^2 \rangle &=(1+\bar{\alpha})\frac{\I_{14}}{\I^2_{12}}+\frac{1}{\I^2_{12}}\int_{\e} |W_{12}(\e)|^2 Z_1^2(\e)P(\epsilon)\\
\langle \d12^d \dng^d \rangle &=\frac{1+\bar{\alpha}}{\I_{10}}+\frac{1}{\I_{12}\I_{10}}\int_{\e} W_{12}(-\e)W_{10}(\e) Z_1^2(\e)P(\epsilon)\, .
\end{split}
\end{equation}
The three point terms can be calculated similar to Eq. (\ref{eq:SNresponse}) as
\begin{equation}
\begin{split} \langle \hat{P}^d(\k_1) \d12^d\rangle \simeq \,  2\frac{\I_{24}}{\I_{12}}Z_1^2(\k_1) P(k_1)+ \langle |\delta_W(\k_1)|^2\d12\rangle\, ,
\end{split}\end{equation}
 where the continuous squeezed bispectrum are similar to Eq. (\ref{eq:3-pointSolnRSD}) as
\beq
\int_{\hat{\k}_{\ell_1}} \langle|\delta_W(\k_1)|^2 \d12\rangle = \frac{P(k_1)}{\I_{12}\I_{22}} \int_{\e}W_{12}(-\e)W_{22}(\e)\left(P(\epsilon)Z_1(\e) \mathcal{Z}_{21}(k_1,\ell_1,\hat{\e}\cdot\hat{\textbf{n}})+P(k_1)b_2\left[\int_{\hat{\k}_{\ell_1}}Z_1^2(\k_1)\right]\right)\, .
\eeq
Substituting the expansions of the two point and three point terms in Eq.~(\ref{eq:C_FKPvsTrueSN}), we have checked that our analytic estimate of the difference in the covariance is consistent with the results from Patchy mocks. To gauge how the true and FKP SN estimators affect the information content of the power spectrum monopole, we show the fractional change in the cumulative signal-to-noise in Fig.~\ref{fig:TrueSN_S_N}.

\section{Predicting the error in a covariance matrix obtained from a finite mock sample}
\label{apx:CovaError}
In Sec. \ref{sec:Patchy_compare} of the main text, we used bootstrapping to find the error in the covariance from a finite sample of Patchy mocks. We perform an analytic calculation of the error on the covariance in this section and confirm that the errors predicted by bootstrapping are consistent with analytic estimates.We do not discuss the errors associated with inverting a finite sample covariance matrix for parameter estimation \cite{HarSimSch0703,DodSch1309,HalTay1902} in this section. For a sample of $\textup{N}_\textup{m}$ mocks, the estimator for covariance is
\begin{equation}
\hat{\textbf{C}}(k_i,k_j) \equiv \frac{1}{\textup{N}_\textup{m}-1} \left[\sum_n^{\textup{N}_\textup{m}}\, [P^{(n)}(k_i)- \bar{P}(k_i)] [P^{(n)}(k_j)- \bar{P}(k_{j})] \right]\, ,
\end{equation}
where the sample mean power spectrum is given by $\bar{P}(k_i)=\sum_n^{\textup{N}_\textup{m}} P^{(n)}(k_i)$.
For the purpose of simplicity, we assume in this section that the number of mocks is large: $\textup{N}_\textup{m}-1 \simeq \textup{N}_\textup{m}$ and $\bar{P}(k)$ is approximately equal to the true power spectrum $P(k)$. The expectation value of the estimator gives the true covariance as
\begin{equation}
\langle \hat{\textbf{C}}(k_i,k_j) \rangle =\frac{1}{\textup{N}_\textup{m}} \left[\sum_n^{\textup{N}_\textup{m}} \Big\langle [P^{(n)}(k_i)- P(k_i)] [P^{(n)}(k_j)- P(k_{j})] \Big\rangle \right] =\frac{1}{\textup{N}_\textup{m}} \left[\sum_n^{\textup{N}_\textup{m}} \textbf{C}(k_i,k_j) \right]= \textbf{C}(k_i,k_j)\, .
\end{equation}

We want to find the error on elements of the covariance matrix and begin by expanding the term

\begin{equation}
\begin{split}
\langle \hat{\textbf{C}}(k_i,k_j)\hat{\textbf{C}}(k_{i},k_{j}) \rangle=\frac{1}{\textup{N}_\textup{m}^2} \bigg\langle \sum_{n,n'}^{\textup{N}_\textup{m}}\, [P^{(n)}(k_i)- P(k_i)] [P^{(n)}(k_j)- P(k_{j})]  [P^{(n')}(k_{i})- P(k_{i})] [P^{(n')}(k_{j})- P(k_{j})]  \bigg\rangle\, .
\end{split}
\end{equation}
We split the error in the covariance $\big[(\Delta \textbf{C}(k_i,k_j))^2 = \langle \hat{\textbf{C}}(k_i,k_j)\hat{\textbf{C}}(k_{i},k_{j}) \rangle-\langle \hat{\textbf{C}}(k_{i},k_{j})\rangle \langle \hat{\textbf{C}}(k_{i},k_{j}) \rangle\big]$ into a disconnected and a connected part as
\begin{equation}
\begin{split}
(\Delta \textbf{C}(k_i,k_j))^2_\textup{disc} =& \frac{1}{\textup{N}_\textup{m}^2} \left[\sum_{n,n'}^{\textup{N}_\textup{m}} \Big\langle [P^{(n)}(k_i)- P(k_i)] [P^{(n')}(k_{i})- P(k_{i})]\Big\rangle \Big\langle [P^{(n)}(k_j)- P(k_{j})]  [P^{(n')}(k_{j})- P(k_{j})] \Big\rangle \right]\\
&+ \frac{1}{\textup{N}_\textup{m}^2} \left[\sum_{n,n'}^{\textup{N}_\textup{m}} \Big\langle [P^{(n)}(k_i)- P(k_i)] [P^{(n')}(k_{j})- P(k_{j})]\Big\rangle \Big\langle [P^{(n)}(k_j)- P(k_{j})]  [P^{(n')}(k_{i})- P(k_{i})] \Big\rangle \right]\, ,\\
(\Delta \textbf{C}(k_i,k_j))^2_\textup{conn} =& (\Delta \textbf{C}(k_i,k_j))^2- (\Delta \textbf{C}(k_i,k_j))^2_\textup{disc}\, .
\end{split}\label{eq:disc+conn}
\end{equation}
Because different mock samples are uncorrelated with each other, $\Big\langle [P^{(n)}(k_i)-P(k_i)] [P^{(n')}(k_{j})-P(k_j)] \Big\rangle=  \delta^{\textup{K}}_{n,n'} \textbf{C}(k_i,k_j)$, and we get

\begin{equation}
\begin{split}
(\Delta \textbf{C}(k_i,k_j))^2_\textup{disc} &=\frac{1}{\textup{N}_\textup{m}^2} \sum_{n}^{\textup{N}_\textup{m}}\Big[\textbf{C}^2(k_i,k_{j}) + \textbf{C}(k_i,k_i)\textbf{C}(k_j,k_j)\Big] = \frac{1}{\textup{N}_\textup{m}}\Big[\textbf{C}^2(k_i,k_{j}) + \textbf{C}(k_i,k_i)\textbf{C}(k_j,k_j)\Big]\, .
\end{split}
\end{equation}
We devote the rest of the section to finding errors in only the diagonal elements of the covariance ($i=j$) because the diagonals are the most important elements for estimation of cosmological parameters. We first calculate the disconnected part of the error in covariance
\begin{equation}
\sqrt{\frac{(\Delta \textbf{C}(k_i,k_i))^2_\textup{disc}}{(\textbf{C}(k_i,k_i))^2}} = \sqrt{\frac{2}{\textup{N}_\textup{m}}}\, ,
\label{eq:Cov-disconn}\end{equation}
which is equivalent to the standard expression for the error in a Gaussian process. The error is $\sim 3.12$\% for the case of 2048 Patchy mocks and agrees with the width of the error bars obtained using  bootstrapping in Fig. (\ref{fig:Ckk}).  
Let us now return to the connected term in Eq. (\ref{eq:disc+conn}) and show that the connected term is negligible as compared to the disconnected term. We only show the calculation for covariance of the monopole power spectrum but the calculation is similar for the covariance of higher-order multipoles. This is because the covariance of higher-order multipoles shows a similar behavior as the monopole covariance as both are dominated by shot noise at high-$k$ (see Fig. \ref{fig:SN_contrib}). We split the connected term into Gaussian (G-conn) and non-Gaussian (NG-conn) parts
\beq
(\Delta \textbf{C}(k_i,k_i))^2_\textup{conn}\equiv (\Delta \textbf{C}(k_i,k_i))^2_\textup{G-conn}+(\Delta \textbf{C}(k_i,k_i))^2_\textup{NG-conn}\, ,
\eeq
where the individual parts are calculated as
\begin{equation}
\begin{split}
(\Delta \textbf{C}(k_i,k_i))^2_\textup{G-conn}= &\frac{48}{\textup{N}_\textup{m}}\int_{\hat{\k}_1,\hat{\k}_2,\hat{\k}_{1}',\hat{\k}_{2}'}
\langle \delta(-\k_1)\delta(\k_2)\rangle \langle \delta(-\k_2) \delta(\k'_1)\rangle\langle\delta(-\k'_1)\delta(\k'_2)\rangle\langle\delta(-\k'_2) \delta(\k_1) \rangle\\
(\Delta \textbf{C}(k_i,k_i))^2_\textup{NG-conn}=&\frac{1}{\textup{N}_\textup{m}}\int_{\hat{\k}_1,\hat{\k}_2,\hat{\k}_{1}',\hat{\k}_{2}'} \langle \delta(\k_1)\delta(-\k_1)\delta(\k_2)\delta(-\k_2)\delta(\k'_1)\delta(-\k'_1)\delta(\k'_2)\delta(-\k'_2) \rangle_c \, ,
\end{split}\label{eq:Cov-conn}
\end{equation}
where $\hat{\k}_1\, ,\hat{\k}_{1}'\, ,\hat{\k}_2\, ,\hat{\k}_{2}'$ are all integrals over the shell $k_i$.
To make the results of the following part of this Appendix straightforward, we forego a refined calculation of the window functions as was presented in the main text Sec.~\ref{sec:GaussCova} and we now approximate the window as a dirac delta function $(W(\k_1+\k_2)\simeq \Ddel(\k_1+\k_2))$. We can then write the shell-averaged two point functions as a sum over discrete $k$-modes as
\beq
\int_{\hat{\k}_1,\hat{\k}_2} \langle \delta(\k_1)\delta(-\k_2)\rangle\simeq \frac{1}{\textup{N}^2(k_i)}\sum_{\hat{\k}_1,\hat{\k}_2} \langle \delta(\k_1)\delta(-\k_2)\rangle= \frac{1}{\textup{N}^2(k_i)}\sum_{\hat{\k}_1,\hat{\k}_2} \delta_\textup{K} (\k_1-\k_2) \bigg(Z^2_1(\k_1)P(\k_1)+\frac{1}{\bar{n}}\bigg) \simeq \frac{1}{\textup{N}(k_i)} \bigg(P_0(k_i)+\frac{1}{\bar{n}}\bigg)
\label{eq:Cov-conn2}\eeq
where the number of independent $k$-modes in a shell of width $\Delta k$ for a survey with volume $V_s$ is given by $\textup{N}(k)= \frac{4 \pi k^2 \Delta k}{(2\pi)^3}\, V_s$ \cite{GriSanSal1604}. We have also denoted the 3D kronecker delta function as $\delta_{\textup{K}}$ and the linear theory power spectrum monopole as $P_0(k) (\equiv (b_1^2+2b_1 f/3 + f^2/5)P(k))$. Using Eq. (\ref{eq:Cov-conn2}), the diagonal elements in the covariance become $\textbf{C}(k_i,k_i)=2\int_{\hat{\k}_1,\hat{\k}_2} \langle \delta(\k_1)\delta(-\k_2)\rangle\langle\delta(\k_2)\delta(-\k_1) \rangle\simeq \frac{2}{\textup{N}(k_i)}\left(P_0(k_i) + \frac{1}{\bar{n}}\right)^2$ and similarly $(\Delta \textbf{C}(k_i,k_i))^2_\textup{G-conn}$ from Eq. (\ref{eq:Cov-conn}) becomes $\frac{48}{\textup{N}_\textup{m}} \frac{1}{\textup{N}^3(k)} \left(P_0(k) + \frac{1}{\bar{n}}\right)^3$.
Using Eq. (\ref{eq:Cov-disconn}), we estimate the ratio:
\beq
\frac{(\Delta \textbf{C}(k_i,k_i))^2_\textup{G-conn}}{(\Delta \textbf{C}(k_i,k_i))^2_\textup{disc}} \sim \frac{12}{\textup{N}(k_i)}\, ,
\eeq
whose value is $\ll 1$ and decays as $k^{-2}$ at high-$k$.
We finally consider the $(\Delta\textbf{C}(k_i,k_i))^2_\textup{NG-conn}$ term which is independent of $\textup{N}(k)$ and is therefore expected to dominate at high-$k$ over both $(\Delta \textbf{C}(k_i,k_i))^2_\textup{G-conn}$ and $(\Delta \textbf{C}(k_i,k_i))^2_\textup{disc}$ because these scale as $\textup{N}(k)^{-2}$ and $\textup{N}(k)^{-3}$ respectively. For $k_i$ in the high-$k$ regime, we are dominated by shot noise: $\bar{n}P_0(k_i) \ll 1$. We therefore need only compute the lowest order shot noise term in $(\Delta \textbf{C}(k_i,k_i))^2_\textup{NG-conn}$, which is shown in following diagram 

\begin{figure}[h]
\includegraphics[scale=0.55,keepaspectratio=true]{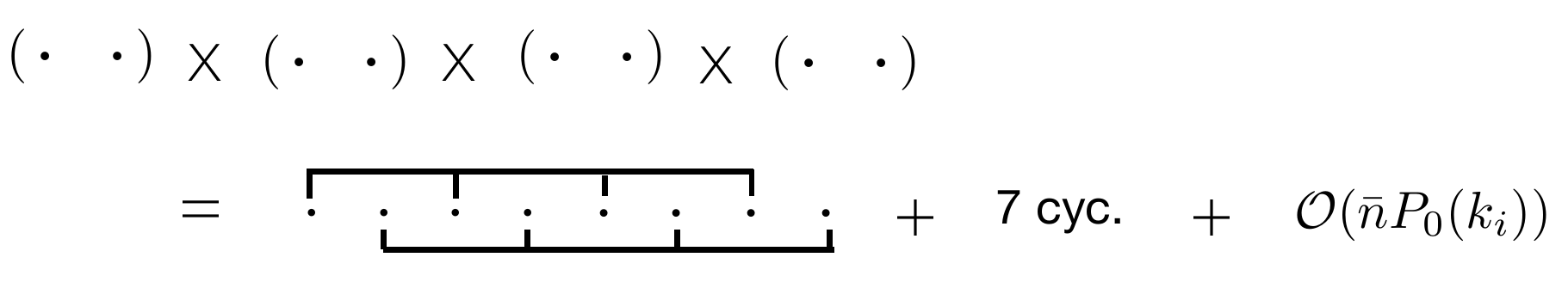}
\label{fig:8ptSN}
\end{figure}
where all the remaining terms are suppressed by factors of ($\bar{n} P_0(k_i)$).
The calculation of the term corresponding to the above diagram is performed similar to Eq. (\ref{eq:SN3}) as
\begin{equation}
\begin{split}
(\Delta& \textbf{C}(k_i,k_i))^2_\textup{NG-conn}\\
\simeq& \frac{8}{\I_{22}^2} \int_{\hat{\k}_1,\hat{\k}_2,\hat{\k}_{1}',\hat{\k}_{2}'} \int_{\x_1, \x_2} W_{14}(\x_1) W_{14}(\x_2) \langle \delta(\x_1)\delta(\x_2) \rangle_c\, e^{-i (\k_1+\k_2+\k'_1+\k'_2)\cdot (\x_1-\x_2)}+\mathcal{O}(\bar{n} P_0(k_i))\\ 
\simeq& \frac{8}{\textup{N}_\textup{m}}\frac{\I_{28}}{\I^4_{22}}\int_{\hat{\k}_1,\hat{\k}_2,\hat{\k}_{1}',\hat{\k}_{2}'} Z^2_1(\k_1+\k_2+\k'_1+\k'_2) P_{\rm L}(\k_1+\k_2+\k'_1+\k'_2)\\
=& \frac{8}{\textup{N}_\textup{m}}\frac{\I_{28}}{\I^4_{22}} \left(1+\frac{2\beta^2}{3}+\frac{\beta^4}{5} \right)\int_{\hat{\k}_1,\hat{\k}_2,\hat{\k}_{1}',\hat{\k}_{2}'}P_{\rm L}(\k_1+\k_2+\k'_1+\k'_2)\\
 \ll& \frac{8}{\textup{N}_\textup{m}} \frac{\I_{28}}{\I^4_{22}} \left(1+\frac{2\beta^2}{3}+\frac{\beta^4}{5} \right)P_\textup{max}\, ,
\end{split}
\end{equation}
where $P_\textup{max}$ is maximum value of the linear power spectrum. 
In the high-$k$ regime, $(\Delta \textbf{C}(k_i,k_i))^2_\textup{disc}\simeq \frac{8}{\textup{N}_\textup{m} \textup{N}(k_i)}\left(P_0(k_i) + \frac{1}{\bar{n}}\right)^2\sim 8/(\textup{N}_\textup{m}\, \textup{N}^2(k_i)\, \bar{n}^4)$. We get the following order-of-magnitude estimate
\beq
\frac{(\Delta \textbf{C}(k_i,k_i))^2_\textup{NG-conn}}{(\Delta \textbf{C}(k_i,k_i))^2_\textup{disc}} \ll \left(\frac{P_\textup{max}}{V_s^3 \bar{n}^6}\right)\Big/\left(\frac{1}{\textup{N}^2(k_i) \bar{n}^4}\right) \sim \left(\frac{k}{10\ \textup{h}/\textup{Mpc}}\right)^4\, 
\eeq
for parameter values corresponding to the high-$z$ bin of BOSS DR12 NGC survey.
Thus we see that for the modes $k_i$ relevant for current and upcoming redshift-space surveys, $(\Delta \textbf{C}(k_i,k_i))^2_\textup{conn}\ll (\Delta \textbf{C}(k_i,k_i))^2_\textup{disc}$ and therefore the relative error on the diagonal elements of the covariance matrix is to a good approximation: $|\Delta \textbf{C}(k_i,k_i)/\textbf{C}(k_i,k_i)| \simeq \sqrt{2/\textup{N}_\textup{m}}$.


\section{Comparison with low-$z$ bin in Patchy mock catalogs}
\label{apx:LowZ_comparison}
\begin{figure}
    \centering
        \includegraphics[scale=0.5,keepaspectratio=true]{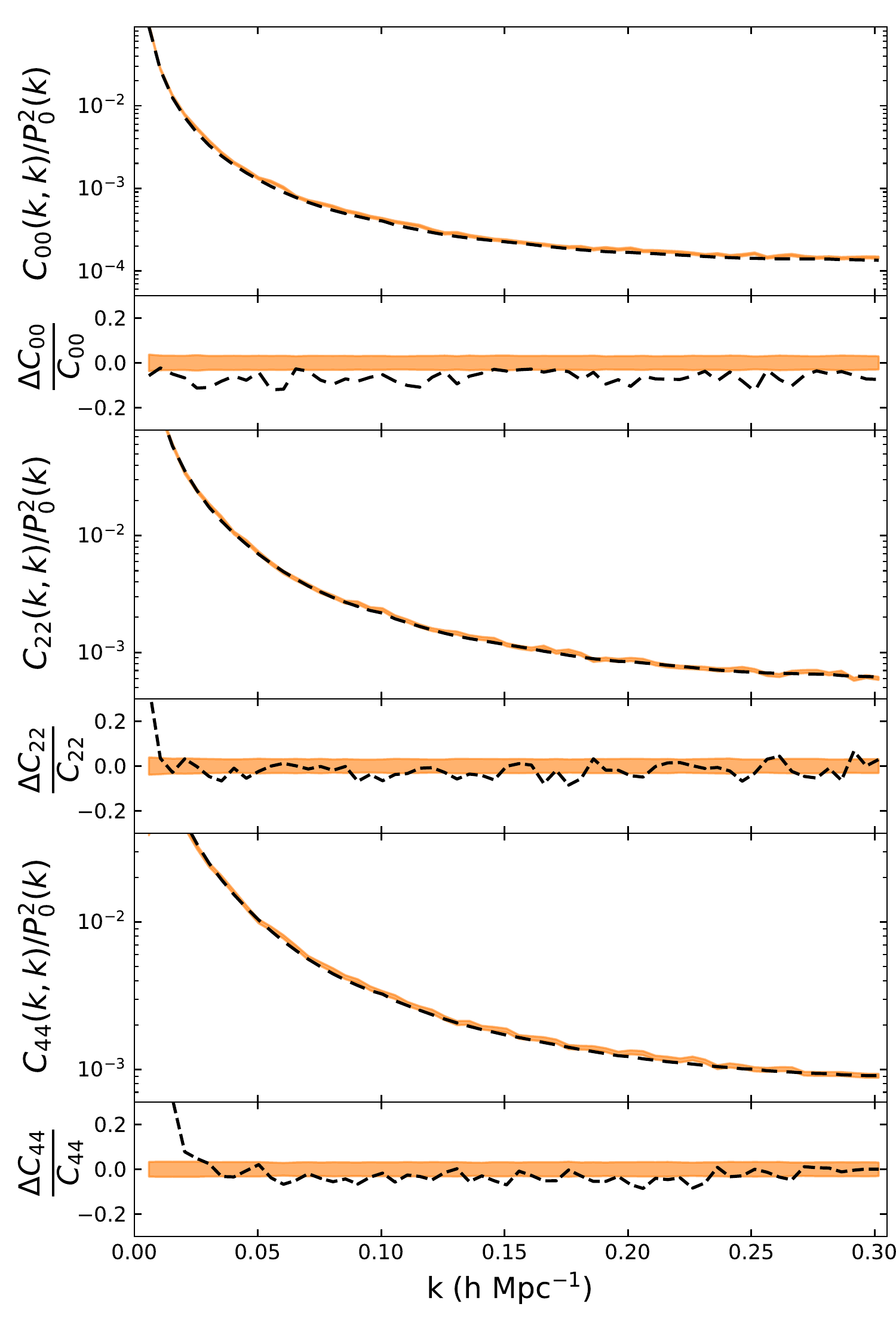}
        \includegraphics[scale=0.5,keepaspectratio=true]{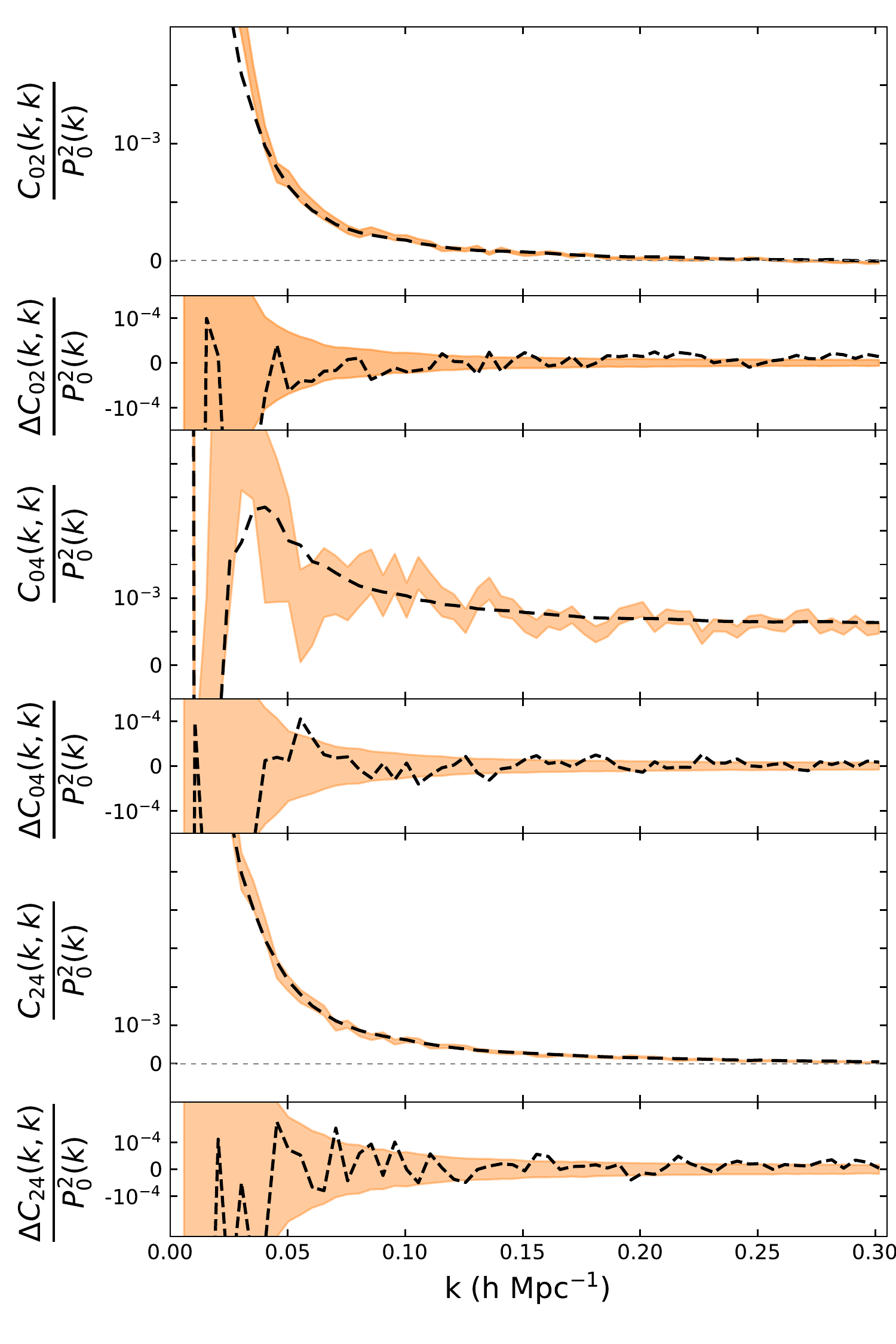}
    \caption{Same as Fig. \ref{fig:Ckk} but for the $0.2< z < 0.5$ bin instead of the $0.5< z < 0.75$ bin. 
    }
    \label{fig:Ckk_LowZ}
\end{figure}

We showed the comparison of our analytic method to the covariance measured from the Patchy mocks for the high redshift bin ($0.5 < z < 0.75$) in Sec.~\ref{sec:Patchy_compare}. W e now do the same for the low-$z$ bin ($0.2 < z < 0.5$). This is interesting because the low-$z$ bin is less shot-noise dominated. Indeed, this has $\bar{n} P=1$ at $k=0.3 \kMpc$, while for the high-$z$ bin results presented in the main text $\bar{n} P=1$ at $k=0.2 \kMpc$. Fig.~\ref{fig:Ckk_LowZ} shows that the level of agreement we get in this case is comparable to that in the high-$z$ bin.

\bibliography{covariance,masterbiblio}

\end{document}